%% file: main.tex
\documentclass[a4paper,fleqn]{cas-sc}

\usepackage{caption}
\usepackage{subcaption}
\usepackage[square,numbers]{natbib}
\usepackage{listings}
\usepackage{censor}
\usepackage[disable]{todonotes}
\usepackage{float}
\usepackage{amsmath} % \checkmark symbol
\usepackage{multicol}
\usepackage{rotating}
\usepackage[commandnameprefix=ifneeded,final]{changes}
\definechangesauthor[name={Reviewer 3 Comment 1}, color=blue]{r3c1}
\definechangesauthor[name={Reviewer 3 Comment 2}, color=blue]{r3c2}
\definechangesauthor[name={Reviewer 3 Comment 3}, color=blue]{r3c3}
\definechangesauthor[name={Reviewer 3 Comment 4}, color=blue]{r3c4}

\usepackage{todonotes}
\newcommand{\com}[2]{}
\newcommand{\censortext}[1]{#1}

\newcommand{\y}[0]{\checkmark}

\def\tsc#1{\csdef{#1}{\textsc{\lowercase{#1}}\xspace}}
\tsc{WGM}
\tsc{QE}
\tsc{EP}
\tsc{PMS}
\tsc{BEC}
\tsc{DE}
\newcounter{reqCounter}
\newcounter{reqRevCounter}[reqCounter]

\DeclareCaptionFormat{table}{\begin{center}#1#2#3\end{center}}
\DeclareCaptionLabelFormat{tablelabel}{\textbf{#1 #2}}

\begin{document}
\let\WriteBookmarks\relax
\def\floatpagepagefraction{1}
\def\textpagefraction{.001}

\shorttitle{Model management for system engineering using ontology and knowledge graphs}

\shortauthors{Ry\'{s}, Lima, Exelmans, Janssens, Vangheluwe}

\title [mode = title]{Model management to support systems engineering workflows using ontology-based knowledge graphs}
%\tnotemark[1,2]

%\tnotetext[1]{This document is the results of the research
%   project funded by the National Science Foundation.}

%\tnotetext[2]{The second title footnote which is a longer text matter
%   to fill through the whole text width and overflow into
%   another line in the footnotes area of the first page.}


\author[1]{Arkadiusz Ry\'{s}}[
    orcid=0000-0002-6225-6330
]
\ead{arkadiusz.rys@uantwerpen.be}
% \cormark[1]
\credit{Conceptualization of this study, Methodology, Software}

\author[1,2]{Lucas Lima}[type=author,
   auid=000,bioid=1,
    % role=Researcher,
   orcid=0000-0003-1859-8437
]
\ead{lucas.albertins@ufrpe.br}
% \cormark[1]
\credit{Conceptualization of this study, Methodology, Software}

\author[1]{Joeri Exelmans}[
    orcid=0000-0002-6916-5140
]
\ead{joeri.exelmans@uantwerpen.be}
\cormark[1] % Corresponding author indication
\credit{Conceptualization of this study, Methodology, Software}

\author[3]{Dennis Janssens}[
    orcid=0000-0003-0549-3775
]
\ead{dennis.janssens@kuleuven.be}
% \cormark[1]
\credit{Case study}

\author[1]{Hans Vangheluwe}[
    orcid=0000-0003-2079-6643
]
\ead{hans.vangheluwe@uantwerpen.be}
% \ead[URL]{http://msdl.uantwerpen.be/people/hv}
%\cormark[1]
\credit{Conceptualization of this study, Methodology}


%\credit{Conceptualization of this study, Methodology, Software}

\affiliation[1]{organization={Flanders Make at University of Antwerp},
   addressline={Middelheimlaan 1}, 
   city={Antwerp},
    citysep={}, % Uncomment if no comma needed between city and postcode
   postcode={2020}, 
   country={Belgium}}

\affiliation[2]{organization={Universidade Federal Rural de Pernambuco},
   addressline={Rua Dom Manuel de Medeiros, s/n, Dois Irmãos}, 
   city={Recife},
   state={PE},
    citysep={}, % Uncomment if no comma needed between city and postcode
   postcode={52171-900}, 
   country={Brazil}}
   
\affiliation[3]{organization={Flanders Make at KU Leuven},
    addressline={Celestijnenlaan 300},
    city={Leuven},
    citysep={},
    postcode={3001},
    country={Belgium}}

%\cortext[cor1]{Principal corresponding author}
%\cortext[cor2]{Corresponding author}

%\fntext[fn1]{This is the first author footnote. but is common to third
%  author as well.}
%\fntext[fn2]{Another author footnote, this is a very long footnote and
%  it should be a really long footnote. But this footnote is not yet
%  sufficiently long enough to make two lines of footnote text.}

%\nonumnote{This note has no numbers. In this work we demonstrate $a_b$
%  the formation Y\_1 of a new type of polariton on the interface
%  between a cuprous oxide slab and a polystyrene micro-sphere placed
%  on the slab.
%  }

\begin{abstract}
System engineering has been shifting from document-centric to model-based approaches, where assets are becoming more and more digital. Although digitisation conveys several benefits, it also brings several concerns (e.g., storage and access) and opportunities. In the context of Cyber-Physical Systems (CPS), we have experts
from various domains executing complex workflows and manipulating models in a plethora of different formalisms, each with their own methods, techniques and tools. Storing knowledge on these workflows can reduce considerable effort during system development not only to allow their repeatability and replicability but also to access and reason on data generated by their execution. In this work, we propose a framework to manage modelling artefacts generated from workflow executions. The basic workflow concepts, related formalisms and artefacts are formally defined in an ontology specified in OML (Ontology Modelling Language). This ontology enables the construction of a knowledge graph that contains system engineering data to which we can apply reasoning. We also developed several tools to support system engineering during the design of workflows, their enactment, and artefact storage, considering versioning, querying and reasoning on the stored data. These tools also hide the complexity of manipulating the knowledge graph directly. Finally, we have applied our proposed framework in a real-world system development scenario of a drivetrain smart sensor system. Results show that our proposal not only helped the system engineer with fundamental difficulties like storage and versioning but also reduced the time needed to access relevant information and new knowledge that can be inferred from the knowledge graph. 
\end{abstract}

% \includegraphics{figs/grabs.pdf}

\begin{highlights}
\item A model management strategy to store and provide access to knowledge about system engineering workflows that uses ontology-based knowledge graphs.
\item An ontology that defines basic concepts and their relationships to support the management of system engineering tasks.
\item A toolchain that provides capabilities for system engineering design, enactment and reasoning on workflows and their related artefacts and data.
\end{highlights}

% Keywords

\maketitle

\input{sections/introduction}
\input{sections/backgroundv2}

\input{sections/framework}
\input{sections/implementation}
\input{sections/casestudy}
\input{sections/relatedworkv2}
\input{sections/conclusion}

\section*{Acknowledgement}
Support for this work by the Flanders Make Strategic Research Centre in the form of the Framework for Systematic Design of Digital Twins (DTDesign) project is gratefully acknowledged.

\printcredits

%\bibliographystyle{cas-model2-names}
\bibliographystyle{elsarticle-num}

\bibliography{cas-refs}

% if necessary, uncomment to show appendix
%\input{sections/appendix}


%Author biography without author photo.
%Author biography. Author biography. Author biography.
%Author biography. Author biography. Author biography.
%Author biography. Author biography. Author biography.
%Author biography. Author biography. Author biography.
%Author biography. Author biography. Author biography.
%Author biography. Author biography. Author biography.
%Author biography. Author biography. Author biography.
%Author biography. Author biography. Author biography.
%Author biography. Author biography. Author biography.

%\bio{figs/pic1}
%Author biography with author photo.
%Author biography. Author biography. Author biography.
%Author biography. Author biography. Author biography.
%Author biography. Author biography. Author biography.
%Author biography. Author biography. Author biography.
%Author biography. Author biography. Author biography.
%Author biography. Author biography. Author biography.
%Author biography. Author biography. Author biography.
%Author biography. Author biography. Author biography.
%Author biography. Author biography. Author biography.

%\bio{figs/pic1}
%Author biography with author photo.
%Author biography. Author biography. Author biography.
%Author biography. Author biography. Author biography.
%Author biography. Author biography. Author biography.
%Author biography. Author biography. Author biography.

\listofchanges[style=<list|summary>]

\end{document}

%% file: sections/introduction.tex
\section{Introduction} %Why

%\begin{itemize}
%\item Motivation: Increasing complexity, heterogeneity, Poor process management, artefact/model storage and versioning, \ldots
%Current solutions are ad hoc and are built on methods, techniques and tools that date back several decades (1980s).
%\textbf{quote what is in 2035 vision report of INCOSE.}

% INCOSE, See page 33:
% https://www.incose.org/docs/default-source/se-vision/incose-se-vision-2035.pdf?sfvrsn=e32063c7_10

%\item Context: PLM, MBSE, Digital Twinning, Knowledge graphs, ontologies, workflows

%\item Overview

%\textbf{Add picture overviewing the strategy}

%\item Contributions

%\item Running Example: Mass-spring-damper system
%\item Structure
%\end{itemize}  

%Example of adding text: \added[id=r3c1]{text added}

%Example of deleting text: \deleted[id=r3c2]{text deleted}.

%Example of replacing text: \replaced[id=r3c3]{old text}{new text}.

Systems engineering processes are becoming increasingly complex. Managing knowledge about processes is crucial for companies. A relevant factor is the heterogeneity of current system development processes. A diverse spectrum of formalisms, languages, and tools are used for different purposes to build better systems. INCOSE's Systems Engineering Vision for 2035 details that knowledge should be considered as a critical asset~\cite{incose}. One of the mentioned challenges in current systems engineering practices is the huge fragmentation of tools and domain-specific approaches to enable collaboration and analysis. The shift from document-centric processes to model-based strategies brings opportunities but also comes with challenges in dealing with all digital information. Therefore, knowledge management is crucial to reaching their vision.
% TODO Add the DES paper as a reference here
% TODO I don't know how to cite: https://man.fas.org/eprint/digeng-2018.pdf

A simple example that several companies face is to keep knowledge of standard procedures or successful cases. Facilitating the repeatability and replicability of workflows is critical to achieving companies' goals. They may lose valuable information due to circumstances like employees changing roles, turnover, and so on. Thus, having a well-formed way to keep this knowledge may reduce costs and effort. Moreover, several models and artefacts are generated during the system development process. Most of the time, several versions of the same documents need to be stored and made available for future reference. These artefacts are generated or required by activities, which in turn, are part of a workflow executed by engineers. Nevertheless, they must be consistent, traceable, stored, and available for different purposes like analysis, testing, simulation, or just as a reference for future activities. 

Experiments play an essential role in systems engineering because they may provide evidence that the system works according to expectations. However, the inherent complexity of executing these procedures brings challenges. For instance, discovering what activities and in which order they should be executed for a given procedure, what artefacts are created during their execution, which choice led to which result (and vice versa), what was the latest ("most correct”) version of a model, and how to find a particular artefact required by a given activity, are some of several possible obstacles that may hinder the development of systems. Therefore, explicitly modelling workflows and their related artefacts, together with a knowledge management mechanism, may optimise the design process.

Product Lifecycle Management (PLM) software records artefact versions and simple relations between them, possibly linked to the explicit enactment of workflows. However, PLM software usually focuses on CAD artefacts, and has limited support for other file formats. Further, the possible links between data are usually ad-hoc and hard-coded. Relations are not captured in a ``meta-model'', and exist only implicitly as ``whatever the code does''. As a result, it is difficult to reason about complex relations between objects that span multiple links. The only way to navigate them, is to write more code, repeating the same patterns. This ultimately makes these systems hard to maintain.
%\com{Arkadiusz}{Ellipsis left in.} % can this comment be deleted?

In~\cite{randy22}, the authors propose a multi-paradigm modelling framework for model-based systems engineering. According to it, formalisms and transformation between them should be specified in a Formalism Transformation Graph (FTG). Workflows are specified in terms of Process Models, and the result of executing them is recorded in a Process Trace, which details events related to workflow activities and artefacts. Although the elements are clearly described, they are not formalised. There is no implementation of the semantics of these workflows, nor concrete mechanisation of the approach. Possible query descriptions are listed, but no concrete implementation is proposed. 

Recently, the industry has shown an increasing interest in Knowledge Graphs (KGs), which are structures to represent data and their relationships for a particular domain. This data can be spread to different sources. Usually, KG data conforms to concepts defined in Ontologies. They provide a means to define properties on elements of a particular domain. 

We take advantage of these concepts to leverage the ideas presented in~\cite{randy22} using an ontology to serve as a formal definition of systems engineering elements. This ontology also allows the construction of a knowledge graph to relate systems engineering process data, which in turn enables several capabilities like versioning, querying and reasoning on this data. Further, the system engineer does not need to worry about the underlying notations and data structure to use these services because we have implemented several tools to support the design and enactment of workflow models, and also to access the information from models, enactments, and artefacts. Finally, we have applied our framework in a real-world context during the development of a drivetrain smart sensor system, where the engineer used our tools to execute experiments during the system development process. Besides the results, while executing these experiments, our tools and methodology also allow these experiments to be repeated and replicated in the future.

% ... contributions:
%\subsection{Contributions}

In summary, our major contributions are the following: 

\begin{itemize}
    \item An extensible ontology for model management of arbitrary artefacts, their internal structure, links between them, and versioning information.

    \item Based on this ontology, an ontology for workflow models, and for traces produced by the enactment thereof. These traces link to the artefacts produced and consumed by the enactment.
    
    % Further, artefact versions can be linked to be produced by the enactment of workflow models (which are also artefacts).
    % An ontology that defines the concepts and relationships to specify workflows and record traces of their enactment. This ontology also registers information about artefacts considering links between them, versioning, and their internal structure. 
    \item A model repository, based on our ontology,
    % that manages knowledge about workflows and their execution using Knowledge Graphs and
    that allows querying, reasoning, and services to be executed on the stored data. 
    \item A modelling environment for the creation of workflows in our custom workflow language. 
    \item An engine for the enactment of these workflows, linked to our repository.
\end{itemize}

The remainder of the paper is structured as follows. Section~\ref{sec:background} presents some background on model management, ontologies and knowledge graphs. Section~\ref{sec:framework} describes requirements for a model management framework using a spring-mass-damper system as the running example. Section~\ref{sec:implementation} explains the architecture and implementation of our framework, and the several tools developed to support model management for systems engineering workflows. Section~\ref{sec:casestudy} shows how we applied our framework in a real-world case for a drivetrain digital twin system. Finally, Section~\ref{sec:relatedwork} discusses related work and Section~\ref{sec:conclusions} gives a summary and presents our final remarks. 

%% file: sections/backgroundv2.tex
\section{Background} % Lucas
\label{sec:background}

In this section, we present topics that are relevant to our proposed framework. First, we discuss in Section~\ref{sec:modelmanagement} the challenges of model management. Section~\ref{sec:kg} describes what knowledge graphs are, their relationships with ontologies, and some standards and languages used to define them.
%Section~\ref{sec:worflow} details the concepts related to workflow modelling.

%\begin{itemize}
%\item Model Management (includes versioning)
%\item Knowledge Graphs and Ontologies 
%\item Workflow Modelling/Management
%\end{itemize}  

\subsection{Model Management}
\label{sec:modelmanagement}

Given the shift from document-centric to model-centric development of systems, the governance over the models is becoming a concern of extreme relevance. Therefore, the availability of tool support to tackle this complexity is crucial. A Model Management System (MMS) comes to provide a means to create, store, manipulate and access models~\cite{APPLEGATE1986}. This term emerged in the early 1980s as an extension to decision support systems to facilitate the use of analytical tools in providing solutions~\cite{Elam1980}.
However, such systems can be used for diverse purposes like model elaboration,
storage, testing, validation, execution, maintenance, integrity, interoperability, abstraction, extraction, and provisions for model security and model sharing. 

Baldwin et al.~\cite{BALDWIN1991} consider three modelling task dimensions: model formulation, model representation and model processing. The first regards understanding the problem and eliciting possible models to help solve it. As our focus is the replicability of experiments, we need an understandable way to represent the steps performed during these experiments together with elements that make artefacts and activities explicit. Therefore, an intuitive language to represent workflows covering both data (artefacts) and control flow is crucial to our goals. 

Model representation refers to the structures, both data and architectural properties, used to represent knowledge relevant to the models. For instance, model parameters, model structure (e.g., relationships between elements), relationships to other models and data. Bharadwaj et al.~\cite{Bharadwaj92} classify model representation into three categories: database approach, graph-based approach, and knowledge-based approach. In the database approach, the models are organised according to a data model hiding from the users the physical details of the model base. Some examples of this approach are the Entity-Relationship (ER) framework~\cite{Chen76} and the Relational model framework~\cite{Codd70}. The graph-based approach represents knowledge in terms of one or more graphs or digraphs. According to Bharadwaj et al.~\cite{Bharadwaj92}, this approach has several advantages, including conceptual clarity, ease of programming and manipulation, and more effective communication between analysts and decision-makers. Some instances of this method include the structured modelling framework~\cite{Geoffrion87}, graph grammars~\cite{Jones90,Jones91}, and logic graphs~\cite{KIMBROUGH198627}. Finally, the knowledge-based approach applies Artificial Intelligence (AI) techniques to model management.
% like semantic networks, first-order predicate calculus and production rules. 
Some advantages of this approach are automatic model formulation, answering user queries with the support of inference methods, and the information about modelling being made explicit and accessible instead of embedding it within an algorithm. 

In a semantic network, objects and relationships between them are represented as nodes and arcs in a directed graph. Inference over the knowledge base can be performed by traversing the edges of the graph. It is a convenient mechanism to express concepts, facts and their relationships. One of the main representatives of this approach is the semantic web~\cite{W3C}, which defines a set of languages and standards to allow the processing of information from the Web by computers (as opposed to most of
the current Web, which is mostly targeted at human consumption) by using machine-understandable formats~\cite{Hitzler21}. This is to be achieved by the usage of metadata incorporated into the web data to allow reasoning over it. We discuss some of these standards in Section~\ref{sec:kg}.

The last modelling dimension is model processing, which discusses the computation of the designed models to deliver an adequate solution. Clearly, this capability is highly dependent on the model formulation and representation because the mechanisms can be triggered for processing according to how the model is created and the used notations. The processing can be related to different purposes, for example, model consistency, property analysis, reasoning capabilities, simulation, and optimisation methods. %From the user perspective, it must be an interactive process where the MMS explains its processing results in a user-friendly language. From the system perspective, the goal is to generate and evaluate a solution to the given model utilising the most appropriate techniques. 

Model management has become more significant nowadays, given the investment in Model-based strategies. Nevertheless, the concern from the 1980s still remains but on a much larger scale. Large amounts of models are being generated and must be managed by MMSs. New functions are emerging to minimise complexity during the system design process. For instance, %pattern identification can be applied to automatically propose models during the early design, 
the usage of available inference tools to reason over model properties, and enabling the traceability between the different elements represented inside the models. 

\added[id=r3c3]{Therefore, having a suitable toolchain is crucial to increase the likelihood of success during the system development process. For instance, Product Lifecycle Management (PLM) software extensively relies on tooling to guide the development of products according to user-defined workflows. In this scenario, model/artefact management plays an important role. It is now a big market where most tools are proprietary. In this scenario, usually, coded extensions must be provided to support new types of artefacts and languages, links between model elements, or reasoning features, which penalises extensibility }

\subsection{Ontologies and Knowledge Graphs}
\label{sec:kg}

Recently, companies considerably increased their interest in semantic applications to use in their specific domains. Model Management has not been different because we need even more sophisticated mechanisms to tackle the evolving complexity and scale of systems. The terms Ontology and Knowledge Graph have been surfacing in the last years as structures to support Model Representation and facilitate Model Processing (dimensions presented earlier in Section~\ref{sec:modelmanagement}). 
%When managing our models, we need sound mechanisms to identify and relate the model elements. Hence, an instrument to classify the possible model elements and their relations is essential. On the other hand, we need a concrete realisation of these elements and their relationships. Ontologies serve the former purpose as they define types of existing things for distinguishing domains and properties to describe them, while Knowledge Graphs can be seen as the materialisation of these elements and relationships. 
In the following, we detail these two topics. 

\subsubsection{Ontologies}

The motivation for the usage of Ontologies came from the necessity of clarifying terms and relationships used in different areas. It is not unusual to have different teams, sometimes in the same company, that have distinct understandings of the same term or use different terms for the same entity. This can lead to lots of confusion and communication overhead. When we think about digitisation, for instance, in automating processes, a uniform understanding is even more important to allow machine-readable structures that can be processed by a computer. 

Ontologies support this necessity by determining an explicit knowledge base of concepts and their relationships. For that, it uses several components like concepts, their relations, attributes, instances and axioms~\cite{GRUBER1993}. Concepts are the types or classes of elements, e.g., "car" and "person". While "is driven by" could be considered a relation between "car" and "person". A car can have an attribute "mass", which specifies the weight of the car. The instance mechanism is a concrete entity of a particular concept, for instance, a car with a given model, car plate and all of its features (attributes) specified. %And axioms (also known as facts) provide properties that are assumed valid for concepts and relations. 

Besides the goal of making the information explicit and improving data interoperability, ontologies are also employed to retrieve additional knowledge from the current base through the use of reasoners. This way, new information can be deduced using logic inference. For instance, "A car X is driven by a person P" + "A car is driven by at most one Driver" => "P is a Driver". Finally, it also allows checking the consistency of partial descriptions of a product from different developers~\cite{Mostefai2006,BOCK2010}. 

During the 2000s, research and applications on ontologies considerably increased due to its pivotal role in the definition of Semantic Web standards~\cite{W3C}. %The meaning of the term "Semantic Web" is quite controversial, and it lacks consensus. Some see it as a field of research~\cite{Hitzler21}, while others regard it as a technological stack. 
%According to W3C's vision, it is the Web of linked data, i.e., a set of data structures (graphs) connected through the Web according to their identifiers and using a machine-readable format to allow its processing. 
%Although these graphs may refer to different domains, they interoperate through the usage of standards. 
In this context, ontologies are a major enabler for data integration, sharing, and discovery. Moreover, they have been envisioned to be reusable, i.e., the same ontology supposedly can be applied in similar environments. %Although the Semantic Web proposes this idea of collaboration between the different networks of data, the technology stack can be used by institutions to internally define how their data is organised and also how to process these data according to their needs. In addition, institutions can decide what parts of their data will be publicly made available.  

\subsubsection{Knowledge Graphs and Semantic Web Standards}

While ontologies provide a way to define concepts and relationships, a \textit{Knowledge Graph} (KG) can be understood as a materialisation of the network of elements that conforms to the ontology definitions. %We consider real data about individuals to build it. 
It is a form of model representation that mixes graph-based and logic-based approaches, as described in Section~\ref{sec:modelmanagement}. The term emerged in 2012 when Google utilised it to describe the searches on notable entities so that relevant information about them is displayed on the result search page. 
%Figure~\ref{fig:tonyhoare} displays what is shown by Google when we search by Tony Hoare. Users also could navigate the nodes of the graph that are connected, e.g., we could access Bill Roscoe's result page because he is listed as one of Hoare's notable students.

%\begin{figure}[!h]
%    \centering
%    \includegraphics[scale=0.2]{figs/tonyHoare.png}
%    \caption{Illustration of Google Knowledge Graph usage.}
%    \label{fig:tonyhoare}
%\end{figure}
After Google, other major companies decided to invest in such a notion, such as Facebook, Microsoft, and IBM~\cite{Noy2019}, and the term has been established in the industry since. However, more than a novel approach, it took well-established research concepts (some mentioned in Section~\ref{sec:modelmanagement}) and articulated them on an industrial scale with different emphases. Among the techniques employed in KGs, \textit{Federation} and \textit{Virtualisation} play relevant roles in boosting this technology. The former provides capabilities for the connection between graphs from different architectures, implementation and data sources. This way, it is possible to navigate these integrated graphs to return a search result. The motivation for Virtual Knowledge Graphs is to maximise the scalability of operations over the graph and improve performance, given that all the data does not need to be materialised, that is, be kept in the original KG notation. Instead, just a conceptual representation of the domain of interest is presented, resulting in a reduced graph. The portions of the virtual data can be materialised on demand. Nevertheless, it brings challenges in integrating the different data sources and the KG, besides assuring the integrity and completeness of all the data~\cite{Xiao2019}. 

The way KGs are structured may also vary. For instance, Google does not disclose its own representation. It provides only an API to access the contents\footnote{https://cloud.google.com/enterprise-knowledge-graph/docs/search-api}. Another alternative is to use off-the-self solutions that provide mechanisms to build your own knowledge graph~\cite{neo4j,ontop}. Although good support is provided by these tools, we are somehow tied to their internal representations and encoded functionalities to manage the KG. Finally, there is the Semantic Web approach standardised by W3C, where graphs are represented using RDF (Resource Description Framework)~\cite{RDF2014}. RDF specifies a syntax to create directed typed graphs structured according to what is called triplestores, which are statements that relate three elements, a subject, a predicate, and an object.
%It is an open and well-defined standard that has been used by a large community.% The Linked Open Data Cloud is an enormous open-access knowledge graph that can serve as a reference for structuring possible data domains\footnote{https://lod-cloud.net/}, and it follows the Semantic Web standards, including RDF. 
%Also, it can be integrated with other standardised languages (e.g., SPARQL for querying and OWL for defining ontologies) to deliver features to the end user.
%\subsubsection{RDF}
%The Resource Description Framework (RDF) is a W3C standard that specifies a syntax to create directed typed graphs~\cite{RDF2014}. These graphs are structured according to what is called RDF triplestores, which are statements that relate three elements, a subject, a predicate, and an object. %Figure~\ref{fig:triplestore} illustrates the structure of an RDF triplestore for the car example previously mentioned. The subject "Car X" is a node of the graph representing some resource, the object "Person P" is also a node in the graph, and the predicate "is driven by" is an arc from the subject to the object to express a relationship between them. All the elements in RDF must be uniquely identified by an IRI (Internationalized Resource Identifier). The main difference between a subject and an object is that the latter can be either a reference to a resource or a literal value.
%\begin{figure}[!h]
%    \centering
%    \includegraphics[scale=0.5]{figs/triplestore.png}
%    \caption{Representation of an RDF triplestore.}
%    \label{fig:triplestore}
%\end{figure}
Although simple, this structure is flexible and generic, with a high level of expressiveness, which is suitable for building graph databases. It has several serialisation formats, e.g., based on XML or JSON. However, the most commonly used one is the Turtle (Terse RDF Triple Language) format. %The RDF/Turtle specification in Listing~\ref{turtle} shows a simplified example of the relationship between "CarX" and "PersonP". Here we use prefixes to improve readability. The subject "CarX" is of type \texttt{<http://example. com/car/Car>}, while "PersonX" is of type \texttt{<http://example.com/people/Person>}. "CarX" has a model "ModelX", which is a string, and it is driven by "PersonX".

%\begin{lstlisting}[language=XML,caption=Turtle Example.,basicstyle=\ttfamily\small,label=turtle]
%@prefix car: <http://example.com/car#> .
%@prefix people: <http://example.com/people#> .
%@prefix rdf:  <http://www.w3.org/1999/02/22-rdf-syntax-ns#> .

%car:CarX rdf:type car:Car ;
%   car:hasModel "ModelX" ;
%   car:isDrivenBy people:PersonX .

%people:PersonX rdf:type people:Person .
%\end{lstlisting}

%The main purpose of these specifications is to have a machine-readable notation to describe how different resources are defined and interact with each other. In addition, it allows integration with different other standards (e.g., OWL and XSD) or different ontologies referenced by the prefixes. 

%\subsubsection{SPARQL}

While RDF supports the specification of elements and their relationships in terms of a graph-like structure, using text matching to search through this structure is not convenient. SPARQL is a SQL-like language that supports querying RDF graphs~\cite{sparql2013}. Queries are defined according to triplestore patterns with the possibility of using a diverse set of operators like filtering, unions, optional patterns, among others. SPARQL queries can be executed from an endpoint that should be made available to execute queries over a graph. One of the main capabilities of SPARQL is that queries can retrieve information from graphs on different data sources available in different endpoints. 

Listing~\ref{sparql} displays a simple illustration of a SPARQL query related to a simple car example. It returns the car and the person that drives the car. The elements prefixed with '?' indicate variables that should match possible values in the search. The statement "?c a car:Car" matches a subject ("?c") that is typed by "car:Car" (the object). The term 'a' between the subject and the object is just a syntactic sugar for the predicate "rdf:type".

\begin{lstlisting}[language=SQL,caption=SPARQL Example.,basicstyle=\ttfamily\small,label=sparql]
PREFIX car: <http://example.com/car#>
SELECT ?c ?p WHERE {
  ?c a car:Car .
  ?c car:isDrivenBy ?p
}
\end{lstlisting}

%\subsubsection{OWL}

Finally, the Web Ontology Language (OWL), or its latest revision OWL 2~\cite{Hitzler2012}, provides capabilities to define ontologies, initially for Web resources, but now it can be applied in a broader context. It describes a domain in terms of classes (types), properties (relations) and individuals (instances), and it supports the detailed description of features from these objects. Its core is based on a description logic designed to support deductive reasoning~\cite{Hitzler2010}. 

The official concrete syntax for OWL is RDF/XML, but it is also common to use the RDF/Turtle. Compared to RDF, OWL is much more expressive, given its larger vocabulary to express classes (types), properties (relations), individuals (instances), and data values. %For instance, we can state class hierarchies, equivalences, or disjoints. It is also possible to specify properties for a relation, e.g., symmetry, transitivity, reflexivity, and restrict their cardinality. 
In addition, there are several tools that support reasoning over OWL specifications~\cite{owlreasoners}. They can be employed to infer new knowledge, to check constraints, and also the consistency of ontologies.  %For instance, if we say that \textit{Robert} is of \textit{rdf:type} \textit{Professor}, and \textit{Professor} is a \textit{subClassOf} \textit{Person}, then I can now infer that \textit{Robert} is also of type \textit{Person}. Moreover, we can check that constraints over properties are also satisfied. Thus, we can check the consistency of our ontology. Finally, it is also possible to make your ontology modular by reusing definitions from other files using the \textit{owl:import} clause.

Although OWL expands RDF in some sense, it is still a notation created for machines. Some tools, like Protégé~\cite{protege}, help in the usage of the language, but it still requires considerable effort to work directly with OWL using RDF/XML or RDF/Turtle serialisation formats.

\subsubsection{OML}
\label{sec:oml}

OML (Ontological Modelling Language) can be seen as a Domain-specific language (DSL) for OWL~\cite{oml}. It provides a more user-friendly syntax that is mapped to several patterns of an OWL2 subset. Therefore, it inherits OWL expressivity, modularity, extensibility, and description logic semantics. It has been developed by the Jet Propulsion Laboratory of NASA with the aim of helping engineers define systems engineering vocabularies and using them to describe systems.

%\com{Joeri}{In the following sentence, I think it would be good to make it very explicit that vocabulary == type level and description == instance level:}

\begin{figure}
     \centering
     \begin{subfigure}[b]{0.55\textwidth}
         \centering
         \includegraphics[scale=0.7]{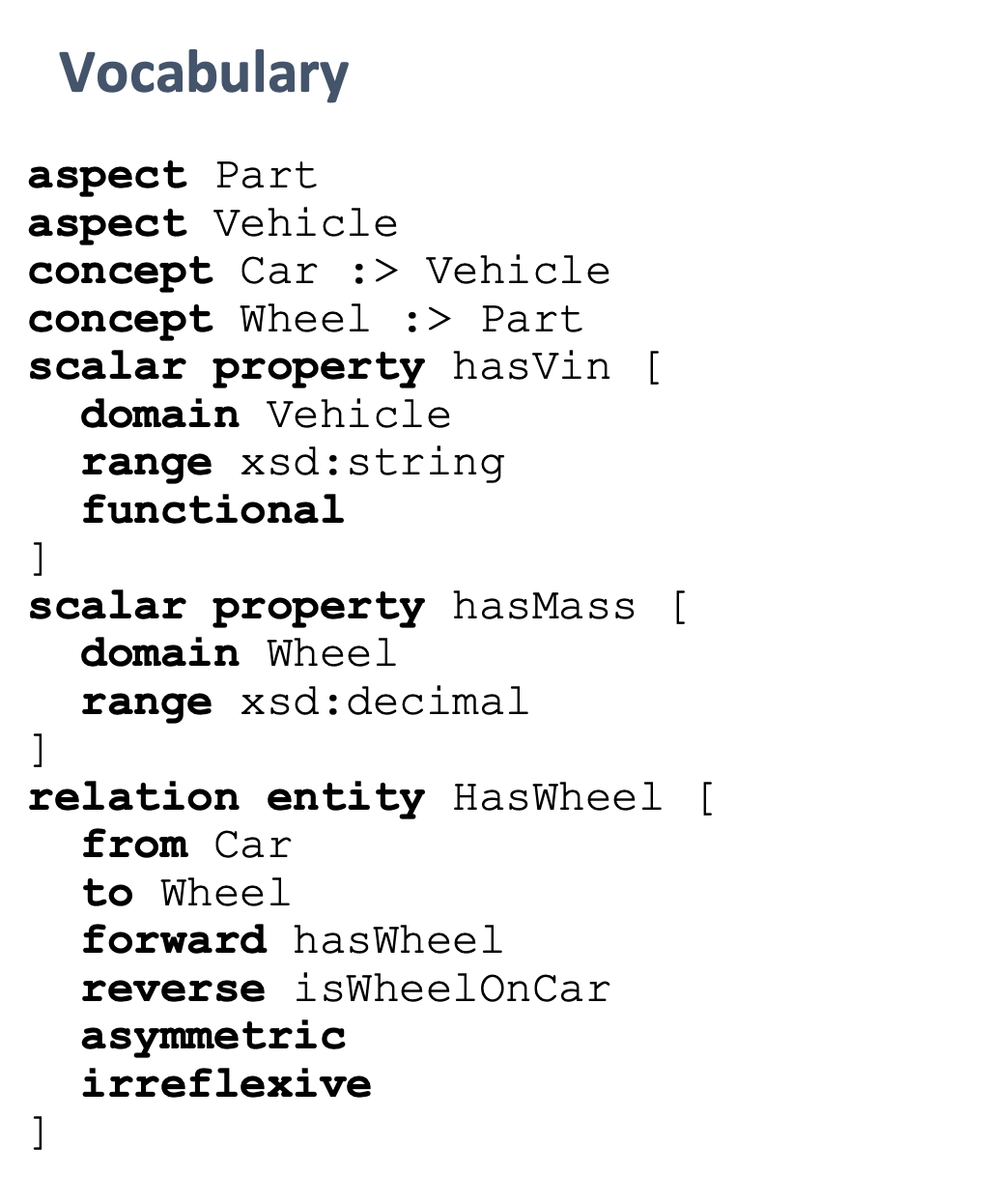}
         \caption{OML vocabulary example.}
         \label{fig:omlvoc}
     \end{subfigure}
     \hfill
     \begin{subfigure}[b]{0.4\textwidth}
         \centering
         \includegraphics[scale=0.7]{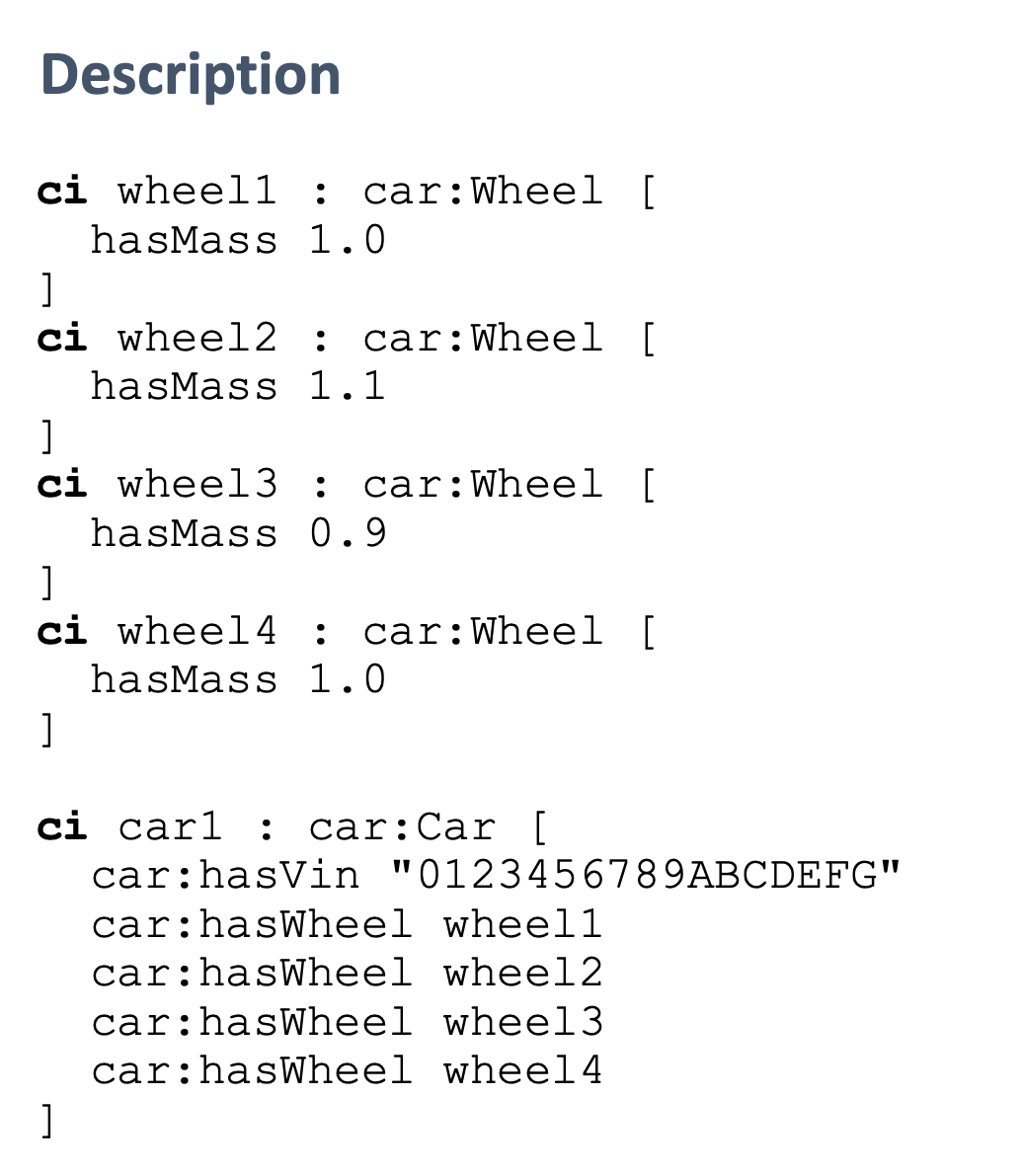}
         \caption{OML description example.}
         \label{fig:omldesc}
     \end{subfigure}
    \caption{Example of OML specification.}
         \label{fig:omlexample}
\end{figure}

OML provides a textual language that allows the creation of vocabularies and descriptions. The former declares concepts, properties and relations, that is, definitions at the type level. The latter contains individuals that conform to the vocabularies, that is, elements at the instance level. Figure~\ref{fig:omlexample} shows an example of a vocabulary for a simple car (\ref{fig:omlvoc}) and a possible instance of this vocabulary~(\ref{fig:omldesc}). A \textit{concept} is the basic mechanism to describe an entity. An \textit{aspect} is a concept that cannot have a related instance. It is similar to the notion of an abstract class in object orientation. Both \textit{aspects} and \textit{concepts} can have \textit{properties} when we want to relate them to a literal. On the other hand, \textit{relations} are used to specify a relationship between two \textit{concepts}. The same constraints that are supported in OWL can be used in the relations. For instance, we declare the relation \textit{HasWheel} from a \textit{Car} to a \textit{Wheel} to be \textit{asymmetric} and \textit{irreflexive} in Figure~\ref{fig:omlvoc}. In the description scenario illustrated in Figure~\ref{fig:omldesc}, we instantiate a car named \textit{car1} and its four wheels with their respective masses and relations. 

%\com{Joeri}{Maybe mention that OML allows you to "close the world" (whereas OWL has an open-world assumption}

OML has an Eclipse-based environment to support the development of ontologies called Rosetta~\cite{rosetta}. There, users can not only design their ontologies but also check their consistency using embedded OWL reasoners. Rosetta translates OML to OWL in order to reason over the ontologies and provide some feedback on possible non-conforming elements. Moreover, it is also possible to load the ontologies and related instances in an RDF-based knowledge graph through its integration with Apache Jena and Apache Fuseki~\cite{jena} to execute queries using SPARQL. 

A particular difference between OWL and OML is that the former considers the open-world assumption, that is, statements can be either true, false or unknown. The absence of information is considered unknown information, so it is not possible to state that something missing is false~\cite{Allemang2020}. This is particularly important in the context of the Semantic Web because the data must exist but is not available in the ontology yet. However, for some applications (e.g., integrity constraints, validation and data structures), closed-world semantics are more suitable. OML allows closed-world semantics via the usage of its vocabulary bundle, which is an ontology that bundles a set of vocabularies~\cite{oml}.

%\subsection{Workflow Modelling}
%\label{sec:worflow}

%\begin{itemize}
%    \item Importance of Workflow Modelling
%    \item Fluxograms
%    \item Activity Diagrams
%    \item BPMN
%\end{itemize}

%% file: sections/framework.tex
\section{Requirements for a Model Management Framework} %What - Joeri
\label{sec:framework}

In this section, we state the key requirements that a model management framework for system engineering must provide in our view. We start by introducing a running example of the engineering of a mass-spring-damper system to illustrate our requirements.
%and also how we implemented them, which will be described in Section~\ref{sec:implementation}.

\subsection{Running example}

\begin{figure}[htbp]
    \centering
    \includegraphics[width=\textwidth/2]{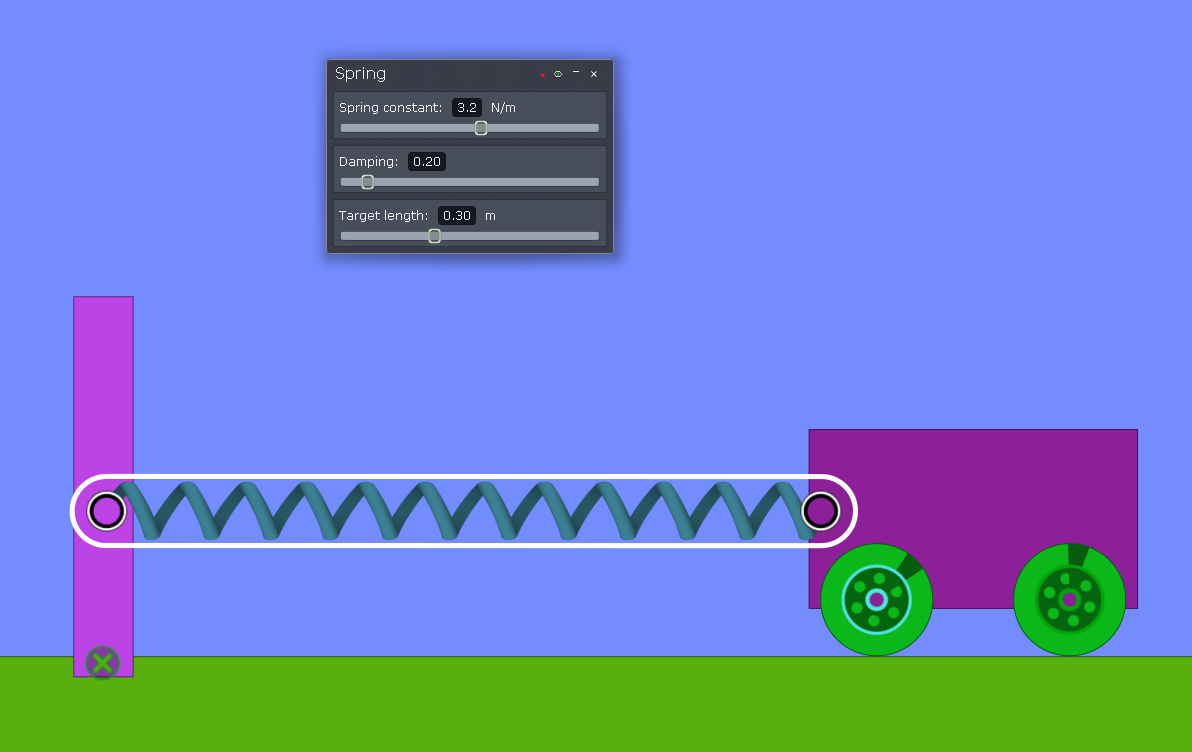}
    \caption{Simple representation of the system.}\label{fig:smd-illustration}
\end{figure}

%\com{Arkadiusz}{Do I need to add Algodoo (the sim program) attribution for this image?}
%Lucas: I don't think it is necessary. 

In order to illustrate the framework, we provide an example consisting of a mass-spring-damper system and all related components\footnote{\censortext{https://msdl.uantwerpen.be/git/arys}/mass-spring-damper} as shown in Figure~\ref{fig:smd-illustration}.
A mass-spring-damper system is a commonly used model in physics and engineering to describe the behaviour of systems that involve the interaction between a mass, a spring, and a damper. It is often used to analyse and predict the motion and response of mechanical systems, such as vibrating structures or suspension systems.
Such systems are described in detail in chapters 3 and 4 of Dynamical Systems for Creative Technology~\cite{dsct}.
%\com{Lucas}{is there any literature we could reference for this classical problem?}
%\com{Joeri}{yes, chapter 4 of the (free) book ``Dynamical Systems for Creative Technology''}

The system consists of three main components:
\begin{itemize}
    \item Mass ($m$): The mass represents the physical object or body in the system. It is typically a point mass, meaning it is concentrated at a single location. The mass interacts with the other components and experiences forces that affect its motion. It slides horizontally in our example and is depicted using a purple rectangle on green wheels.
    \item Spring ($k$): The spring represents an elastic element in the system. It provides a restoring force proportional to the displacement of the mass from its equilibrium position. This force is directed opposite to the displacement, attempting to bring the mass back to its resting position. The stiffness of the spring is quantified by the spring constant (k), which determines how strong the restoring force is for a given displacement.
    \item Damper ($b$): The damper represents a dissipative or damping element in the system. It exerts a resistive force that opposes the motion of the mass. The damping force is proportional to the velocity of the mass, with the constant of proportionality being the damping coefficient ($b$). The damping coefficient determines how quickly the system dissipates energy and reduces oscillations. We don't depict a damper explicitly. In our case, the damping force is provided by friction.
\end{itemize}

%\com{Lucas}{I think we should add a figure illustrating the problem. Maybe the one for xournal?}

When the system is at rest, the mass is in its equilibrium position, where the net force acting on it is zero. If the mass is displaced from this position, the spring exerts a force proportional to the displacement, trying to restore the mass to its equilibrium. Simultaneously, the damper opposes the motion by exerting a force proportional to the velocity.

The behaviour of the system depends on the parameters mass, spring constant, and damping coefficient, as well as the initial conditions. When subject to an external force or disturbance, the system exhibits various types of motion, including oscillations, transient responses, or steady-state responses\cite{ryan11}.
%\com{Arkadiusz}{I need to add the examples initial condtition, maybe here?}
The dynamics of the system can be described by a differential equation. For example, the equation of motion for the mass-spring-damper system is typically represented by Newton's second law:
\[ F(t) = m * \frac{d^2x}{dt^2} + b * \frac{dx}{dt} + k * x \]
Here, \(x\) is the displacement of the mass from its equilibrium position, \(t\) is time, and \(F(t)\) represents any applied external forces. The terms involving the derivatives $\frac{dx}{dt}$ and $\frac{d^2x}{dt^2}$ account for the velocity and acceleration of the mass, respectively.

By solving these equations, engineers and scientists can analyse the response of the system, predict its behaviour under different conditions, and design appropriate control strategies to achieve desired outcomes, such as minimising vibrations or optimising performance.

\begin{figure}[htbp]
    \centering
    \includegraphics[width=0.7\columnwidth]{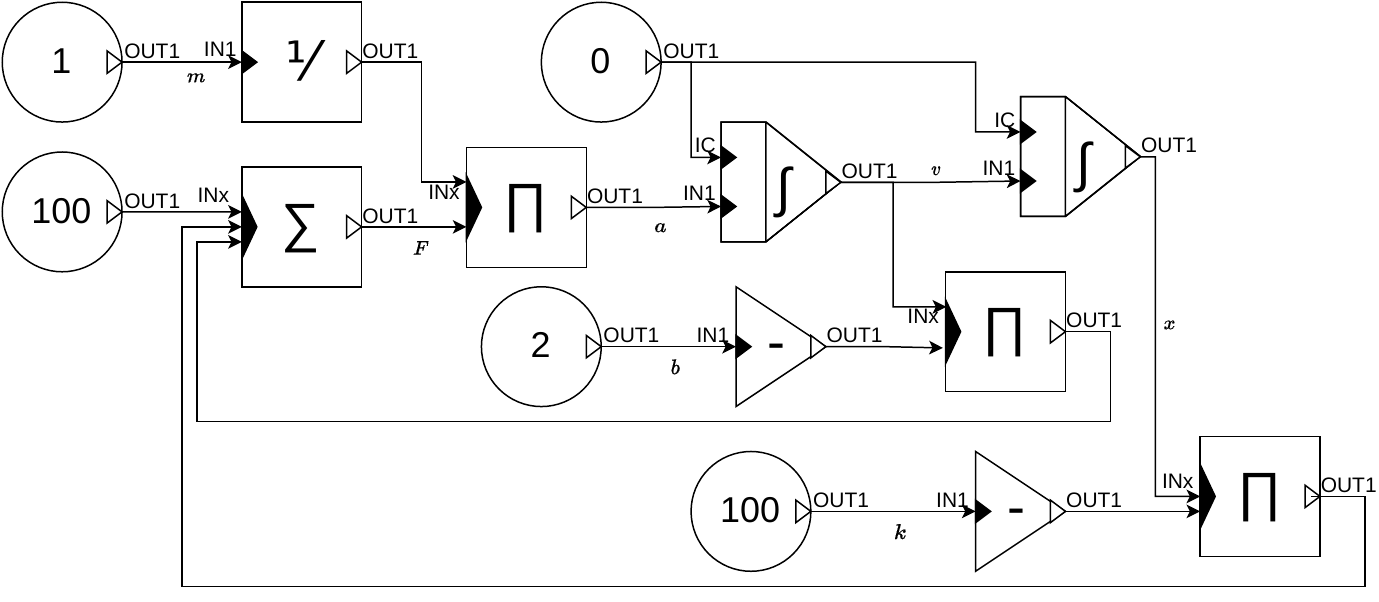}
    \caption{Causal Block Diagram model of the mass-spring-damper system.}\label{fig:cbd-smd}
\end{figure}

For our example, we want to find a spring constant \(  k \) such that an initially excited system is damped enough for the oscillation to have an amplitude lower than \( 0.05 \) between 5 and 7 seconds after the experiment starts.
We can model this system as a Causal Block Diagram model, as seen in Figure~\ref{fig:cbd-smd}.
Each of the integrator blocks represents part of the equation \( y^{(n)} = f - a_1y^{n-1} - ... -a_{n-1}y' - a_ny \) \cite{juan07}. % JE: I don't get this equation...
By repeating the simulation of this model with different parameter values for $k$, a solution can be found.

\begin{figure}[htbp]
    \centering
    \includegraphics[width=\textwidth/2]{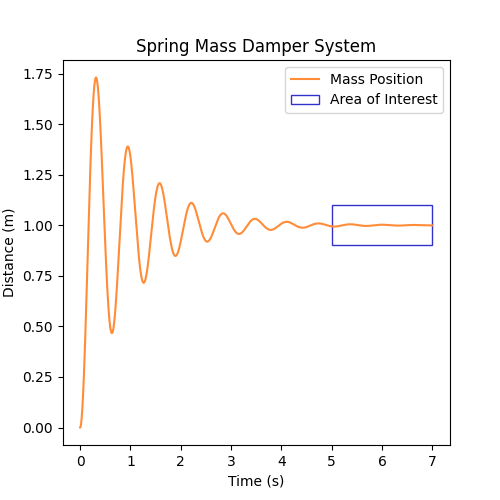}
    \caption{System trace of a single mass-spring-damper causal block diagram simulation.}\label{fig:smd-cbd-trace}
\end{figure}

A plot of an execution trace is shown in Figure~\ref{fig:smd-cbd-trace}.
We want to find a spring constant for which the oscillations are sufficiently damped after 5 seconds.
The blue rectangle in the figure illustrates this area of interest.

%\com{Lucas}{In our case, what we need to find out? It would be interesting to have a description of the problem. I think, in the case of our experiment, it is to find the more adequate constant k for a given scenario. Is that it?}

% \begin{itemize}
%     \item Requirements 
%     \item FTG+PM++ + language + execution engine 
%     \item Model repository 
%     \item Consistency Checking (reasoning) 
%     \item Queries % Arkadiusz?
% \end{itemize}

\subsection{Workflow modelling and enactment}
\label{sec:workflow}

\begin{figure}
    \centering
    \includegraphics[width=\columnwidth]{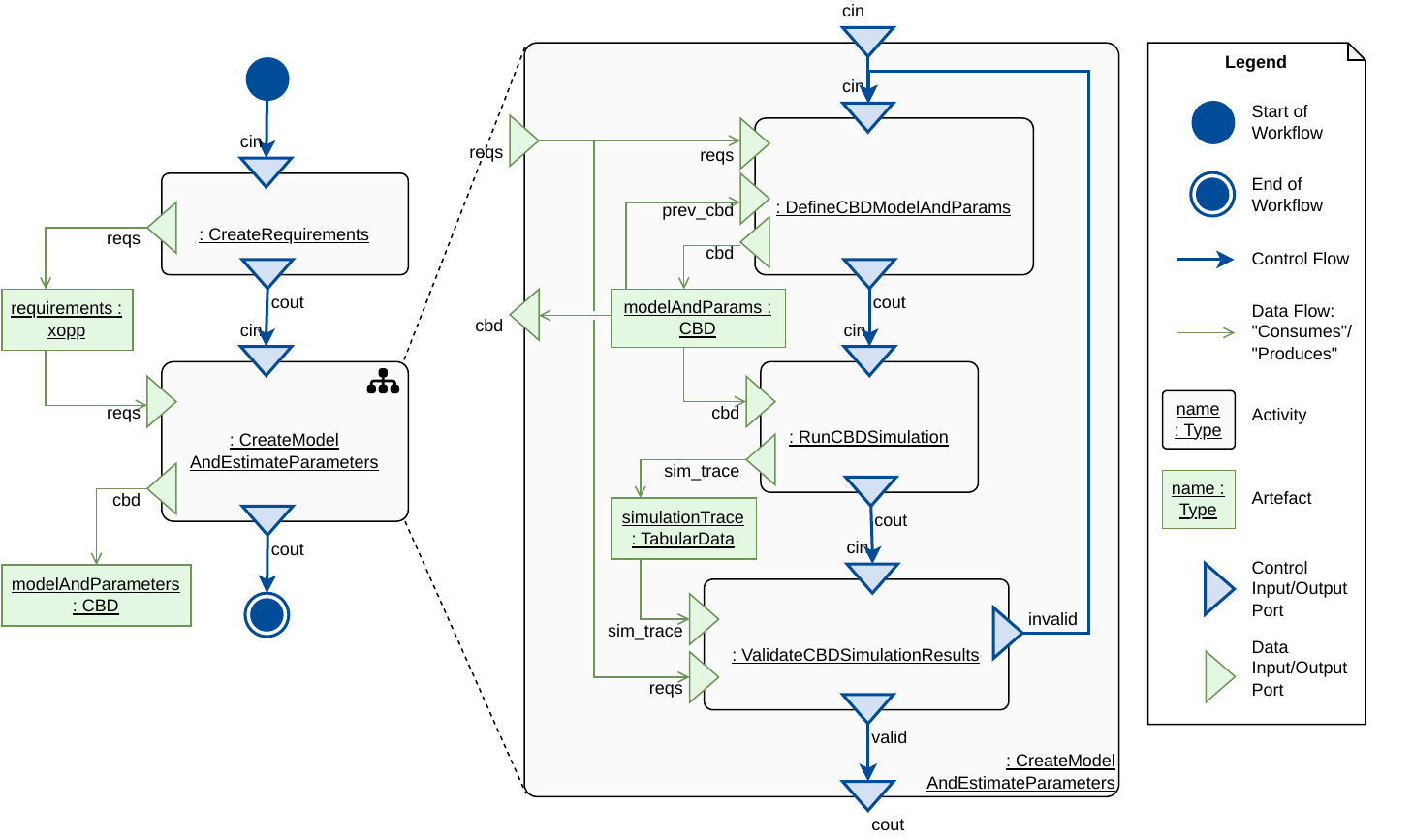}
    \caption{Running Example: Mass-spring-damper Workflow Model.}
    \label{fig:springdamper_pm}
\end{figure}

% \paragraph{Motivation}
In the previous section, we have introduced a simple engineering problem.
% In this section, we will introduce a relatively simple workflow for solving this problem.
% \paragraph{Solution}
% Our solution for dealing with this complexity is to model workflows explicitly in an executable workflow language. By interpreting a workflow model, a workflow enactment engine knows the activities that should happen, and the artefacts and their file types that should be produced.
% \paragraph{Example}
We now look at an example of an explicitly modelled workflow intended to solve the ``mass-spring-damper'' engineering problem.
The workflow model is shown in Figure~\ref{fig:springdamper_pm}, on the left. It is a model in the FTG+PM language \cite{randy22}, a very simple workflow language that follows Moody's ``Physics of Notation'' \cite{moody}. While explaining the workflow model, we give a very rough overview of the FTG+PM language. For a complete description, we point the reader to \cite{randy22}.

At the highest level, the model consists of two \emph{activities}: \textsf{CreateRequirements} and \textsf{CreateModelAndEstimateParameters}.
Activities connect to each other via \emph{control flow} (thick blue arrows) and \emph{data flow} (thin green arrows). Control flow specifies the order in which activities can be executed. Data flow specifies the artefacts (and their types) that are expected to be produced and consumed by the activities.

% Activities have input and output ports for \emph{control flow} (thick blue) and \emph{data flow} (thin green). In the example, both activities have one control input and one control output port, the minimum for any activity. An activity starts when it receives a token on one of its control flow input ports, and when it ends, it produces a control flow token on one of its output ports. When an activity starts, it \emph{consumes} (a version of) an artefact on each of its data input ports, and when it ends, it \emph{produces} (a version of) an artefact on each of its data output ports. 

Our engineering workflow starts with writing down the requirements of the system.
This is modelled by the first activity \textsf{CreateRequirements}, which has no data input ports (it consumes no data), and one data output port. When it ends, it must produce via this port an artefact named `requirements' of type `xopp' (Xournal++\footnote{https://xournalpp.github.io/}, a sketching app --- an example of a requirements document for the mass-spring-damper problem is shown in Figure~\ref{fig:smd-requirements-xournal}), and passes control to the next activity, \textsf{CreateModelAndEstimateParameters}, which consumes the requirements artefact. When this activity ends, it produces a Causal Block Diagram (CBD) model with the right parameters (example for mass-spring-damper: Figure~\ref{fig:cbd-smd}), that describes the behaviour of a system that is valid with respect to the requirements. The workflow ends here.

To make it more interesting, we can refine the last activity further with another (nested) workflow model, shown on the right of Figure~\ref{fig:springdamper_pm}. This model consists of three activities:

% Initially, a set of requirements is created in the activity \textsf{CreateRequirements}. The expected file type of the requirements is Xournal++ (`xopp'), a sketching app that allows freehand drawing. An example of a requirements document for the mass-spring-damper problem is shown in Figure~\ref{fig:smd-requirements-xournal}. It states that all parameters of the system are fixed, except for the spring constant $k$, which must be such that the oscillation ends within the time frame of 5-7 seconds after the initial impulse. We will use an iterative workflow to discover the right spring constant, consisting of the following activities:
\begin{enumerate}
    \item The activity \textsf{DefineCBDModelAndParams}, consumes the requirements, all previous versions of the CBD model (if there are any, will be explained later), and produces a new version of the CBD model.
    %, that represents the discretised formula for describing the evolution of the variables position ($x$), velocity ($v$) and acceleration ($a$) of the moving mass every next timestep.
    % For simplicity, we assume that the system's parameters $m$, $k$ and $c$ are part of the CBD model.

    % The activity \emph{consumes} as input the requirements document, and \emph{produces} as output (new version of) a CBD model.
    
    \item The activity \textsf{RunCBDSimulation} runs the new version of the CBD model produced by the previous activity for a duration of 7 seconds. As output, it produces a table (e.g., a CSV file) containing a trace of the values of the simulation's variables (for mass-spring-damper, these will be $x$, $v$ and $a$) at every timestep.
    
    % As input, it consumes the CBD model from the previous activity, and produces

    \item The activity \textsf{ValidateCBDSimulationResults} checks if the simulation trace that was produced by the previous activity is consistent with the requirements. The two possible outcomes (\textsf{valid} or \textsf{invalid}) are modelled in our workflow model as two respective control flow output ports.
    If the model's behaviour is valid, the workflow ends. Otherwise, we perform another iteration of creating a new \emph{version} of the CBD model, simulating it, and validating it against the requirements. The requirements are left unaltered.
\end{enumerate}

Even this simple workflow comes with artefacts of heterogeneous file formats and will result in multiple versions of these artefacts (because of the iteration cycle).
Obviously, more complex engineering problems will lead to much more complex workflows, with multiple domain experts working in parallel, performing real-world experiments and simulations involving even more different file formats and artefact versions.
Often, these workflows exist only in the minds of engineers, and they are informally communicated. We instead propose to model them explicitly, like we have done here.

% \paragraph{Reuse of activities and workflow models}
Workflow models in the FTG+PM language are highly reusable and hierarchically composable. For instance, the activity \textsf{CreateModelAndEstimateParams} is reusable in any context where a CBD model needs to be created and its parameters estimated in an iterative fashion according to a set of requirements.
The enactment of a workflow model produces an \emph{event trace}, consisting of events like ``activity started'' and ``activity ended''. These events can be linked to the artefact versions consumed/produced respectively.
% An ongoing trace is an \emph{append-only} sequence: only new events can be appended to it. A completed trace becomes \emph{immutable}.

\subsection{Model Repository}

In order to persist event traces and artefacts produced during the enactment of workflow models like from our previous example, we need a model repository. We now motivate a number of requirements for this repository to satisfy: \emph{versioning}, \emph{traceability}, \emph{typing/introspection} and finally \emph{querying}.

\subsubsection{Versioning}\label{sec:versioning}
% \com{Joeri}{Fulfilled by: our custom OML vocabulary for versioning information}

% \paragraph{Motivation}
A lot of engineering work is done in iterative, incremental cycles. When a workflow model contains cycles, multiple versions of artefacts are produced. For instance, in our running example (Figure~\ref{fig:springdamper_pm}), each time the activity \textsf{DefineCBDModelAndParams} ends, a new version of a CBD model is produced. Not only do we want to link the new CBD model to the ``end activity'' event, we also want to link it to the previous version of the CBD model. This versioning information needs to be recorded, and be queryable.

\added[id=r3c1]{Further, not only artefacts evolve, but also the languages/formalisms (that the artefacts conform to). It should be possible to also record the versioning history of language definitions. When a language evolves, it may introduce breaking or non-breaking changes. In the case of breaking changes, existing artefacts may not conform to the new version of the language. In the case of non-breaking changes, existing artefacts may conform to both the old and the new version of the language.}

% \paragraph{Solution}
% As a general pattern, we never want to throw any data away. Our repository should be an \emph{append-only} data structure. This applies to event traces (events can only be added) and to artefacts (we can only add new versions).

\subsubsection{(Fine-Grained) Traceability}\label{sec:traceability}
% \com{Joeri}{Fulfilled by: The use of OML/OWL/RDF}

% \paragraph{Motivation}
Our overall goal is to increase \emph{traceability} in data produced during systems engineering, and be able to \emph{query} this data in a powerful manner. Traceability is a broad concept, that takes many forms \cite{randy22}.
One form of traceability is linking artefact versions to the activities that produced them during workflow enactment.
% Further, we want to support \emph{fine-grained} traceability, i.e., between \emph{elements} of artefacts.
Other kinds of traceability are links between artefacts and their meta-models (which are also just artefacts). These links can also exist at a \emph{finer} level, between the \emph{elements} of artefacts and the elements of meta-models.
Another example of fine-grained traceability is the correspondence links between the concrete and abstract syntax of a model. 
%(example will be given in Section~\ref{sec:conversion_to_oml}).

\begin{figure}
    \centering
    \includegraphics[scale=0.5,trim={0 12cm 0 0},clip]{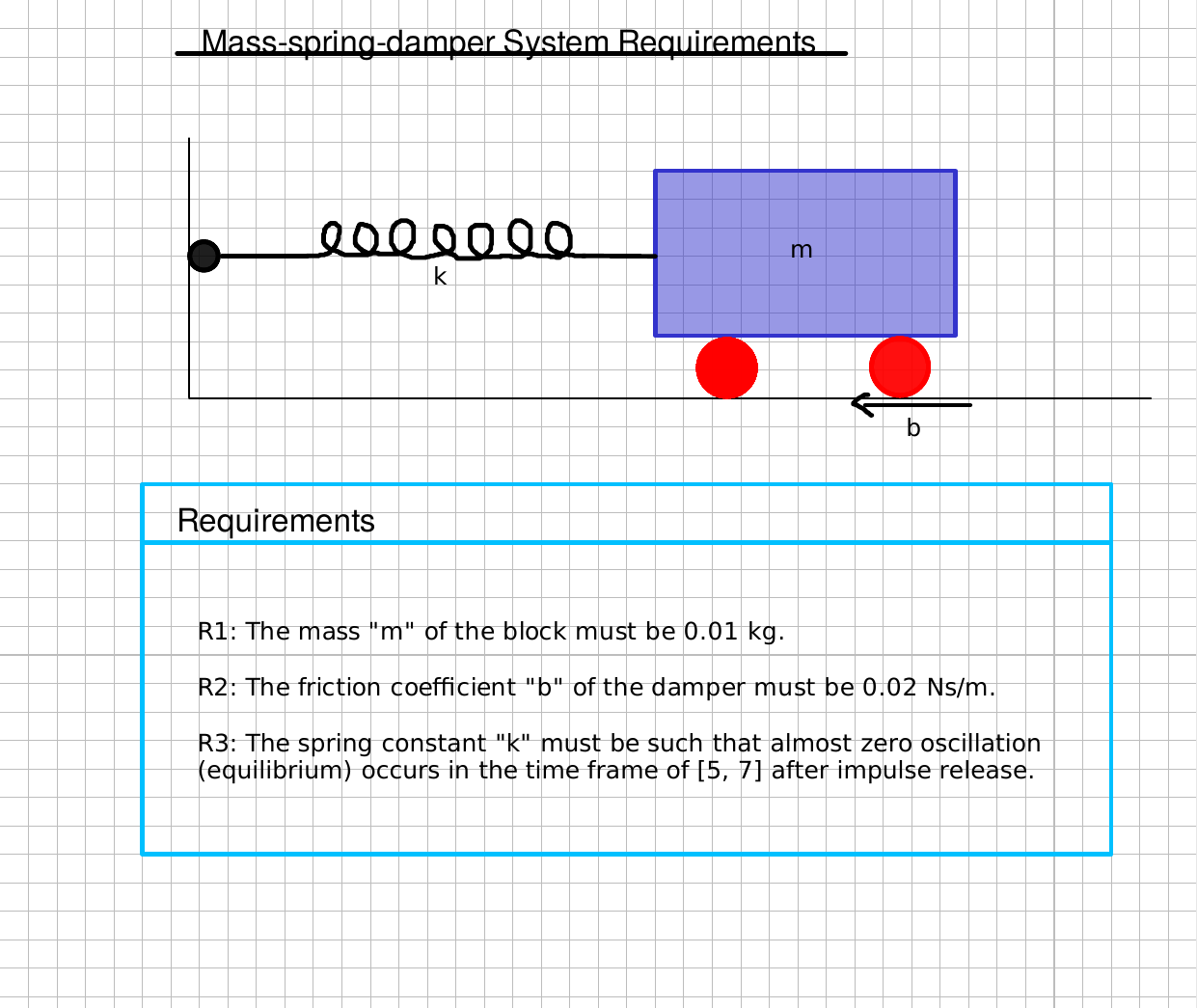}
    \caption{Running Example: mass-spring-damper Requirements (as Xournal++ document).}
    \label{fig:smd-requirements-xournal}
\end{figure}

For yet another example of fine-grained traceability, consider the CBD model in Figure~\ref{fig:cbd-smd}. Every connection (line) between the different blocks represents a signal that changes with every timestep during the simulation. The activity \textsf{RunCBDSimulation} in Figure~\ref{fig:springdamper_pm} performs this simulation and produces a table with the signal values at every timestep. Every column in this table can be linked (``caused by'') to a signal in the CBD model. Further, we could link the CBD's constant blocks representing the parameters  $m$, $b$ and $k$ to the text labels ``R1: ...'', ``R2: ...'' and ``R3: ...'' of the requirements in the Xournal++ document (Figure~\ref{fig:smd-requirements-xournal}).

% \paragraph{Solution}
In order to persist all these different kinds of traceability links between arbitrary artefact elements, we want to treat all data in our repository in a unified manner, as one large \emph{graph}.

\subsubsection{Typing/Introspection}\label{sec:typing}
% \com{Joeri}{Fulfilled by: The use of OML (descriptions are typed by vocabularies)}

% \paragraph{Motivation}
By knowing the types of artefacts and their elements (called \emph{introspection} in programming languages), interactive exploration and query-building of our model repository become greatly enhanced. The system can answer questions like ``What is the structure of objects of type X?'', and ``What operations can be performed on it?'', and ``Give me all objects of type X that implement requirement Y''.

For instance, given a CBD model, the system knows that it can be simulated, producing a table of signal values (via the activity \textsf{RunCBDSimulation}). Hence, for a given CBD model, the system can automatically suggest a query for the user to retrieve all simulation traces produced by that CBD model.

% \paragraph{Solution}
To support introspection, we explicitly store the types of artefacts and their elements. They are just one kind of traceability link.

% \subsubsection{Adaptors for third-party tools}
% \com{Hans}{This is too much ``how'', does not belong here.}
% \com{Joeri}{Fulfilled by: Custom adaptors written in Python, part of our OML Generator Service}

% \paragraph{Motivation}
% Systems engineering involves a large number of heterogeneous tools and file formats. These file formats are incompatible with each other, and surely they cannot be natively mapped onto some generic graph structure.

% \paragraph{Solution}
% Third-party file formats somehow need to be adapted to appear as graph structures towards reasoners, constraint checkers and query engines. There are two complementary ways to make this happen: (1) We transform data from third-party file formats to our `native' graphs structure, before doing anything with the data. (2) We use a \emph{federated} approach where the `source of truth' remains the original file in the third-party format, and we only make it available as a graph structure (towards a reasoner, query engine, ...) \emph{on demand}. The latter approach involves more complexity, but scales better. For this reason, we use it only for dense data (e.g., large tables).

\subsubsection{Consistency checking}\label{sec:consistencychecking}
% \com{Joeri}{Fulfilled by: OWL reasoner + SHACL}

% \paragraph{Motivation}
Many kinds of inconsistencies can occur in model-based systems engineering. In the general case, we want to be able to express arbitrary consistency rules over arbitrary model elements. One special case of consistency checking is conformance checking, where it is verified whether an artefact really conforms to the constraints (this includes multiplicities) imposed by its type (defined in e.g., a meta-model).

As an example of conformance checking, the meta-model of CBD could have a rule saying that for every block, all of its input ports must be connected to some output port. This makes sense because, otherwise, no value could be computed for this block. Suppose the user forgot to connect the \textsf{IC} port of one of the integrator blocks. When parsing the CBD model, it should be detected that the model is invalid.

% \paragraph{Solution}
% By storing artefacts, their types, and (a subset of) the constraints imposed by the types not just as graphs, but as ontologies (which are also graphs), a \emph{reasoner} can perform a consistency check (using only the ontology data as input). More complex constraints can be expressed and checked by using a language like SHACL.

%\com{Arkadiusz}{Should we add a reference to Xournal?}

\subsubsection{Query language}\label{sec:querying}
\label{sec:queries}
% \com{Joeri}{Fulfilled by: Fuseki's SPARQL engine}
% \com{Joeri}{I think it's not necessary to explain how to use SPARQL (instead, we can just cite a resource that explains this). It is better to show specific examples of SPARQL queries in the context of our running example (but in section \ref{sec:implementation}}
%\com{Lucas}{I think it is important here to differentiate the categories of queries, horizontal/vertical traceability, fine-grained traceability, queries to execute services, and so on. This way we can reference them in the implementation section (I am doing that in the WEE section).}

A final requirement for our model repository is to support a query language. With a query language, relatively complex data retrieval operations can be expressed in a relatively simple manner. Query languages can be powerful: for instance, existing queries may be composed into new queries. Query languages are usually easier to understand for non-technical users than having to use an API and write code.

Queries can be broadly categorised into 3 categories:
\begin{enumerate}
    \item Queries aiding in traceability
    \begin{itemize}
        \item Queries for horizontal traceability
        \item Queries for vertical traceability
        \item Queries for fine-grained traceability
        \item Queries for versioning
    \end{itemize}
    \item Queries to execute services
    \item Queries to explore the structure
\end{enumerate}

\paragraph{Horizontal traceability}
Horizontal traceability refers to traversing the FTG+PM++ from left to right or vice versa.
Each artefact belongs to a type, and all parts of the workflow are typed.
For example, we know the \textsf{requirements} artefact is typed by a Xournal++ file formalism \textsf{xopp}.

\paragraph{Vertical traceability}
Vertical traceability allows for following the workflow. We can clearly see how the process progresses when looking at the spring-mass-damper process model in Figure~\ref{fig:springdamper_pm}. By following the blue control flow facilities, we relate each of the activities and achieve full vertical traceability within the process model.
Here, we know that the \textsf{RunCBDSimulation} activity is always preceded by the \textsf{DefineCBDModelAndParams} activity and succeeded by the \textsf{ValidateCBDSimulationResults} activity.

\paragraph{Fine-grained traceability}
Whenever a relation refers not directly to an artefact but to the inner structure of one, we consider it a fine-grained relation. Any query using these relations is, therefore, a fine-grained traceability query.
An example here could be the relation between a column of the \textsf{simulationTrace} \textsf{TabularData} artefact instance and a specific block in the corresponding \textsf{modelAndParams} {CBD} artefact instance. This would give us a better understanding as to which blocks are responsible for generating which data.

\paragraph{Versioning}
Enactment of a workflow containing a circular dependency will create multiple versions of an artefact for an activity. These artefacts are related to each other, so earlier versions can be returned in a query.
If we enacted multiple iterations of the workflow, we would end up with multiple \textsf{simulationTrace} artefact instances. Each of these artefacts would have a direct relation to its precursors from previous workflow iterations.

\paragraph{Services}
By using special keywords, we can call external services which perform part of the query.
These service calls could include looking for outliers in the \textsf{simulationTrace} or even running, validating, or generating traces of the \textsf{CBD}.

\paragraph{Exploration}
Some queries are performed to get the structure of data. This includes knowing which types are available and what their properties are.
We are able to check which formalisms and/or transformations are available to us.
For the spring-mass-damper running example, these would include, but are not limited to, exploring the structure of \textsf{CBD}, \textsf{xopp}, \textsf{TabularData} and \textsf{ProcessModel}\footnote{process models are also just artefacts} artefacts.
% and \textsf{CreateRequirements}, \textsf{CreateModelAndEstimateParameters}, and \textsf{ValidateCBDSimulationResults}.

\paragraph{Composing categories}
Simple queries can be composed into (more) complex queries.
For instance, we can construct a query on the spring-mass-damper system to figure out the initial model (\textit{first version}) whose later versions led to the correct parameter estimation and its value.

%% file: sections/implementation.tex
\section{Implementation} %How - Arkadiusz + Joeri + Lucas
\label{sec:implementation}

In this section, we explain our prototype implementation and explain how it satisfies the requirements described in Section~\ref{sec:framework}.
% In order to provide mechanised support for the end-user of our framework, we have developed a toolchain that considers those requirements for our framework described in Section~\ref{sec:framework}. 

%\com{Arkadiusz}{Do we want to add a references to Moodys paper? https://ieeexplore.ieee.org/document/5353439
%It's about how to design good visual communication, I %think the FTG+PM++ follows this approach.
%}
%\com{Lucas}{Maybe add it in the related work section?}
%\com{Joeri}{Citation to Moody's paper added in previous section on FTG+PM}

% \begin{itemize}
%     \item Framework Architecture % Joeri
%     \item Ontology (OML) % Lucas
%     \item Federation services % Arkadiusz
%     \item Drawio Plugin % Joeri
%     \item WEE % Lucas
%     \item Graph Explorer % Arkadiusz
%     \item Ad-hoc querying using Fuseki? % Arkadiusz?
% \end{itemize}

\subsection{Architecture Overview} % Joeri

To implement our model repository, we decided to use an OML/OWL/RDF stack. By using RDF, we have a unified graph-like data structure for all our models and the traceability links between them (solving \ref{sec:traceability}), that can be queried by the existing SPARQL language (solving \ref{sec:querying}). By using OWL on top of RDF, we can define simple consistency rules (solving \ref{sec:consistencychecking}) and typing information for introspection (solving \ref{sec:typing}). We do not use OWL directly but instead generate it from OML documents, for reasons explained in Section~\ref{sec:oml}. OML distinguishes between type-level (``vocabularies'') and instance-level (``descriptions'') documents. The descriptions are generated from artefacts in different file formats via the use of adaptors.  \replaced[id=r3c1]{We manually defined their respective vocabulary in advance, and they serve as the metamodels for the artefacts. Our framework is agnostic in terms of modelling methods (e.g., ARCADIA~\cite{roques2016}, IBM Harmony~\cite{ibmharmony}). Any language or artefact that needs to be considered should have its vocabulary (metamodel) defined in OML first, and then, workflows that manage these artefacts and are related to a particular modelling method can be specified.}{Their respective vocabularies were written by hand.}

OML has no built-in solution for versioning. Users of OML are recommended to use a text-based versioning system like Git for versioning the OML documents themselves. The main problem with this approach is that the versioning information remains external to OML and, therefore, also to the generated OWL/RDF, making it impossible to access this information with (SPARQL) queries. Therefore, we use the ``poor man's approach'' of simply adding for every new artefact version a new OML description document and OWL/RDF data generated from it to our model repository. \added[id=r3c1]{We do not support language evolution (partially solving \ref{sec:versioning}). The reason for not supporting language evolution will be explained in Section~\ref{sec:notypeevolution}.}

\subsubsection{Components}

\begin{figure}
    \centering
    \includegraphics[width=0.9\columnwidth]{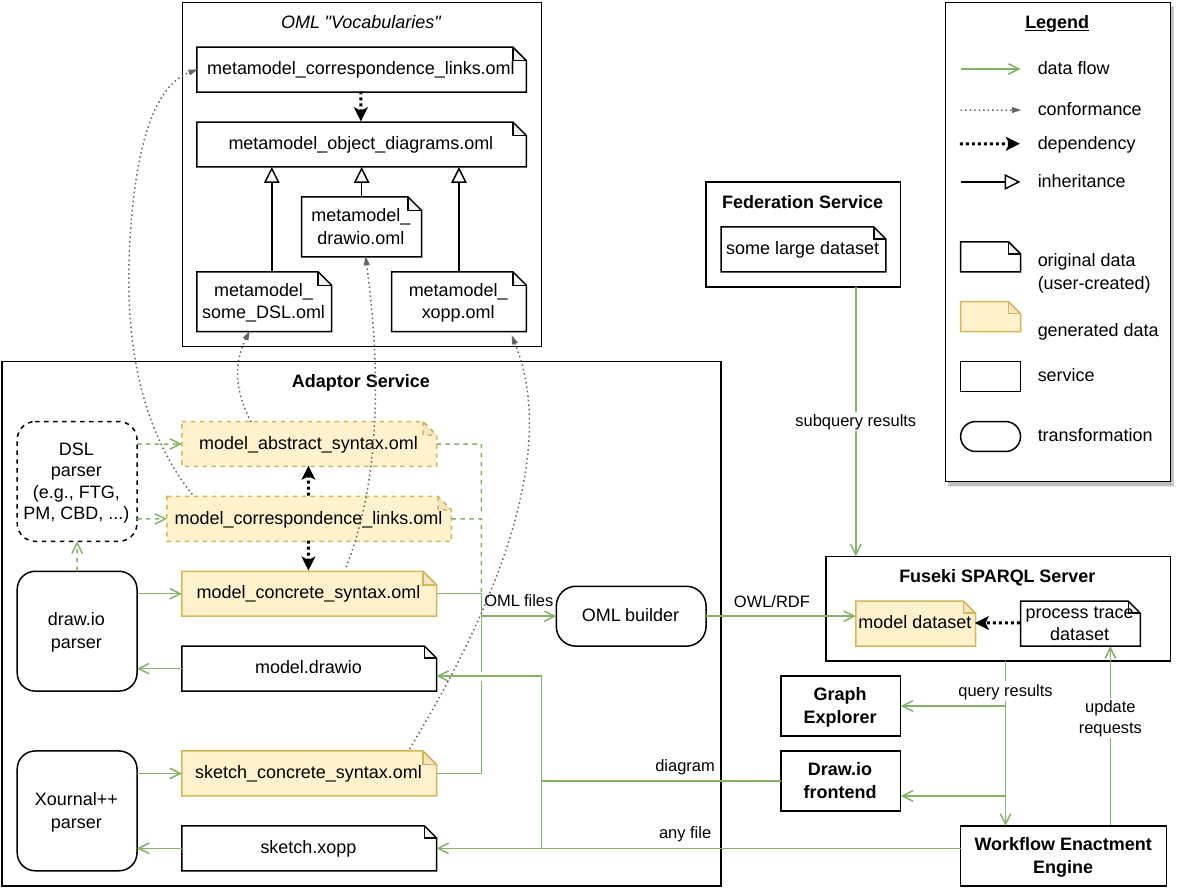}
    \caption{Architecture.}
    \label{fig:services_architecture}
\end{figure}

Our implementation consists of the following components (Figure~\ref{fig:services_architecture}):

\begin{description}
    \item [OML Vocabularies] are a set of hand-written ``meta-models'' that all the instance-level data (artefacts and workflow event traces) must conform to.
    \item [Adaptor Service] transforms artefacts from different file formats to RDF (by first transforming to OML).
    \item [Fuseki] is an RDF database and SPARQL query engine. It stores all the RDF data.
    \item [Workflow Enactment Engine] is a web service that interprets workflow models in our FTG+PM language (Section~\ref{sec:workflow}) and saves execution traces directly in Fuseki.
    \item [Draw.io Frontend] is a plugin for the diagramming tool draw.io that allows the user to navigate traceability links (of the model repository) from/to diagrams for which we have parsers.
    \item [Federation Services] allow large datasets to be queryable via SPARQL without having to transform their data to RDF in advance (this would not scale).
    \item [Graph Explorer] is an application for exploring the knowledge graph with the use of SPARQL queries.
\end{description}

We will now explain these components in more detail.

\subsection{OML Vocabularies}

We have defined an ontology using OML (described in Section~\ref{sec:oml}) that supports the design of diagrams in a context based on two generic vocabularies for \textit{objects} and \textit{links}. \textit{Objects} represent nodes used in diagrams, and \textit{links} are the edges connecting two objects. According to these two vocabularies, support for other languages can be provided. Figure~\ref{fig:services_architecture} depicts how these two OML vocabularies are used to create language-specific vocabularies. The \textit{FTG+PM} abstract syntax can be described using the \textit{metamodel\_object\_diagrams.oml} vocabulary, as well as their concrete syntax in \textit{Draw.io}. We use this same mechanism to describe the metamodels for formalisms used in our framework. For instance, causal block diagrams, FTG, process models, process traces, textual documents, tabular data and Xournal++ (xopp) are currently supported formalisms. For those formalisms, we do not have a metamodel yet. We have created a generic formalism file, which can represent any type of file to be managed by our framework. However, fine-grained traceability will not be supported for these formalisms as we do not have knowledge about their internal structure.

\begin{figure}
    \centering
    \includegraphics[scale=0.5]{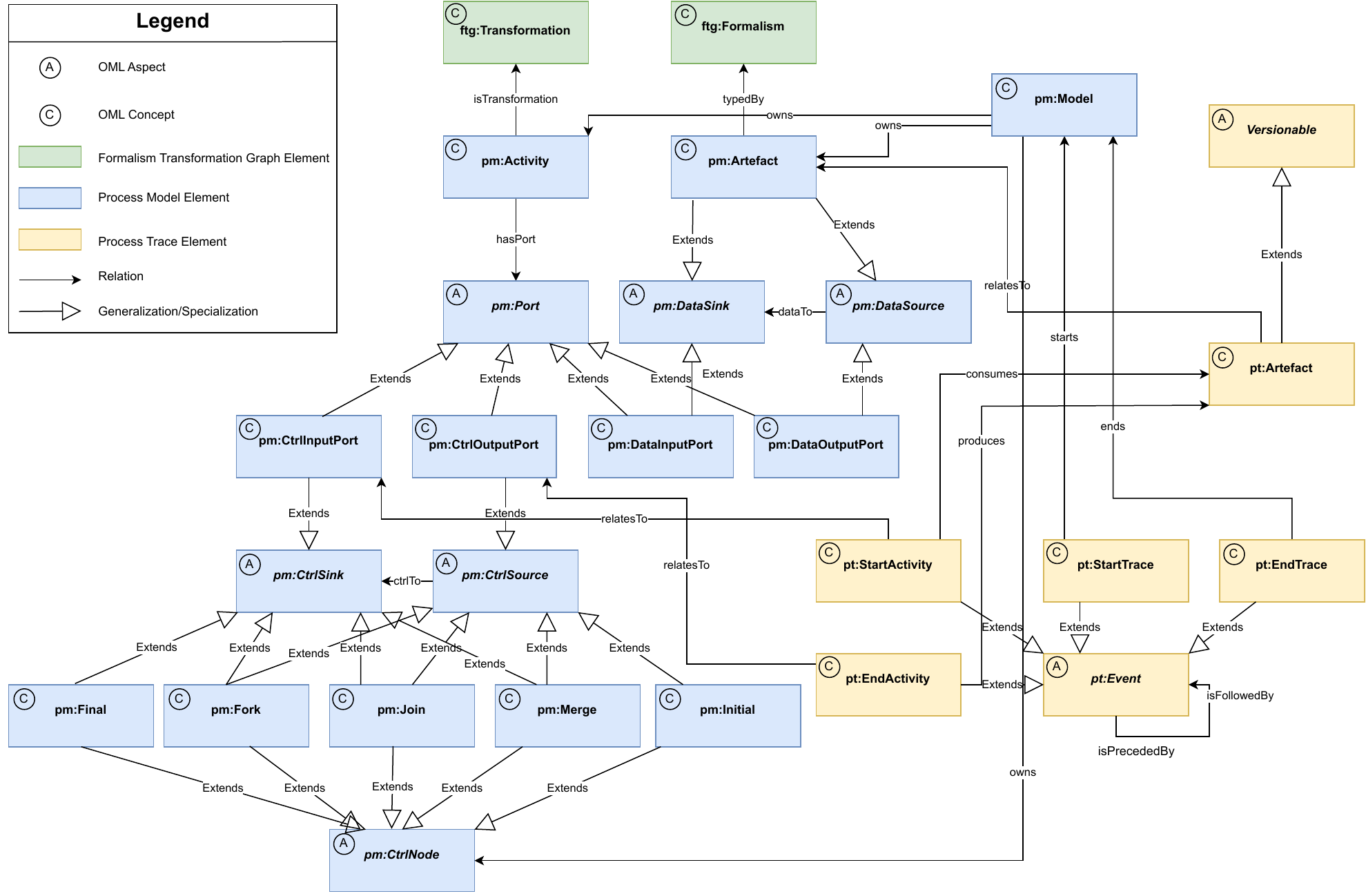}
    \caption{Part of the OML Vocabularies for the FTG+PM language.}
    \label{fig:ontology}
\end{figure}

Especially formalism transformation graphs, process models and process traces are vital to support the description of experiment workflows in our framework. In Figure~\ref{fig:ontology} we show a graphical illustration of part of the OML vocabularies created to define process models and traces that were presented in Section~\ref{sec:workflow}.

 A \textit{pm:Model} represents a workflow, and it is composed of \textit{pm:Activity}, \textit{pm:Artefact} and \textit{pm:ControlNode} elements. An \textit{pm:Activity} describes a task to be executed, and it conforms to one \textit{ftg:Transformation}, which is part of the FTG vocabulary and represents a generic description of an action that can receive inputs and provide outputs. A \textit{pm:Activity} has ports, which can be control or data ports. Ports can also be input or output to an activity. For instance, a \textit{pm:CtrlInputPort} is an entry point to execute a \textit{pm:Activity}, while a \textit{pm:CtrlOutputPort} indicates the control flow to be followed when a \textit{pm:Activity} terminates. While control ports indicate possible flows of execution, data ports indicate the production or consumption of artefacts by activities. A \textit{pm:Artefact} indicates a resource utilised during the workflow, and it is typed by a \textit{ftg:Formalism}, which represents a class of elements that conform to the same metamodel in the FTG vocabulary. A \textit{pm:ControlNode} represent an element used to coordinate the control flow of the diagram. For instance, \textit{pm:Initial} indicates the point where the workflow starts while \textit{pm:Final} determines its termination. A \textit{pm:Fork} can create concurrent flows while \textit{pm:Join} synchronises several input concurrent flows in a single output control flow. 
 
 Elements of process traces can be a \textit{pt:Event} or a \textit{pt:Artefact}. A \textit{pt:StartTrace} event represents the beginning of executing a trace described by a \textit{pm:Model}, and \textit{pt:EndTrace} represents the end of the trace. A \textit{pt:StartActivity} indicates the moment an activity starts by a control flow received in its \textit{pm:CtrlInputPort}, while the \textit{pt:EndActivity} means the termination of a particular activity that will continue the flow by its \textit{pm:CtrlOutuputPort}. Finally, a \textit{pt: Artefact} is a concrete version of a resource detailed by \textit{pm:Artefact}, and that is generated/consumed during a \textit{pm:Model} enactment. Given that the workflow can have loops, the same artefact can be generated but with different versions (\textit{Versionable}). 

 Although Figure~\ref{fig:ontology} offers an intuition about the entities and relationships we have defined, it does not provide the complete details and properties specified in our ontology. For example, to determine that a \textit{pt:StartTrace} event has not any other event before, we specify the property \texttt{restricts relation isPrecededBy to max 0} in the context of a \textit{pt:StartTrace}. In this case, describing that the relation \textit{isPrecededBy} does not happen. The opposite happens for \textit{pt:EndTrace} where we use the property \textit{restricts relation isFollowedBy to max 0}. The complete OML ontology that supports our framework is available in our Git repository\footnote{\censortext{https://msdl.uantwerpen.be/git/lucasalbertins}/DTDesign/src/main/examples/oml}. There, the vocabularies for the FTG, PM, and PT will be available, but also the ones for the previously mentioned supported formalisms.

\begin{figure}
     \centering
     \includegraphics[scale=0.8]{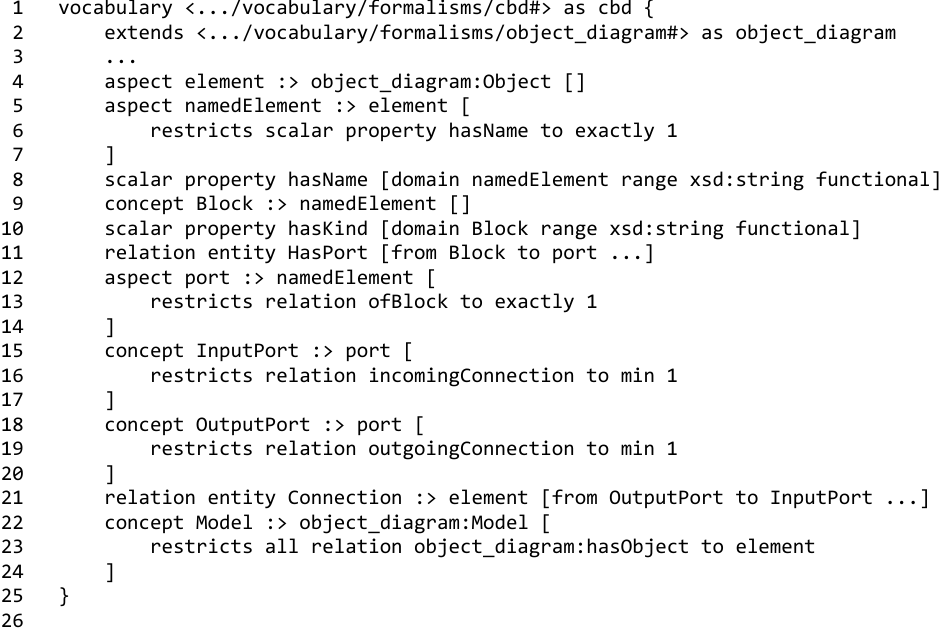}
     \caption{OML textual representation of the ontology for the causal block diagram.}
     \label{fig:cbd-vocabulary-text}
\end{figure}

\added[id=r3c1]{
Our complete implementation of the ontology includes the vocabularies required for the artefacts. The causal block diagram is one of such vocabularies. It is based on \textit{links} and \textit{objects} in the same way the process model ontology is. This manifests in the OML vocabulary definition in Figure~\ref{fig:cbd-vocabulary-text} in the form of an extension of the object diagram ontology.
A causal block diagram like the one in Figure~\ref{fig:cbd-smd} has blocks which are connected by means of ports and their relations. This simple representation is visible in the vocabulary.
It has 4 concepts, one for the full model, another for the blocks themselves, and two more for representing the input and output ports. We augment these with two relations, the first to connect between output and input ports and the second to keep track of which ports belong on which block, and we are almost ready.
The last required action to complete the vocabulary is adding the properties and perhaps some restrictions.
Regarding the CBD example, it includes the name for all named elements and a few cardinality restrictions.
We provide a more in-depth explanation of the OML syntax itself in Section~\ref{sec:oml} with an example in Figure~\ref{fig:omlexample}.
}

\subsection{Adaptor Service}
\label{sec:omlgeneratorservice}

% \paragraph{Purpose}
\added[id=r3c2]{
    The Adaptor Service's main task is the translation of arbitrary artefacts to OML.
}
The Adaptor Service exposes a REST endpoint where other services (most importantly, the Workflow Enactment Service) can ``upload'' files. Uploaded files are parsed, converted to OML, which in turn is converted to an OWL/RDF knowledge graph, and added to Fuseki's dataset. The originally uploaded files are also stored (in the adaptor service).

\added[id=r3c2]{
Although artefact conversion to OML could also be done locally, we chose to make this functionality available as a central service, so multiple services could use it.
\textit{REST} can be implemented with minimal effort on both the client and server side in all major programming languages.
}

%Different approaches include using \textit{websockets}, \textit{gRPC}, or perhaps even \textit{MQTT}.
%These all meet our requirement of being accessible over a network and have their benefits.
%Using \textit{websockets} would allow us to perform real time communication whereas our approach requires the client to initiate the request. The second contender; \textit{gRPC} would allow us to have increased throughput. This however, comes at the cost of having the define an interface beforehand and generating bindings for any required languages.
%\textit{MQTT}, the last alternative, is more suited for devices with limited resources. It follows a different model by having a server where clients can subscribe to a topic or publish messages. A client only receives messages it is subscribed to.

%But ultimately we did not require any of the specialised features.
%%In the end these are implementation details and all of them would have been suitable for the framework.

% \paragraph{Conversion to OML}\label{sec:conversion_to_oml}
For every file type to be converted to OML, we need a dedicated parser and OML generator. For some file types (e.g., Xournal++, a sketching application), we only convert the file's contents to OML. In this case, the OML representation contains exactly the same amount of information as the original file, and in theory, we could restore the original file from the OML data (not implemented). For other file types (e.g., draw.io), the contents of the file (e.g., a diagram) may be additionally interpreted as an instance of a DSL. For instance, a draw.io diagram representing a workflow model can be seen as just a diagram consisting of shapes and connections (concrete syntax) but also as a process model with activities, control and data flow (abstract syntax). In this case, we not only convert the original file contents to OML but also invoke a second DSL-specific parser that attempts to construct an instance of the DSL's abstract syntax in OML. Finally, we also store traceability links between elements of the generated OML files for concrete and abstract syntax. These traceability links are stored in yet another OML file.
All generated OML files are so-called \emph{descriptions} (i.e., OML's instance-level) that conform to (manually written) \emph{vocabularies} (i.e., OML's type-level).

% \paragraph{Technical details}
All of our parsers were written in Python, and OML generation makes use of the Jinja2 template library for Python. The OML to OWL conversion is taken care of by a Gradle build script included with the Rosetta IDE for OML. During OML to OWL conversion, the generated OWL output is checked for inconsistencies. Inconsistencies usually indicate non-conformance between a description and its vocabulary. When an inconsistency occurs, the procedure fails. In addition, SHACL (Shapes Constraint Language)~\cite{bibSHACL} can be employed to specify arbitrary consistency rules to check the generated OWL. 
The source code of the OML Generator Service is available in our Git repository\footnote{\censortext{https://msdl.uantwerpen.be/git/jexelmans}/drawio2oml/src/master/backend}.

\added[id=r3c2]{
This storage solution can be replaced by any similar option as long as it supports the translation of artefacts to OML.
The translation is a minimum requirement as we use the generated OML models to, not only, reason about the system, but also create relations between internal concepts.
% Converting from RDF to drawio PT. Everything (-PT) dra -> OML.
Our implementation supports converting incoming artefacts to OML.
We also support converting the RDF representation of a process trace back into a process trace which can be visualised as a \textit{drawio} diagram.
Ideally this conversion process would be bi-directional for all entities.
}

\subsection{WEE - Workflow Enactment Engine}
\label{subsec: WEE}

Once we have an environment to design workflows, it should be possible to execute them and record the traces and their respective data. Manually designing the traces while executing the workflows would require too much effort. Moreover, the generated process traces should be available to other tools for visualisation and information recovery. Therefore, WEE has been developed to support both matters. First, it provides an environment to guide users through the possible flows of execution related to previously designed process models, recording not only the tasks being performed but also artefacts generated and used during the workflow enactments. WEE's second purpose is to provide access to the recorded knowledge on process traces and process models as a service. The latter is currently used by the Drawio plugin to render information about the traces. However, any other tool that could benefit from these data could use this service.

\begin{figure}
    \centering
    \includegraphics[scale=0.4]{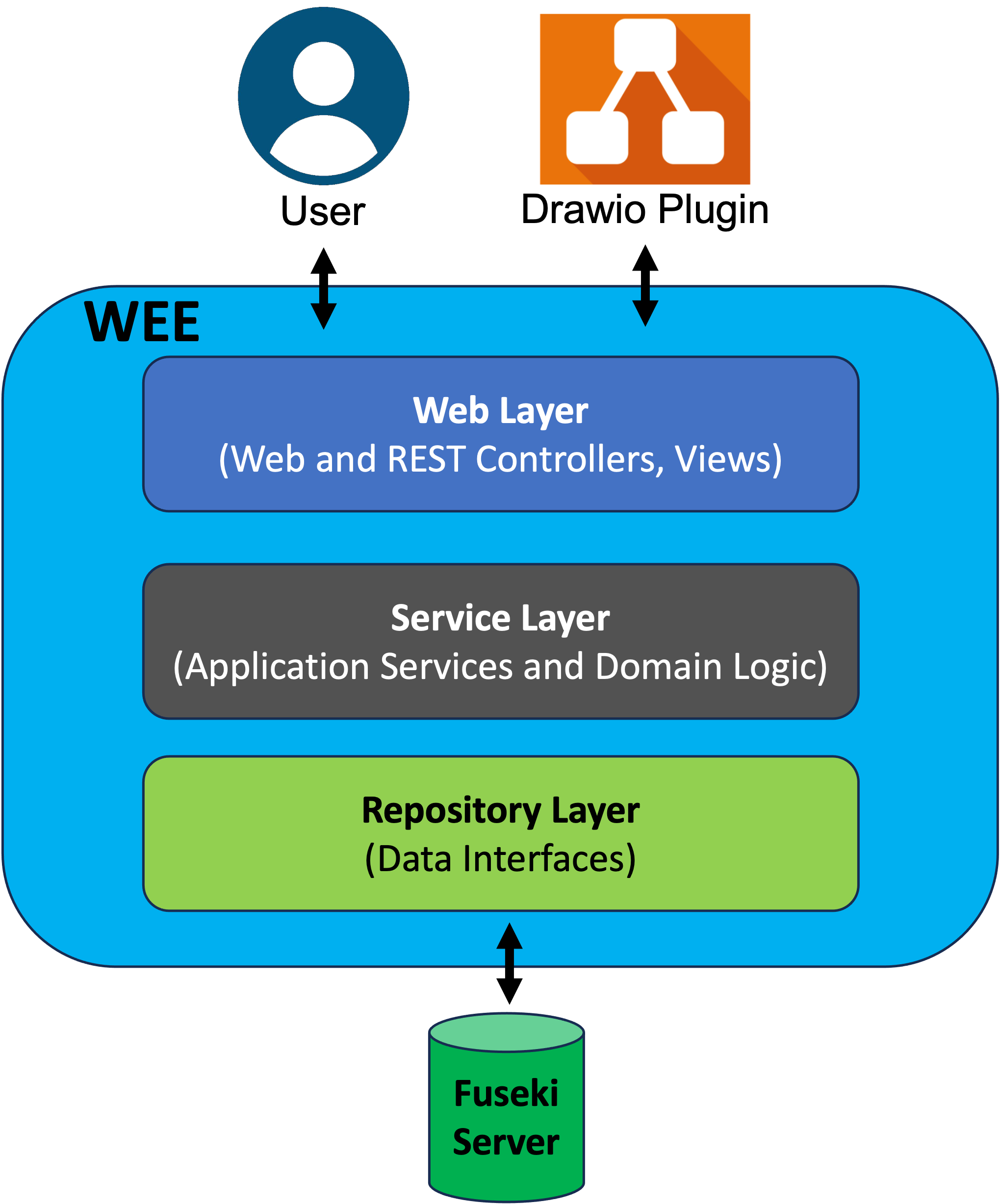}
    \caption{WEE Architecture.}
    \label{fig:wee_arch}
\end{figure}

WEE has been conceived as a Web application and can be accessed via browser or through the REST API services. Figure~\ref{fig:wee_arch} shows WEE architecture, which is detailed in three layers. The \textit{Web Layer} is responsible for the interaction between users and other systems with the tool. It has a Web front-end that allows users to interactively enact selected process models. This layer communicates with the \textit{Service Layer} to invoke the appropriate services. The \textit{Service Layer} owns the business logic used to manipulate process models and traces. It also can request services from the \textit{Repository Layer}, which encodes the statements to query and update the knowledge graph stored in the \textit{Fuseki Server} with data from all process models and traces.

\begin{figure}
    \centering
    \includegraphics[scale=0.4]{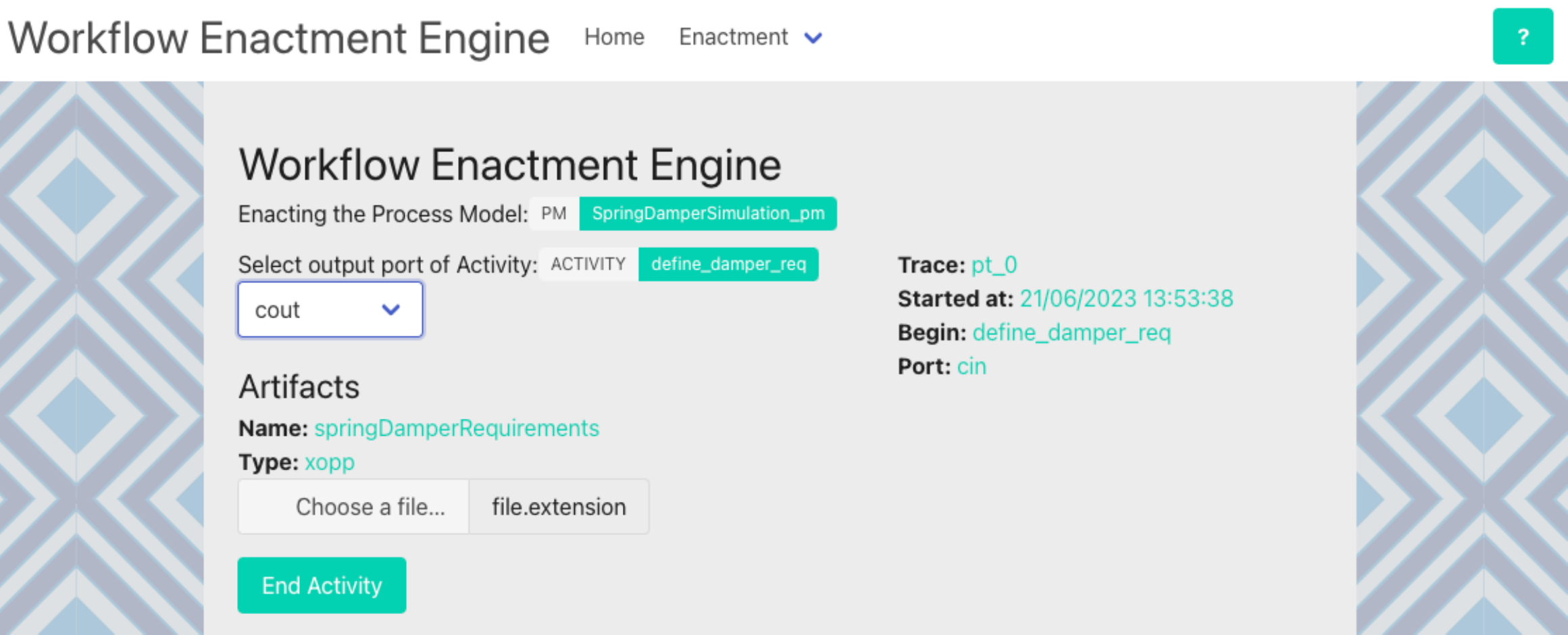}
    \caption{Example of activity conclusion in WEE.}
    \label{fig:wee_upload}
\end{figure}

Using the \textit{Web Layer}, the user can interact with the tool to enact a previously designed process model. First, the user must select a PM to enact from a list returned from the knowledge graph. Then, the enactment process starts and the tool displays the activities available based on the possible flows of the PM. For instance, considering our running example of the Mass-Spring-Damper PM shown in Figure~\ref{fig:springdamper_pm}, the first activity to be executed is \textit{define\_damper\_req}. After concluding this task, the user must indicate its termination and upload the artefact \textit{springDamperRequirements} of type \textit{xopp}, in conformance to what is described in the process model. Figure~\ref{fig:wee_upload} displays the screen for this task. It shows the process being enacted, the current activity, and the artefact that is required. 
After uploading the requested file, the user can click the \textit{End Activity} button. After starting or ending an activity, we update the current process trace in the knowledge graph with the related events and artefacts provided by the user. The current events performed so far are outlined on the right-hand side of the screen, as can be seen in Figure~\ref{fig:wee_upload}. The files are also stored using a local storage service we have developed. Some types of artefacts are described in our ontology in OML. It can be seen as meta-model descriptions of these types. For instance, if a CBD model is uploaded, we convert it to OML, and we perform consistency checks before uploading their RDF representation to the knowledge graph in the Fuseki server (as detailed in Section~\ref{sec:omlgeneratorservice}). This allows executing queries that not only link different CBDs, but also fine-grained queries that relate a CBD to information from different artefacts, as discussed in Section~\ref{sec:queries}.

\begin{figure}
    \centering
    \includegraphics[scale=0.37]{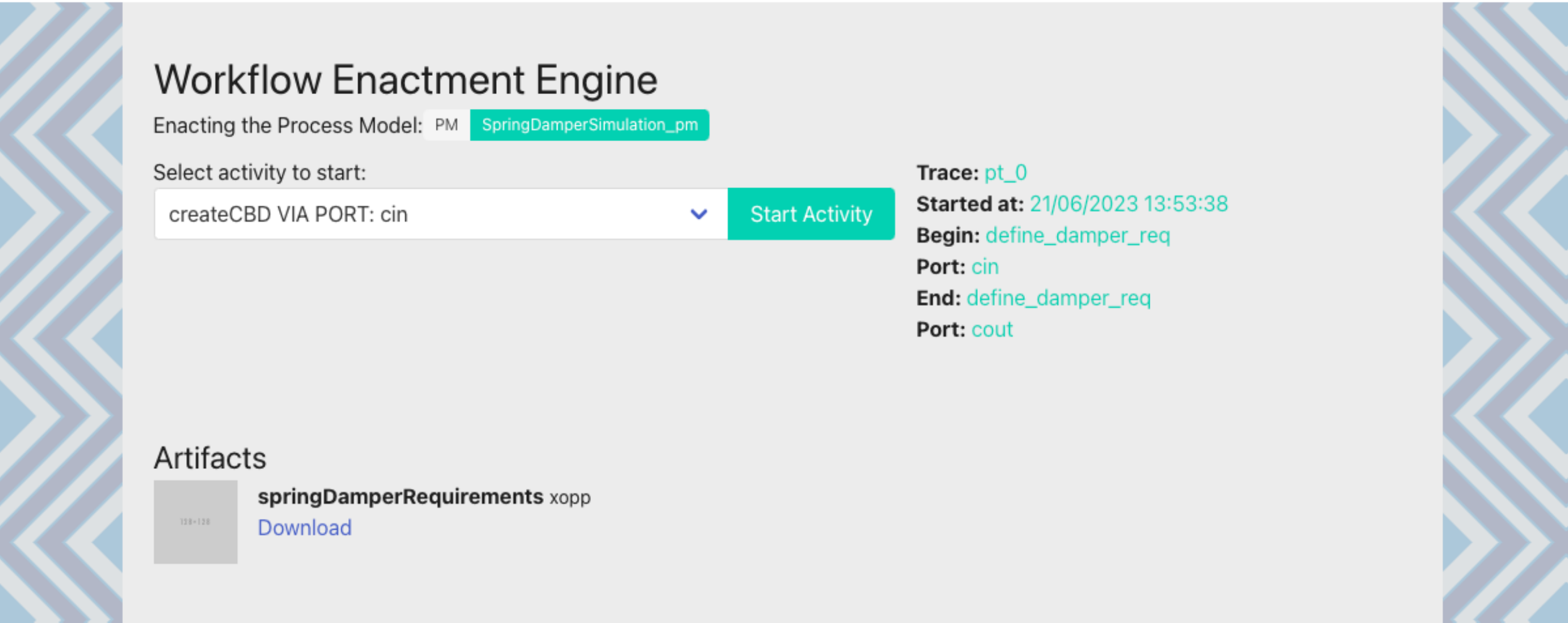}
    \caption{Example of starting an activity in WEE.}
    \label{fig:wee_download}
\end{figure}

Artefacts that have been uploaded by activities are also made available for other activities that require them. Figure~\ref{fig:wee_download} shows the screen to start the following activity \textit{createCBD}. As it needs the \textit{springDamperRequirements} artefact previously uploaded, we made it available for the user to download here. Although, in this case, the uploaded artefact was required by the next activity, it may not always be the case. Artefacts may be available in different phases of the workflow. However, the user does not need to worry about gathering artefacts to start an activity because WEE makes them available in the expected task, which is quite convenient. 

The process model enactment follows until the user reaches a final node, and the trace terminates. Nevertheless, it is possible to stop an enactment at any time and resume it later at the point the user pauses it. For instance, the user may close the WEE window during an enactment, and on the next day, after selecting the PM he was enacting, WEE shows a list of non-concluded enactments. Then, the user can choose the one he wants to continue enacting until its termination. Once the traces are stored in the knowledge graph, it is possible to query them to search for specific information, for example, the latest version of an artefact, or discover the activity that generated a particular artefact, but also reason on the data from the enactment or from artefacts. This can be executed using Graph Explorer (Section~\ref{sec:graphexplorer}) or executing queries directly on Fuseki. However, for the latter, it is required knowledge not only of SPARQL but also of how our ontology is structured. Nevertheless, WEE also provides a REST API to read data from process models and traces. Via HTTP calls, it is possible to obtain all the traces of a particular process model, only finished traces, only unfinished traces, and also get the nodes of a process model. The returned result can be JSON or XML descriptions of the elements. Such a mechanism is useful for tool integration. For instance, it is used by the Draw.io plugin to render visualisations of the process traces generated by WEE. Other tools could also benefit from the same API. More details about WEE can be seen in its git repository, including tutorials on how to use the tool\footnote{\censortext{https://msdl.uantwerpen.be/git/lucasalbertins}/wee/wiki}.

\subsection{Draw.io Frontend}

\begin{figure}[htbp]
    \centering
    \includegraphics[width=\textwidth]{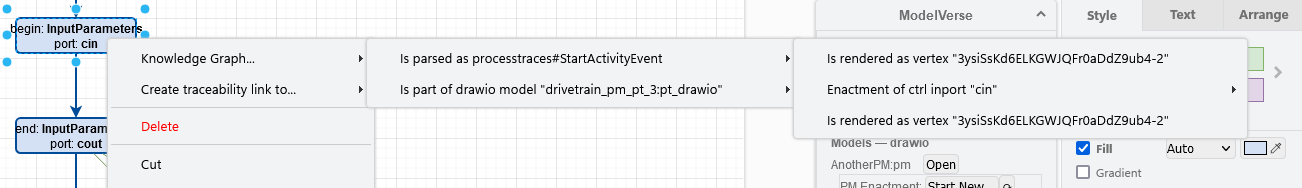}
    \caption{Draw.io plugin extended context menu.}\label{fig:drawio-context-menu}
\end{figure}

Draw.io Frontend is a plugin written for the open-source diagramming tool draw.io. We use draw.io for the authoring of various visual DSLs, and the plugin queries Fuseki in the background to find out whether a diagram is an instance of such a DSL. When this is the case, it uses traceability links from concrete syntax (draw.io shapes) to abstract syntax (DSL model elements) to inform the user about the types of elements and possibly additional inter-model traceability links. Exploration of these links happens via context menus. For instance, it is possible to navigate from a ``StartActivity''-event in a process trace to the control flow input port that was used to start the activity (see screenshot in Figure~\ref{fig:drawio-context-menu}). Its source code can be found in its Git repository\footnote{\censortext{https://msdl.uantwerpen.be/git/jexelmans}/drawio/src/master/src/main/webapp/myPlugins/dtdesign.js}.

% [INSERT SCREENSHOT OF CONTEXT MENU]

\subsection{Fuseki}

Fuseki is an RDF graph database and SPARQL query engine.
In our implementation, Fuseki contains two separate datasets: (1) the \emph{model dataset}, which contains the data of all model versions and traceability links between them, and (2) the \emph{process trace dataset}, which contains (possibly ongoing) event traces of enacted workflows, and all traceability links between this data. The model dataset is generated via (again generated) OML data, whereas the process trace dataset is inserted directly into Fuseki as OWL/RDF data by WEE.

\begin{figure}[htbp]
    \centering
    \includegraphics[width=\textwidth/2]{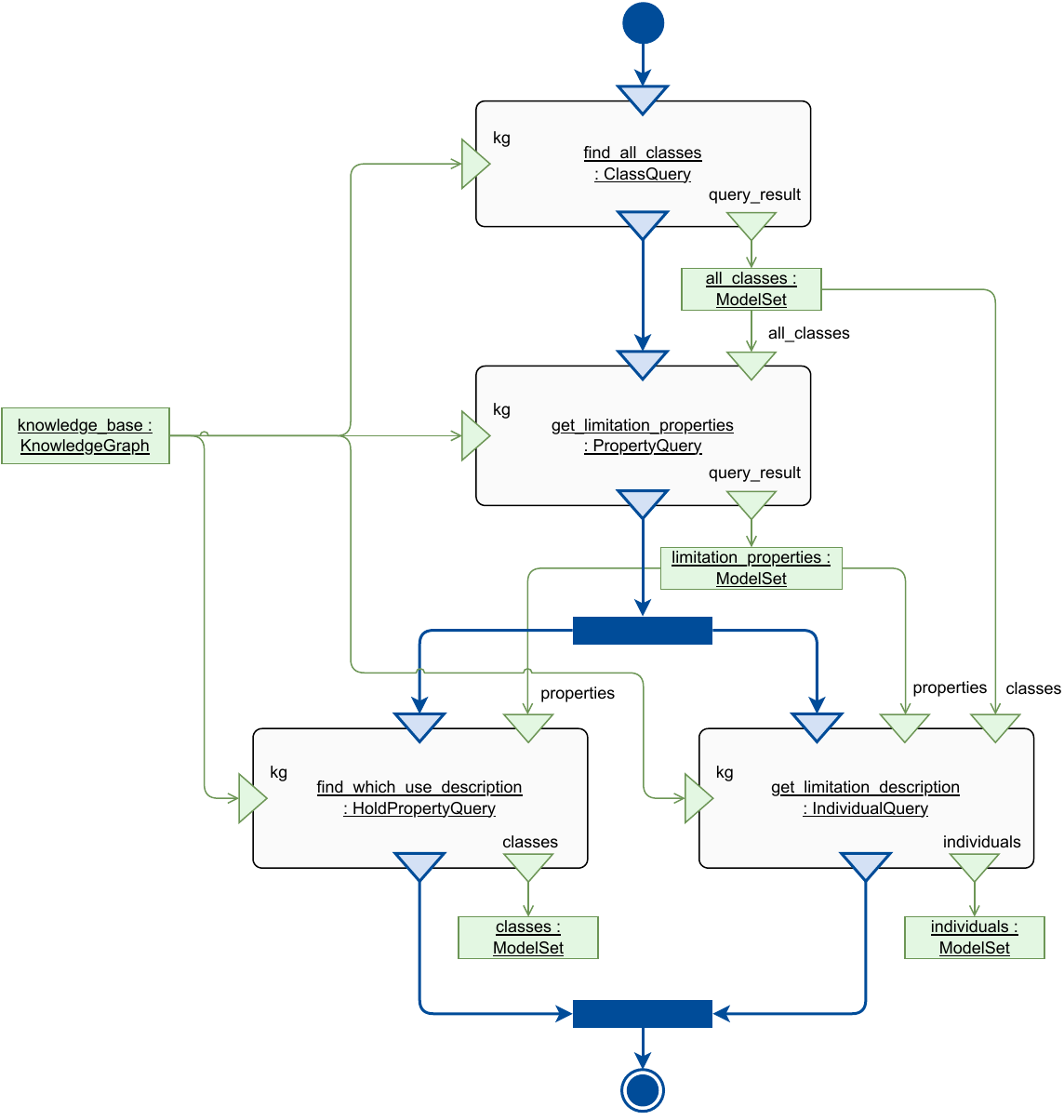}
    \caption{Workflow model for extracting information from the Knowledge Graph using SPARQL queries.}\label{fig:extraction-workflow}
\end{figure}

Other components of our architecture only \emph{read} Fuseki's data via its SPARQL query engine.
We provide example SPARQL queries to alleviate the traversal of the FTG+PM. Prefixes are left out for the sake of brevity.
Figure~\ref{fig:extraction-workflow} is a workflow which can be followed to get a general feel of the knowledge graph in question, and all its queries fall into the structure exploration queries from Section~\ref{sec:querying}.

%We use SPARQL queries to perform all the data extraction as explained in Section~\ref{sec:querying}.
Getting all the types is a good to place to start when encountering an unfamiliar knowledge graph.
The types related to the FTG+PM are of interest here.
The query in Listing~\ref{code:sparql-types} will return all class types as \emph{a} is a shortcut for \emph{rdf:type}.

\begin{lstlisting}[language=SPARQL, caption=SPARQL query which returns all possible types., basicstyle=\ttfamily\small, label=code:sparql-types]
SELECT DISTINCT ?type
WHERE {
  ?entity a ?type .
}
\end{lstlisting}

We proceed to delve deeper into one specific type once we have an overview of all available types in the Knowledge Graph.
Here, in Listing~\ref{code:sparql-type-properties}, we chose to find all available properties of the Formalism type.
For our ontology, this returns 8 values:
\begin{enumerate}
    \item http://www.w3.org/1999/02/22-rdf-syntax-ns\#type
    \item http://ua.be/sdo2l/vocabulary/base/base\#hasGUID
    \item http://ua.be/sdo2l/vocabulary/formalisms/pm\#occursAsArtifact
    \item http://ua.be/sdo2l/vocabulary/formalisms/cs\_as\#renderedAs
    \item http://ua.be/sdo2l/vocabulary/formalisms/traceability\_model\#traceLinkFrom
    \item http://www.w3.org/2002/07/owl\#sameAs
    \item http://ua.be/sdo2l/vocabulary/formalisms/ftg\#isInputOf
    \item http://ua.be/sdo2l/vocabulary/formalisms/ftg\#isOutputOf
\end{enumerate}

\noindent\begin{minipage}{.45\textwidth}
\begin{lstlisting}[language=SPARQL, caption=SPARQL query returning all properties of a certain type., basicstyle=\ttfamily\small, label=code:sparql-type-properties]
SELECT DISTINCT ?property
WHERE { 
    ?limitation a ftg:Formalism .
    ?limitation ?property ?value .
}
\end{lstlisting}
\end{minipage}\hfill
\begin{minipage}{.45\textwidth}
\begin{lstlisting}[language=SPARQL, caption=SPARQL query returning instance of a type and their name property., basicstyle=\ttfamily\small, label=code:sparql-instances-of-types]
SELECT DISTINCT ?individual ?name
WHERE {
   ?individual a ftg:Formalism .
   ?individual ftg:isInputOf ?name .
}
\end{lstlisting}
\end{minipage}

As we now have a list of all available properties for the \emph{ftg:Formalism} we can proceed to query the parameters which values we want.

To come back to the composite query example from Section~\ref{sec:querying}, we present the query which in our system will deliver on that promise.
Listing~\ref{code:sparql-composite} returns the artefact location.
We do this by first looking for the activity type which creates requirements. We then follow its link to the formalism it generates and filter for requirement formalisms.
From here on, we move to the process trace by finding the artefact which corresponds to the formalism type we just found.
The process trace we are looking for will contain an instance of this artefact. So, we match and filter for the requirement artefact instances in the process trace.
% We look for the last version of this artefact.
We then return its location property to the end user.

% Using these query concepts, we can construct a query on the spring-mass-damper system
% to figure out the initial model (first version) whose later versions led to the correct parameter estimation and its value.

\begin{lstlisting}[language=SPARQL, caption=Composite SPARQL query showing multiple categories being used in conjuntion., basicstyle=\ttfamily\small, label=code:sparql-composite]
SELECT DISTINCT ?prelement ?tag ?location
WHERE {
   ?individual a ?type .
   ?type rdfs:subClassOf* ftg:Transformation .
   ?individual base:hasGUID "CreateModelAndEstimateParameters" .
   ?individual ftg:hasOutput ?formalism .
   ?formalism base:hasGUID "CBD" .
   ?formalism pm:occursAsArtifact ?iri .
   ?element pt:relatesTo ?iri .
   ?element a ?atype .
   ?atype rdfs:subClassOf pt:Artifact .
   ?element base:nextVersionOf* ?prelement .
   ?prelement pt:tag ?tag .
   ?prelement pt:hasLocation ?location .
}
\end{lstlisting}

\subsection{Federation services}
\label{subsec: Federation services}
When working with existing large-scale systems, we do not want to convert all data to RDF, as this would lead to a blowup in data size.
Therefore, we implemented middleware\footnote{\censortext{https://msdl.uantwerpen.be/git/arys}/spendpoint} and services to solve the problem of having to store all the data in the knowledge graph.

\begin{figure}[htbp]
    \centering
    \includegraphics[width=0.9\textwidth]{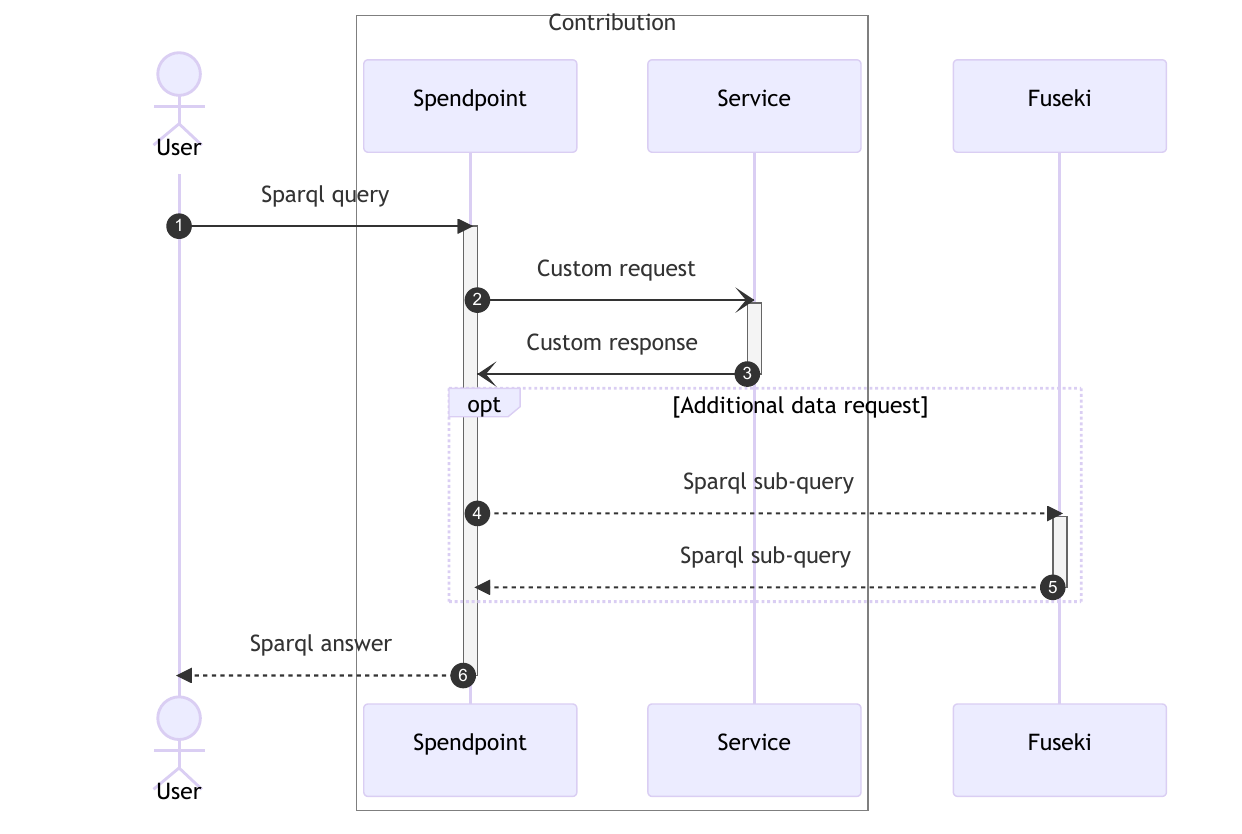}
    \caption{Order of operations for a custom service call.}\label{fig:service-sequence-diagram}
\end{figure}

The middleware intercepts SPARQL queries and performs intermediate operations, like calling a service.
This process is shown in Figure~\ref{fig:service-sequence-diagram}.
We start with a user who wants to perform a query or service call.
The query itself is structured as in Listing~\ref{code:outliers}.
Note the \emph{SERVICE} and \emph{BIND} keywords which are used to perform the sub-query.
Once the query arrives at the middleware, it is processed, and the relevant parts are extracted.
The parameters required for the subsequent service call are all present within the \emph{BIND} instruction.
Spendpoint then performs a call to the specified service, passing on all the parameters.
Once the service response is received, we proceed to parse it and convert it into a graph representation so it can be sent to the user in the form of a SPARQL response.
Note that steps 1 and 4 can also be reversed: the main query can be sent to the Fuseki server, which will forward the relevant part towards the Spendpoint middleware.

This way, we can create models of the data, which is then populated on-demand by a service call.
The middleware can operate on any type of endpoint and converts the acquired data into SPARQL-compliant resources.
We can add service endpoints to expand the possible list of operations.
For the proof of concept, we provide 3 services:
\begin{enumerate}
    \item An outlier service
    \item A CSV to PARQUET converter
    \item An example service which returns predefined data
\end{enumerate}

\begin{lstlisting}[language=SPARQL, caption=SPARQL query for outlier detection., basicstyle=\ttfamily\small, label=code:outliers]
SELECT ?outlier ?outlier_relation ?outlier_value WHERE {
  SERVICE <{{ service_endpoint }}> {
    SELECT ?outlier ?outlier_relation ?outlier_value WHERE {
      BIND(
        dtf:outlier(
          "{{ outlier_file }}",
          "{{ outlier_column }}",
          "{{ outlier_artefact }}"
        ) AS ?outlier
      ) .
      FILTER (?outlier_relation IN (tabular:holdsContent, tabular:hasRowPosition))
    }
  }
}
\end{lstlisting}

\subsection{Graph Explorer}
\label{sec:graphexplorer}

The Graph Exploring Tool (or GET)\footnote{\censortext{https://msdl.uantwerpen.be/git/arys}/graph-exploring-tool} is a desktop application that allows exploring the knowledge graph using SPARQL queries. The tool includes templates for quick traversal of compliant knowledge graphs. The tool separates actions into multiple panels (Figure~\ref{fig:get-annotated-ui}):

\begin{figure}[htbp]
    \centering
    \includegraphics[width=\textwidth]{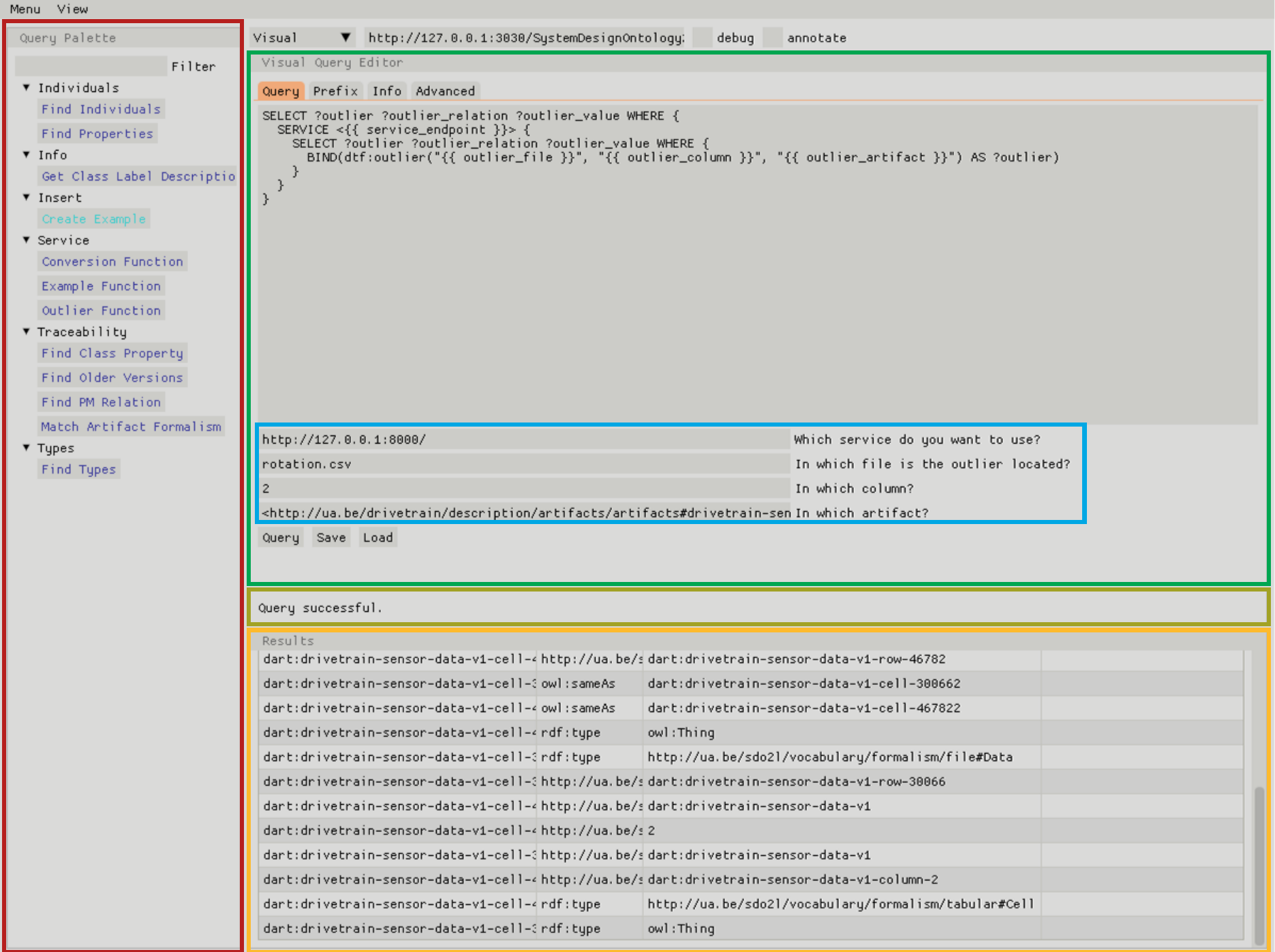}
    \caption{Annotated Graph Exploring Tool user interface.}\label{fig:get-annotated-ui}
\end{figure}

\begin{itemize}
    \item The left Query Palette panel (red, leftmost) allows loading a predefined template. Hovering over the name of the template will provide a tooltip with the description of the action carried out by that template.
    \item The centre and main Query Editor panel (green, upper) is where the template is shown once it has been chosen. The template itself can be edited or the provided fields (blue) can be used to insert references. The reference fields support dragging and dropping.
    The query can then be performed by clicking the \textit{Query} button.
    \item The small one-line panel (yellow, below green) shows the status. This includes messages on whether a query was successful. Any error will also be displayed right here.
    \item The most important panel is the Results panel (orange, bottom). Any information returned by the query will be displayed here. You can drag and drop any of the fields in the result onto the parameters fields in the template.
\end{itemize}

Listing~\ref{code:outliers} is an example of a template which is used by the tool.
The templated values are filled in using the GUI and can be dragged and dropped from the results panel.
Advanced users are also able to define their own templates or perform custom queries.
All templates need the query, its parameters, and an explanation of what the template does.

% JE: I've outcommented this, because I don't think it adds much:
% \paragraph{Ad-hoc querying with Fuseki}
% Fuseki also provides a web-interface for performing SPARQL queries.
% This interface can be used by anyone familiar with SPARQL.
% The Spendpoint middleware we developed works with any SPARQL compliant query. It is therefore possible to use this interface for service calls.

%% file: sections/casestudy.tex
\section{Case Study} % Dennis
\label{sec:casestudy}
In this section the obtained model management strategy is applied to a study looking for alternative measurement strategies in a drivetrain setup, in the field of mechanical engineering. Section \ref{subsec: Motivation Drivetrain} explains the motivation to explore alternative measurement strategies on existing drivetrains. Section \ref{subsec: Difficulties drivetrain} indicates the issues and difficulties that were encountered in the study without the help of an external model management strategy. Section \ref{subsec: Improvement by environment} explains how the obtained model management strategy and environment can improve the workflow for exploring different models and strategies for alternative measurement solutions. Section \ref{subsec: Sub-application} describes how the outlier detection service can be applied to the drivetrain case and indicates which type of aspects can be given to a data retrieval query. Finally section \ref{subsec: summary case study} summarises the conclusions of using this tool in this type of applications.

\subsection{Motivation for alternative measurement strategies}
\label{subsec: Motivation Drivetrain}

\begin{figure}[!t]\centering
	\includegraphics[width=9cm]{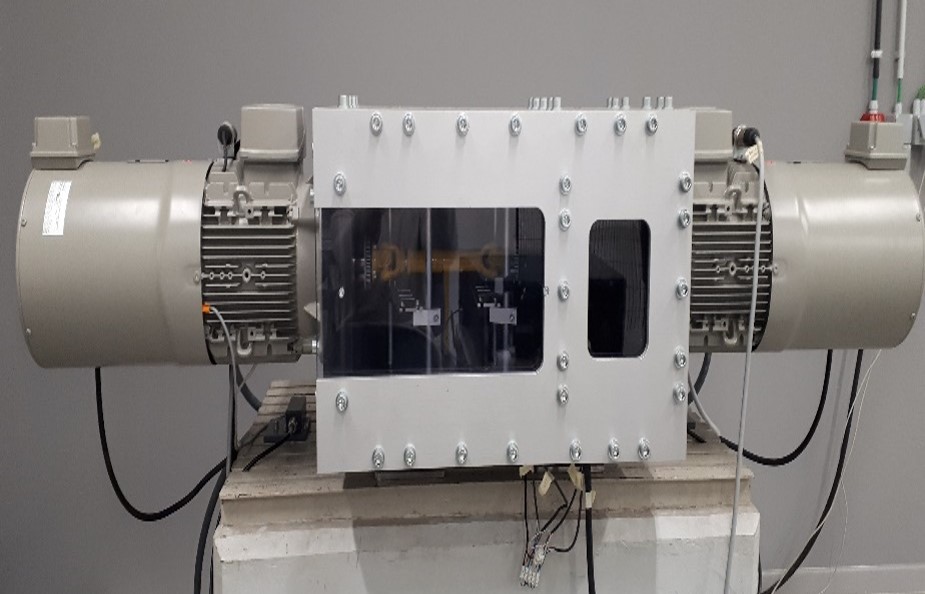}
	\caption{Drivetrain setup.}\label{FIG_drivetrain}
\end{figure}

The case study focuses on a study performed on a drivetrain setup consisting of two induction motors positioned in a back-to-back configuration, with an interchangeable driveshaft \cite{8010352}. In this setup, one motor serves as the drive motor responsible for controlling the speed, while the other motor acts as a load motor, mimicking a specific torque profile. Both motors are equipped with rotational position and acceleration sensors, and a torque sensor is mounted between the driveshaft and the load motor. The setup is shown in Fig. \ref{FIG_drivetrain}.

The objective of the study was to explore an alternative approach for measuring the rotational states of motion, and if possible, the torque as well. The reason to look for alternative solutions is the intrusiveness and high cost of the current state-of-the-art sensors. One potential method involves using accelerometers placed on the periphery of the driveshaft, and combining the readings from these sensors with a physical model that describes the accelerometers' rotation within the gravitational field. This method enables flexible sensor placement on existing drivelines, as the sensors can be added on the driveshaft without having to dismount it.

During the model exploration phase, various aspects were considered, including sensitivity analyses of different model parameters, ranging from basic to more complex models. Optimisation methods were employed to determine unknown parameters of both the model and sensors. Furthermore, the study evaluated whether the increased model complexity and increased challenges in identifying the required parameters are justified by the improved accuracy. Both simulation and experimental validation were conducted, which presented important challenges in managing both the models and the obtained experimental data.

\subsection{Difficulties in model management for research}
\label{subsec: Difficulties drivetrain}

Throughout the model exploration phase, extensive investigations were conducted involving various model combinations and data sources, each yielding distinct results and conclusions. This process resulted in the creation of an extensive folder structure with numerous subfolders, each containing a different approach. However, due to the absence of a comprehensive structured framework to organise the required information, consistency was lost between the different folders. This made it increasingly difficult to navigate through them and locate the necessary models. Furthermore, since the direction in which the models would evolve was uncertain, a consistent naming convention was not implemented. Consequently, the file names became more inconsistent with the content as the model exploration process progressed. The following sections outline the issues encountered and explain how the lack of proper tools to manage the relevant information resulted in significant time loss.
\subsubsection{Folder structure and file naming}
\begin{figure}[!t]\centering
	\includegraphics[scale=0.6]{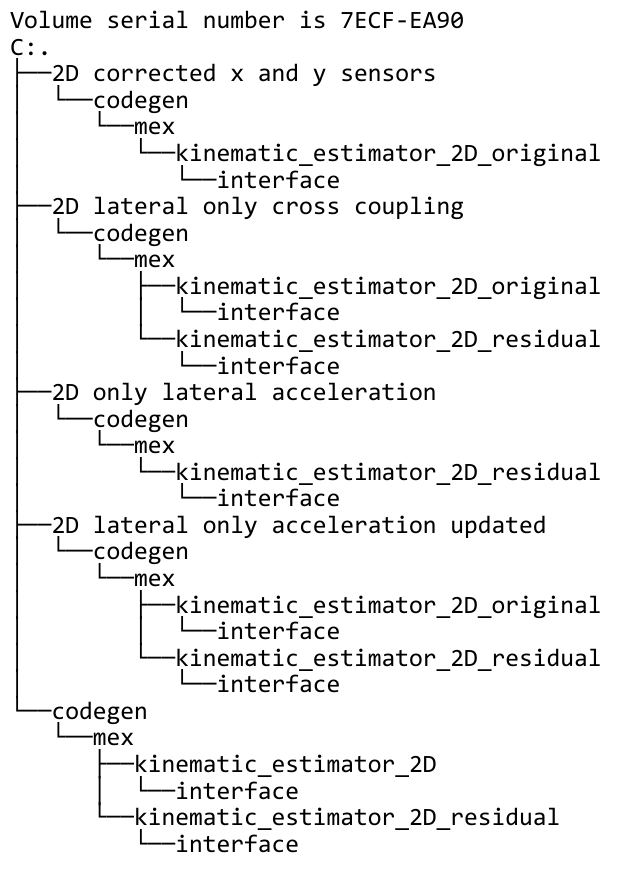}
	\caption{Original folder structure.}\label{FIG_folder_struct_2}
\end{figure}

\begin{figure}[!t]\centering
	\includegraphics[scale=0.6]{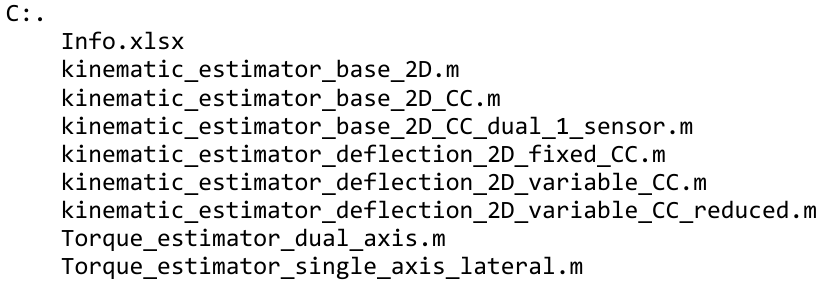}
	\caption{Original folder structure.}\label{FIG_core_models}
\end{figure}

To show an example, Fig. \ref{FIG_folder_struct_2} shows the folder structure in an early phase of the study, where different aspects were investigated throughout the different folders. In the folder named '2D lateral only cross coupling', only the influence of cross coupling in accelerometers in the lateral direction was taken into account. In the folder named '2D only lateral acceleration', both the cross coupling and mounting angle were taken into account, also only in the lateral direction. At the time, this way of naming appeared convenient and clear, but as more combinations of terms in the models were investigated, new ways of naming were needed. Those were not consistent with the original way of naming, making it more difficult to keep track of the contents. Within both folders the goal was to investigate whether the unknown parameters could be obtained through optimisation. An important mistake here was that the model file named 'kinematic\_estimator\_2D\_residual' had different contents in the different main folders, being different parameters, requiring different inputs, etc. Due to this, a large obstacle encountered during the exploration phase was that a file with the correct model name was used, but it originated from a wrong folder that explored a certain strategy different from the one that was envisioned. The reason this was not noticed was because of different models having the same file name in different main folders, making it unclear from which folder this model eventually originated. When selecting this specific model for further deployment, unexpected results were observed, leading to a time consuming search for the error in the model, as it was not clear what caused these unexpected results. Had all of the model files had a clear naming convention, or a consistent versioning strategy tracing each version to its corresponding conclusions and validation strategy, this could have been avoided.
\newline
To deal with this issue at the time, after the error was found, all models were put together in one folder location, and the path to these folders is added to the scripts that call for these models when evaluating them, preventing accidental copy-paste errors. An example of this folder is shown in Fig. \ref{FIG_core_models}.

\subsubsection{Model versions}
\begin{figure}[!t]\centering
	\includegraphics[width=12cm]{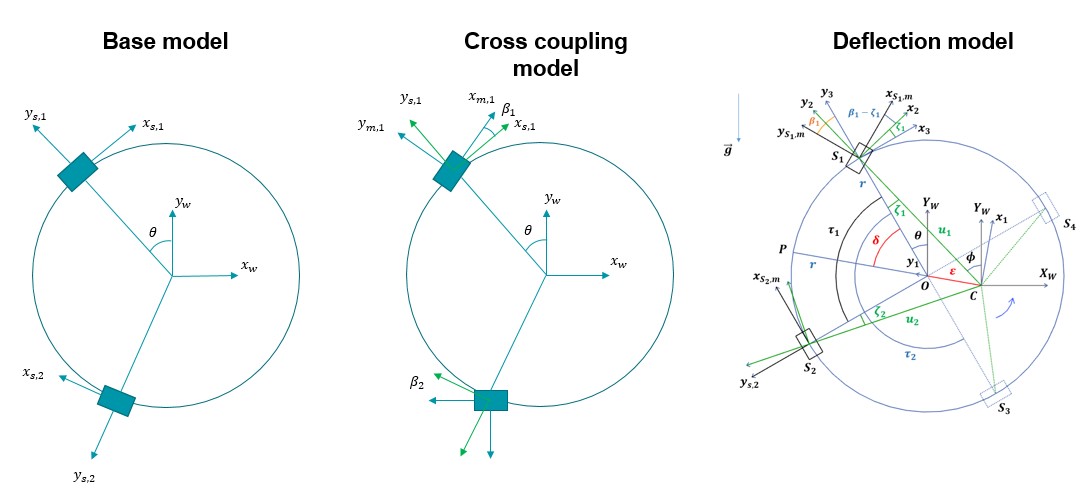}
	\caption{Kinematic models of different complexity.}\label{FIG_model_complexity}
\end{figure}

An important issue with above mentioned strategy however is that one specific model can also have different versions. Fig. \ref{FIG_model_complexity} shows 3 different kinematic models each showing a different level of complexity in representing rotating accelerometers.
For each of these models, a different version can for example use a different discretisation method for the same physical model, or an adaptive covariance update instead of a constant one, etc. This also calls for an external strategy to keep track of all versions with their respective conclusions. As a first solution to this issue, a simple Excel file containing the version updates was used together with their conclusions and important remarks. This however might not be sufficient in the long run, as not all relevant information to be able to navigate the model versions later on might be available. Using this method to go back to a certain version, the model implementation itself needs to be updated each time, where it would be more efficient to have copies of every version that can be accessed at any time. However, doing this in an ad-hoc manner, while having no consistent naming system or information storage tool available, it can quickly become complicated and error-prone.

\subsubsection{The need for experiment modelling}

Another important problem source was the experimental data used to evaluate the models. To ensure accurate parameter identification, specific criteria had to be met by the data associated with each model. These criteria encompassed various aspects such as a minimum sampling frequency, the frequency spectrum content of the acceleration, a threshold for rotational acceleration in constant speed profiles, and more. Moreover, not only the requirements for the data but also the data source itself could vary. This variation included the use of different accelerometers or even a distinct driveshaft, which had the potential to impact both the model and sensor parameters. Consequently, it became essential to carefully document relevant details of the experimental data, such as its content, the objectives of the conducted experiments, setup configurations, the actual equipment used, sampling frequencies, and so on. When moving on to a model of different complexity, the data requirements may or may not change. Possibly the required data might already exist from previous experiments. However, locating it can be a tedious task if it hasn't been appropriately linked to the necessary information, especially when dealing with a multitude of experiments. To avoid having to inspect the data manually or redoing experiments of which the data already exists, it is essential that all the required information is well documented for each experiment. All of the above mentioned aspects also apply to simulations that emulate the sensor data. For simulations, scripts with parameter settings were used to emulate the setup configuration. By keeping track of the exact content of the scripts, all these simulations are repeatable and can be reused when needed. Section \ref{subsec: Improvement by environment} describes how the developed tools can improve the model design workflow by providing an environment that allows for uploading the essential files together with their documentation for each activity in performing an experiment on the drivetrain.

\subsection{Design flow improvement by integration of the obtained environment}
\label{subsec: Improvement by environment}

\begin{figure}[!t]\centering
	\includegraphics[width=9cm]{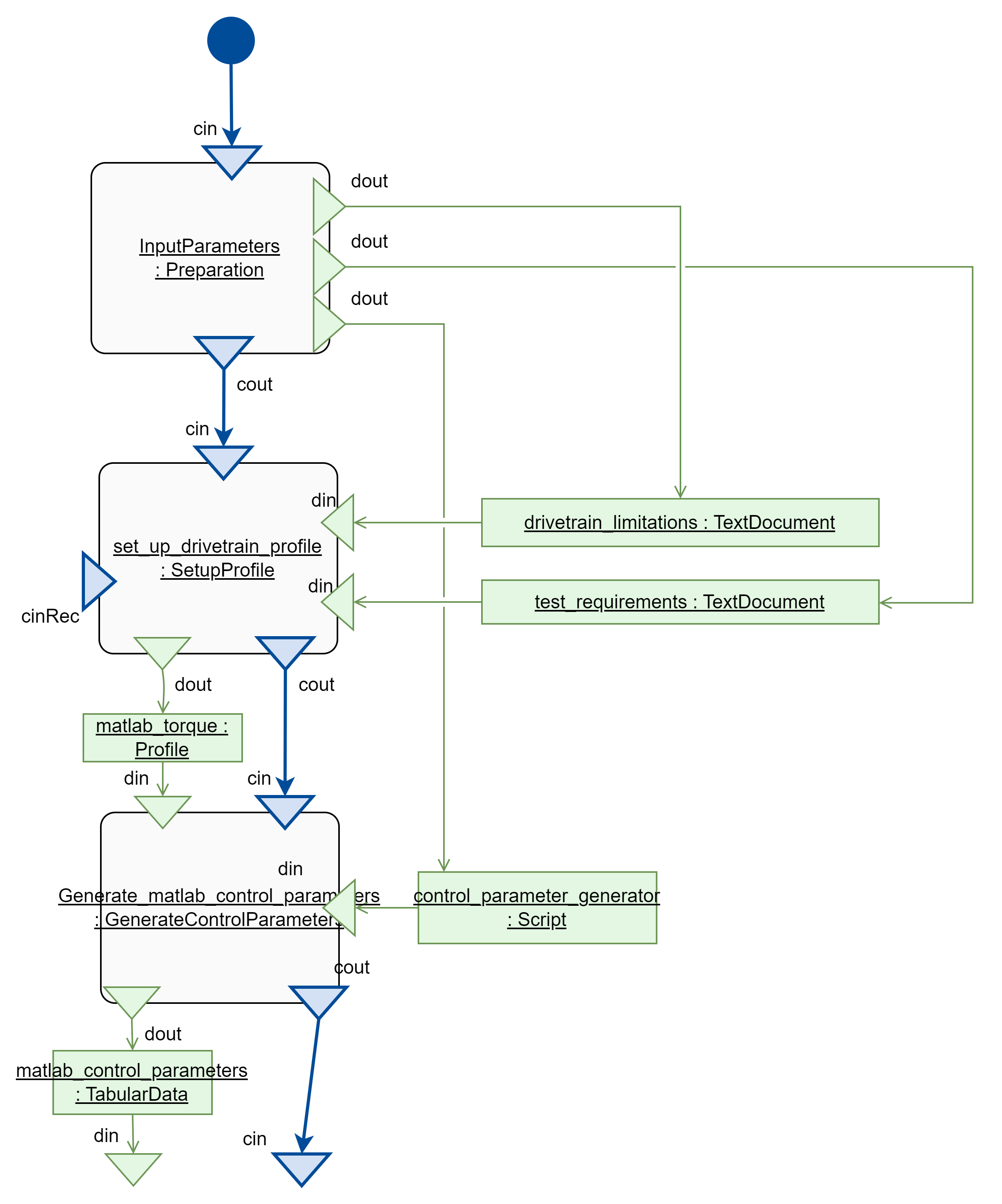}
	\caption{Part of the process model representing the workflow for conducting experiments on the drivetrain.}\label{FIG_PM_drivetrain_exp}
\end{figure}

The issues described in section \ref{subsec: Difficulties drivetrain} can be tackled by applying the model management framework as discussed in the previous sections. For each step in the workflow, the reasoning behind it, the conclusions, and the exact data used can all be documented in a structured way, where it is later accessible by executing queries on the resulting knowledge graph. Fig. \ref{FIG_PM_drivetrain_exp} shows part of the process model for conducting experiments on the drivetrain which was mentioned in section \ref{subsec: Motivation Drivetrain}. The artefact 'drivetrain\textunderscore limitations' can be a document summarising the setup and its limitations, a datasheet, or other documentation. Given the limitations, a torque profile is created that matches the test requirements. Both files and the generated torque profile can then be uploaded through the WEE environment as described in section \ref{subsec: WEE} and linked to the activity 'set\textunderscore up\textunderscore drivetrain\textunderscore profile'. In this way, the data generated during this experiment will always be linked to the setup, the test requirements and the actual torque profile. In the next activity 'Generate\textunderscore matlab\textunderscore control\textunderscore parameters', the coefficients for a 10\textsuperscript{th} order linear controller are generated from the torque profile and a script that generates these coefficients. By uploading all artefacts in WEE, including the controller coefficients which are saved in tabular form, all settings and correct files will again be linked to the data that will result from this experiment. The rest of the process model is not shown, but follows a similar approach that links all relevant artefacts to the corresponding activity. A similar process model can also be set up for simulations, where all relevant settings and simulated models can then be included as artefacts.
Such a process model can also be built for developing and evaluating a new physical model of certain complexity. The WEE environment then allows for uploading each model version, with its associated reasoning, conclusions and requirements for validation data, be it experimental or simulation based. This ensures that all conclusions match with the correct model version and used data. This approach effectively deals with all issues described in section \ref{subsec: Difficulties drivetrain}. The Graph Exploring Tool (GET) can afterwards be used to efficiently query the resulting Knowledge Graph (KG) for the required model or data, as well as execute services on the content of the KG. In this way, a lot of time can be saved in retrieving both a required model, or data that meets specific requirements. The next section briefly describes the outlier detection service in context of the drivetrain study. It also indicates some requirements that can be given to a query to retrieve certain data.

\subsection{Sub-application: outlier detection and querying for specific data}
\label{subsec: Sub-application}
The developed framework not only allows for keeping track of all essential information, it also provides services that can be executed on the existing data. For example, outlier detection as described in section \ref{subsec: Federation services} can be used to check if a certain data set respects the limitations for outliers. Should a new physical model require a limitation on the outliers of the rotational speed data for parameter identification, and this has not been included as information during the experiment, this service allows for querying for a data set that complies with this limitation. This eliminates the need to manually inspect the available data or to redo an experiment for which the data already exists. 

The GET also allows the user to retrieve experimental data that meets specific requirements by querying the resulting KG. Such a query can for example demand the data to comply with all of the following requirements:
\begin{enumerate}
    \item Time series data
    \item Acceleration data $(m/s^2)$
    \item Measured with an accelerometer of type X
    \item Measured on test setup Y
    \item Motor speed of setup Y constant at 1000rpm
    \item Sampling frequency of 2kHz
    \item Minimum data length of 10s
\end{enumerate}
This allows for efficient retrieval of this data, without having to inspect the data manually.

\subsection{Summary case study}
\label{subsec: summary case study}
To summarise, the provided framework allows for significant improvements on following levels, compared to ad-hoc methods:
\begin{enumerate}
    \item Traceable link between the experimental data and the developed models: Establishing a traceable link between experimental data and the developed models is essential for understanding how the models were constructed and validated. By documenting the data sources, experimental setup, and measurement techniques used during the data collection process, researchers can ensure that the models are based on reliable and representative data. This traceability enables transparency and reproducibility, allowing others to verify and validate the existing models or even newly developed models by referring back to the original experimental data.
    \item Traceable link between the experimental data and the equipment used to obtain it: In model development for this kind of applications, the equipment used during experiments, such as the sensors, test rigs, or even simulation tools, can significantly impact the quality and accuracy of the collected data. Therefore, it is crucial to keep track of the link between the experimental data and the specific equipment used. This traceability provides insights into the reliability, calibration, and limitations of the equipment, ensuring that the data is interpreted within the appropriate context. It also helps in troubleshooting and identifying potential sources of error or bias in the used experimental setup.
    \item Model versioning with a link to a detailed description of the reasoning behind it: As the development of the physical models evolves over time, it becomes necessary to manage different versions or iterations of the models. Model versioning with the developed tools allows to track and document changes made to the models, including modifications, improvements, or updates. By attaching traces to the documentation of the underlying reasoning, assumptions, or methodologies to each version, the development process becomes more transparent and understandable. This documentation facilitates collaboration, knowledge sharing, and informed decision-making by providing a historical context for each of the model versions. It can also help future researchers to build further upon the reasoning behind the models that were developed by other researchers.
\end{enumerate}

All of above aspects are then part of an overarching KG. By having this KG available, together with a tool that allows for domain specific querying, a lot of time can be saved during the model development process. As for the drivetrain study described earlier for example, the mistake of using a file with the correct name from a wrong folder would not have happened had all models been correctly linked to their reasoning and validation conclusions. Instead of looking for the error in the model itself, or validating the model with new data, the KG could have been searched based on the conclusions or reasoning behind this model to find the correct corresponding model. This mistake ultimately led to a time loss of approximately two weeks, which could have been avoided had the proposed framework been available.
\newline
\newline
The main limiting factor of this framework is that for each new workflow the FTG and PM will need to be set up and linked to the correct environment. However, this well outweighs the potential time loss by not having above mentioned traceability available.

%% file: sections/relatedworkv2.tex
\section{Related Work}
\label{sec:relatedwork}

Although model management is a concept we can trace back to the 80s, it has evolved considerably over the last decades. However, we considered it important to discuss works closely related to ours according to their relevance for the industry and that are contextualised in the current trends of system engineering. We have categorised them into three main groups: MBSE methods and tools, digital engineering strategies and (cognitive) digital twins. Afterwards, we present a comparison between the most relevant approaches compared to ours in Table~\ref{tab:comp}.

\subsection{MBSE methods and Tools}

Given the shift from document-centric to model-based systems engineering, the interest in model management strategies and frameworks has increased in recent years. However, the variety of disciplines performed during system engineering activities makes use of a plethora of different notations, languages and formalisms. Therefore, it is essential that model management frameworks provide support for multi-paradigm modelling. In addition to that, the description of workflows for the large number of processes executed is also essential to context the usage of artefacts from different formalisms in an organised flow of activities. 

\subsubsection{Model repositories}

DesignSpace \cite{designspace} is a model repository and \emph{operation-based} versioning system for MBSE. It brings artefacts from third-party tools together in a single graph-like data structure. It integrates with these tools via ``live'' adaptors: the adaptors record edit operations as they happen. As a result, any editor for which a DesignSpace adaptor is available can be used in synchronous collaboration (like Google Docs).
% DesignSpace has hierarchical ``workspaces'', that provide branching and access control: a team may share a workspace that is only visible to the team, branched from a public workspace. A team member may have her own workspace that is only visible to her, branched from the team's workspace.
To the best of our knowledge, DesignSpace does not support data federation or virtualisation; instead, the repository takes ownership of all data. DesignSpace also does not include a workflow enactment service, although it could still be implemented on top of its architecture. DesignSpace supports consistency checking of arbitrary rules over arbitrary elements and has demonstrated to do this in a scalable manner \cite{designspace-efficient-consistency}. DesignSpace does not have its own query language.

OpenMBEE \cite{openmbee} is another model repository and versioning system. It also integrates with third-party tools via adapters. OpenMBEE's adapters are not ``live'': they record snapshots rather than deltas. In the current stable version of OpenMBEE with MMS 4, the model repository uses an SQL database. In their next version of the repository (MMS 5), the developers want to use an OWL/RDF stack. OpenMBEE relies on Syndeia (discussed later) to provide federation. OpenMBEE does not include a workflow enactment engine.

WebGME \cite{webgme} is a web-based modelling and meta-modelling tool. It includes a model repository and a versioning system inspired by Git, but tailored to graphs with \emph{spanning trees}, rather than directory trees of text files. \deleted{WebGME's model repository has a hierarchical ``containment'' relationship: the repository contains models, which contain model elements, which contain other model elements, etc. This containment relation defines a \emph{spanning tree}: every model element is (transitively) contained by the entire repository. A change to an element creates a new version of all its ancestors (including a new version of the model repository). Elements that did not change are re-used, which is similar to how Git works.} WebGME does not include third-party tool adaptors or workflow enactment.

Regarding workflow management and enactment, Product Lifecycle Management (PLM) software proposes to manage a product and its associated data through its development~\cite{plm}. Predominantly, it is used by design and engineering teams working with CAD data and related artefacts. Its primary purpose is to integrate processes for each stage of a product's lifecycle to support supply chain management, but it is also used to improve team collaboration and provide insights on product performance, customer feedback, and market trends. Usually, these tools are commercial-off-the-shelf (COTS) developed by companies that also create CAD software, like Siemens, Dassault and PTC~\cite{siemens,oracle,ptc}.
\added[id=r3c3]{PLM overlaps with our approach, because it usually supports (simple, file-based) versioning (with locking), as well as workflow modelling and enactment. In PLM, this is typically implemented in an ad-hoc manner. Meta-models of artifacts are implicit (``hardcoded''), as opposed to our solution, which has explicit meta-models for all models, including workflow models and traces.
In PLM, relations between data in different artefacts may exist, but usually, these are also implemented in an ad-hoc manner (leading to poor inter-operability between tools of different vendors), as opposed to our solution, where anything can be linked with anything, as long as the data is first translated to OML.
We included 2 PLM tools that seem to push a bit further: Aras Innovator \cite{ArasPLM} has an explicit and open data model \cite{ArasArchitecture}, and 3DEXPERIENCE \cite{3DX} has a unified graph-based data model, like our solution. An deep comparison of PLM tools is beyond our reach, because these tools are proprietary, and vendors are reluctant to share implementation details.
}

\subsubsection{Federation helpers}

Syndeia \cite{syndeia} is a commercial data federation middleware\added[id=r3c3]{, intended not for end-users, but for large industrial tool builders (including PLM vendors).} It brings data from different sources together in a virtual graph. Syndeia does not take ownership of any data --- it only provides federation. If the user wants to have versioning, she needs to bring her own versioning system. If Syndeia has an adaptor for the versioning system, in principle, the complete history of versioned artefacts should become part of the virtual graph. Syndeia can create arbitrary links and carry out automatic synchronisations between arbitrary graph elements.
Syndeia also implements OSLC (see below), so it can ``see'' data from any tool that acts as an OSLC provider.
Syndeia supports Gremlin \cite{gremlin} as a graph query language.
\deleted{Syndeia is intended for software developers to integrate it into their own tools, not for end-users.}
Syndeia does not include workflow enactment.

OpenFlexo \cite{openflexo, OpenFlexoFederation} is another data federation solution. It includes its own meta-modelling language FML, which is similar to class diagrams. It allows for creating domain-specific languages whose instances obtain their data from federated sources. Synchronisation rules between FML instances and federated sources are specified in an operational manner. Synchronisation can be automated and bi-directional. Just like Syndeia, OpenFlexo has no notion of versioning: artefacts themselves have no history, and to our knowledge, OpenFlexo currently has no adaptors for versioning systems. OpenFlexo itself does not come with a workflow language or workflow enactment engine. In \cite{openflexo_multi}, the process language MULTI \cite{multi} was implemented in FML at the level of concrete syntax, abstract syntax, and operational semantics. However, produced/consumed artefacts were not recorded as part of the process traces, and the federation capabilities of OpenFlexo were not used.

Open Services for Lifecycle Collaboration (OSLC) \cite{oslcwebsite, oslcpaper14} is a standard for the exchange and inter-linking of data authored by different MBSE tools. Its ``core'' includes service discovery, a CRUD protocol based on REST and RDF(S), and the ability to define HTML-based ``views'' for data. It further includes a large set of domain-specific schemas, such as requirements, architecture, quality, etc. OSLC implementations can act as producers and/or consumers of data. An OSLC artefact is identified by a URI that is not expected to change. Artefacts can refer to other artefact's URIs. OSLC does not include a repository, versioning or a schema for versioning information. It also does not include workflow enactment.

\subsection{Digital Engineering Strategies}

Noguchi~\cite{Noguchi2020} elaborates on the Digital Engineering Strategy provided by The Department of Defense (DoD). 
The DoD has embarked on a journey towards digital engineering, driven by the goal to enhance system resilience and efficiency~\cite{DOD2018}. Central to this transformation is the DoD's Digital Engineering Strategy (DES), which underscores the integration of advanced modelling and simulation techniques throughout the system lifecycle.
The DoD's Digital Engineering Strategy, introduced to modernise defence acquisition processes, emphasises the use of digital models and simulations to create authoritative and traceable engineering artefacts. This strategy aligns with the principles of model-driven engineering, enabling the creation, exchange, and analysis of digital models to foster collaboration and informed decision-making. The DES advocates for the integration of diverse disciplines, such as systems engineering, software engineering, and cybersecurity.
% TODO I don't know how to cite this: https://man.fas.org/eprint/digeng-2018.pdf

\subsection{Cognitive Digital Twins}

Digital twinning has become a topic of interest for industry and academia in the last decades. It is considered one of the key enablers of industry digitisation~\cite{Tao2019}, which aims at bringing the real (physical) world to digital environments. A Digital Twin (DT) is usually seen as a digital counterpart of a physical system focusing on specific features, for example, monitoring, analysing, experimenting, and simulating physical assets and processes~\cite{Shiva2019}. Given the considerable evolution of Cyber Physical Systems and supporting technologies like the Industrial Internet of Things (IIoT), Cloud Computing, and specialised software, DTs have been applied extensively in several industrial sectors~\cite{Tao2019}. 

More recently, the Cognitive Digital Twin (CDT), or just Cognitive Twin (CT), has emerged as an extended version of DTs with cognitive capabilities. However, there is no consensus on what kind of capabilities a CDT should provide~\cite{Abburu2020a}. Due to the neonate status of CDT research, most of the works do not present concrete implementation focusing on establishing concepts, challenges, and prospective technology to be used for implementing these systems~\cite{Faruque2021,Ali2021,Eirinakis2022}. 

Abburu et al.~\cite{Abburu2020a} propose a three-layered architecture where, at the bottom, we have DTs comprising data and model repositories together with services. In the middle, there is a Hybrid Twin, which is responsible for the orchestration of the DTs. On the top, the Cognitive Twin layer is composed of a knowledge repository and modules responsible for cognitive tasks. Adapters and brokers are also discussed to transfer and transform data from different sources into the layers. Challenges are discussed, and ontology and knowledge graphs are mentioned as technologies to be used in the cognitive twin layer.  In~\cite{Abburu2022b}, more details are provided about the architecture. It is also described as part of the COGNITWIN project, where some industry use case scenarios are described and standards to be used are presented, but no concrete implementation is discussed. 

Eirinakis et al.~\cite{Eirinakis2020} introduces the concept of Enhanced Cognitive Twins in the context of the FACTLOG project. They propose a conceptual architecture composed of a cognition pipeline responsible for learning, event detection, prediction and reasoning over data streams from the physical systems. A second pipeline offers reasoning services by using ML, data analytics techniques, and knowledge graph reasoning approaches. Both pipelines interact using knowledge graphs.  

We have identified two systematic literature reviews published in the area. D'Amico et al.~\cite{DAMICO2022} present a thematic review on semantic DTs, which are also called Cognitive Twins, used in the maintenance context. This paper also discusses an ontological approach to developing this kind of twin. Zheng et al.~\cite{Zheng2022} present a wide survey about CDTs and the common features of the available works, e.g., a CDT is an augmented version of a DT, which contains multiple DT models with unified semantics definitions. Some examples of their capabilities include autonomy, continuous evolution, and intelligent tasks such as attention, perception, comprehension, memory, reasoning, prediction, decision-making, problem-solving, and so on. They also enumerate enabling technologies for CDTs, which include ontology engineering, knowledge graphs, MBSE, PLM, IIoT and cloud computing. They also discuss a reference architecture where the elements considered in the model management layer match what we have implemented in our framework. For instance, the model repository and managers for versioning, consistency, traceability and process management. Although this paper presents a conceptual architecture, it is important to show that our implementation is aligned with the relevant aspects that should be considered for model management. 

Jinzhi et al.~\cite{Jinzhi2022a} distinguishes from the previous works in the sense that a concrete proof-of-concept is presented for CDT in the context of MBSE. Its architecture follows the similar principles discussed in the previous works and also considered in our work, e.g., the usage of ontologies and knowledge graphs as enablers for model and knowledge management and process management to guide the execution of system engineering workflows. This work does not discuss in detail some important aspects like versioning and consistency. They are considered as given from the usage of ontology allied to IT techniques with no further details. For us, versioning is a relevant concern to enable traceability between models. Workflow enactment is supported by a plugin that has been implemented for the Camunda\footnote{https://camunda.com/platform-7/workflow-engine/} process management platform~\cite{Jinzhi2018}. The proposed toolchain translates SysML models to OWL, which is used as input together with simulation results to train a neural network. Afterwards, the generated AI models can support automated parameter selection for system developers instead of using the traditional simulation approach.

The similarities between the CDT architectures and our proposal are evident. For instance, the usage of Semantic Web technologies (OWL, RDF), ontologies to describe domains and relationships between the different models and artefacts, and knowledge graphs to materialise the relevant information to support the desired processes. Although our work focuses on model management, a possible future direction is to extend our framework to provide an extra layer of cognition on top of the knowledge repository. Another crucial difference is that we already describe a toolchain to support these capabilities, while most of the works stay at the conceptual level. Another distinguishable aspect is the importance of workflow (process) enactment as an enabler for model management.  

Table~\ref{tab:comp} classifies related work. When \y happens, it indicates that the approach supports a given feature, while not having it means the opposite. Having a question mark (?) means it is unclear if the work supports the feature or not.
All approaches have a notion of bringing data together in a \textbf{Unified Data Model}, which is typically graph-like (except in the case of Git).
Next, there are two ways, possibly complementary, of bringing data together:
A \textbf{(Model) Repository} acts as a data (model) storage mechanism that takes ownership of the data, \textbf{Federation}, on the other hand, we define as facilitating access to data stored and owned by (possibly multiple) external repositories, without taking ownership.
Model repositories often include \textbf{Versioning} of the data they store. Versioning is itself a deep topic, with many possible approaches, which we do not describe further in this paper.
For model repositories to handle data from third-party tools, this data needs to be translated to the repository's data model. \textbf{Adaptors} fulfil this role. Some solutions, like WebGME, are not intended to integrate with third-party tools (they have their own built-in modelling environment) and therefore do not use adaptors.
With data federation, we differentiate between solutions that virtualise data \textbf{Transparent}ly or not. We define transparent virtualisation as the end-user not having to be aware of the fact that some of the data may be federated. Our own solution is currently not transparent because, when writing a query, the user needs to specify explicitly whether some of the source data must be fetched from an external source. The root of this issue is that the query engine we use works at the level of our (RDF) repository, whereas it should sit at a higher (federated) level. Solutions that focus only on providing federation, such as Syndeia, do not suffer from this issue.
\textbf{Workflow enactment} indicates whether an enactment engine is part of the solution or not. A workflow engine typically requires a model repository to store artefacts produced during enactment.
\textbf{Typing, Abstract Syntax} refers to the ability to define new (model) types, for instance, by meta-modelling. This makes the solution easier to extend with new adaptors. If new types can be defined, \textbf{Conformance Checking} is verifying whether models conform to all the constraints imposed at the type level. A more generic feature is the ability to verify \textbf{Arbitrary (consistency) Rules}, which may describe constraints that include model elements even from different models.
Finally, \textbf{Querying Data} is the ability to access data, whether in a repository or federated, in a uniform manner. Solutions may provide an \textbf{API} and/or a dedicated \textbf{Query Language} (e.g., SPARQL). The latter is usually much more powerful.

\begin{table}[H]
    \centering
    % \begin{tabular}{l||l|l|l|l|l|l|l|l|l|l|l|l|m{3cm}} % Uncomment this line to add the 'purpose' column
    \begin{tabular}{l||l|l|l|l|l|l|l|l|l|l|l|l|l}
    \multicolumn{2}{c|}{}& \multicolumn{3}{c|}{Repository} & \multicolumn{3}{c|}{Federation} & \multicolumn{5}{c}{} \\
    \hline
    \multicolumn{2}{c|}{}&  \rotatebox{90}{Model repository} &  \rotatebox{90}{Versioning}     &  \rotatebox{90}{Adaptors - repository} &  \rotatebox{90}{Data federation}      & \multicolumn{2}{c|}{ \rotatebox{90}{Virtualisation}} &  \rotatebox{90}{Workflow enactment} &  \rotatebox{90}{Typing}      & \multicolumn{2}{c|}{ \rotatebox{90}{Consistency management}} &  \multicolumn{2}{c}{ \rotatebox{90}{Querying data} }
        % &  \multicolumn{1}{c}{\rotatebox{90}{Purpose}}
    \\
    
    \hline
    &  \multicolumn{1}{c|}{\rotatebox{90}{Unified data model}} & & & & & \rotatebox{90}{Non-transparent} &  \rotatebox{90}{Transparent} &  \rotatebox{90}{Included} &  \rotatebox{90}{Abstract syntax} &  \rotatebox{90}{Conform. checking} &  \rotatebox{90}{Arbitrary rules} & \rotatebox{90}{API} & \rotatebox{90}{Query Language} \\
    
    \hline
    \textit{(Model) repositories} \\
    \hline
    
    This paper     & OWL/RDF          & \y     & \y & \y & \y & \y &    & \y & \y & \y & \y &    & \y
        % & Model repository and data federation with versioning, workflow enactment and query language
    \\
    Lu Jinzhi et al.~\cite{Jinzhi2022a,Jinzhi2022b} & OWL/RDF          & \y     &        & \y     & \y     &   &   & \y & \y &        &        &    & \y
        % & Cognitive twin for automated parameter selection
    \\
    DesignSpace \cite{designspace}
                  & graph            & \y     & \y     & \y     &        &    &    &        & \y     & \y     & \y     & \y    
        % & Model repository with versioning and consistency checking
    \\
    Aras PLM \cite{ArasPLM} & RDBMS-based      & \y     & \y & \y     & \y     & ?  & ?  & \y & \y & ?      & ?      & \y    
        % & PLM
    \\
    3DEXPERIENCE \cite{3DX} & graph            & \y     & \y & ?     & \y  & \y & \y & \y & ? & ? & ? & \y & ?
    \\
    OpenMBEE \cite{openmbee}
                  & RDBMS-based                & \y & \y & \y     & \y$^{*}$ & \y$^{*}$ & \y$^{*}$ &        & \y & ?      & \y$^{*}$ & \y & \y$^{*}$
        % & MBSE infrastructure
    \\
    WebGME \cite{webgme}
                  & graph        & \y     & \y     &        &        &    &    &        & \y & \y     & ?      & \y    
        % & Web-based (meta-)modeling environment 
    \\
    Git           & filesystem tree  & \y     & \y     &        &        &    &    &        &        &        &        &        
        % & Text-based versioning system
    \\

    \hline
    \textit{Federation helpers} \\
    \hline

    Syndeia \cite{syndeia}
                  & graph            &        &        &        & \y     & \y & \y &        &    &        & \y     & \y    & \y
        % & Data federation (commercial product)
    \\
    OpenFlexo \cite{openflexo}
                  & FML              &        &        &        & \y     & \y & \y &        & \y     &        & \y     & \y     & \y
        % & Federation
    \\
    OSLC \cite{oslcwebsite}
                  & RDF              &        &        &        & \y     & \y & \y &        & \y & \y     & \y &    & \y
        % & Protocol for data exchange between MBSE tools + set of domain-specific schemas
    \\

    \hline
    \textit{Conceptual works} \\
    \hline

    %\multicolumn{14}{l}{Conceptual Works} \\
    Abburu et al.~\cite{Abburu2020a,Abburu2022b}    & -- & \y &    & \y     & \y & ? & ? &    &    &    &    & \y    
        % & Architecture for implementation of Hybrid and Cognitive Twins
    \\
    Eirinakis et al.~\cite{Eirinakis2020} & -- &    &    & \y & \y &    &    &    &    &    &    &        
        % & Architecture for Enhanced Cognitive Twin to support decision-making
    \\
    \end{tabular}
    
    \vspace{0.3cm}
    %\begin{minipage}[t]{0.30\linewidth}
    %    \begin{description}
            %\item[$^1$] MMS 4
            %\item[$^2$] Poor man's versioning
            %\item[$^3$] Transformation to OML
            %\item[$^4$] Data communication layer
            %\item[$^5$] e.g., CSV adapter
    %    \end{description}
    %\end{minipage}
    \begin{minipage}[t]{\linewidth}
        \begin{description}
            \item[$^{*}$] uses Syndeia
            %\item[$^6$] FTG+PM, WEE
            %\item[$^7$] CAMUNDA + OSLC
            %\item[$^8$] BPMN            
        \end{description}
    \end{minipage}
    %\begin{minipage}[t]{0.25\linewidth}
    %    \begin{description}
            %\item[$^9$] OML
            %\item[$^{10}$] SysML/KARMA
            %\item[$^{11}$] RDBMS schema-like
            %\item[$^{12}$] MDK
            %\item[$^{13}$] Class Diagrams
            %\item[$^{14}$] Not for end-users
            %\item[$^{15}$] RDFS
    %    \end{description}
    %\end{minipage}
    %\begin{minipage}[t]{0.15\linewidth}
    %    \begin{description}
            %\item[$^{16}$] OWL
            %\item[$^{17}$] SHACL
            %\item[$^{18}$] SPARQL
            %\item[$^{19}$] Gremlin
    %    \end{description}
    %\end{minipage}
    % TODO I can't get this to work.
    %\captionsetup{format=table, labelformat=tablelabel, labelsep=colon, justification=centering, singlelinecheck=off}
    \caption{Classification of related work.}\label{tab:comp}
\end{table}
%\end{sidewaystable}

%% file: sections/conclusion.tex
\section{Conclusions}
\label{sec:conclusions}

%\com{Arkadiusz}{Describe main problem again, so why.}
%\com{Arkadiusz}{Describe how the tools solve this.}
%\com{Arkadiusz}{Make text more 'flow'able.}

Companies face specific challenges, such as retaining knowledge of standard procedures and successful cases, and managing the generation and storage of multiple versions of documents and artefacts. The role of experiments in system engineering is significant in providing evidence of system performance, but we also acknowledge the complexity involved in executing them.

We leveraged ontologies to define system engineering elements formally. This approach enables the construction of a knowledge graph to manage system engineering data and allows us to link artefact versions to executed workflows.
Our ontology-based framework consists of seven components, together solving our requirements for explicitly modelled workflows and their automated enactment, and a model repository with versioning, fine-grained traceability, typing, consistency checking and a query language.

We showed how our framework is applied to a spring-mass-damper running example and to a larger engineering use case involving the creation of smart sensors for a drivetrain.

\subsection{Limitations and future work} % Joeri
There are a few limitations to our solution. Many are inherent to our technology choices of OML and OWL/RDF.

\paragraph{Poor man's versioning does not scale.}
As already explained in Section~\ref{sec:implementation}, OML has no built-in solution for versioning. OML users are advised to use Git for versioning the OML documents (types and instances) themselves. However, this makes it impossible to write queries that reason about an artefact's history, because the versioning information is \emph{external} to OML and the generated RDF data. For instance, a query such as ``in which version was this element added?'' cannot be performed. To make versioning information available from within OML/OWL/RDF, we use the ``poor man's approach'' of creating a new copy of the artefact for every new version. This solution is sufficient to demonstrate our proof-of-concept, but ultimately doesn't scale. An optimised versioning system for graph data (e.g., one that only stores \emph{deltas}, as in \cite{ExelmansPRVT23}) would be necessary, and would need to integrate with the SPARQL server, such that the version history can also be queried. We are not aware of any existing versioning system for RDF data that integrates with a SPARQL server, and this could be an interesting ally for further research.

\paragraph{\replaced[]{No language evolution.}{Types are not instances.}}\label{sec:notypeevolution}
\added[]{We already mentioned that we currently do not support language evolution. We see two possible ways to add support in the future:}
\begin{itemize}
    \item \added[]{We could apply ``poor man's versioning'' also to the OML vocabularies: for every new vocabulary version, we simply add a new vocabulary (with a new, unique name that indicates the version). However, a limitation then needs to be worked around: in OML, everything is either a type, or an instance, but never both. At the type level, a \emph{relation} between two types \emph{concepts} is not a link itself, but denotes the \emph{possibility} of having links at the instance level. It is therefore impossible to create versioning links at the type level. A possible workaround is to create a singleton instance for every vocubulary version, and to create versioning links between these singletons.}
    
    \item \added[id=r3c1]{A more radical way would be to define our own meta-meta-model as an OML vocabulary, such that meta-models become instances themselves. Conformance between instance and type can be stored as a simple link. A benefit of this approach is that conformance can be checked \emph{a posteriori}: if a language evolves in a non-destructive manner, existing models will conform to both the old and the new language version. Conformance links to the types in the new language can simply be added when new language versions are added. The drawback of this approach is that we would have to implement our own conformance checking algorithm (e.g., as a set of SHACL constraints), as we can no longer rely on OML's.}
\end{itemize}

\paragraph{OML lacks reasoner and query language.}
OML needs to be transformed to OWL/RDF, before it can be used by a reasoner (to perform a consistency check) or query engine. As a result, any detected inconsistency is presented to the user in the form of a (somewhat cryptic) error at the level of OWL/RDF, which the user must manually trace back to the original OML document(s). Likewise, SPARQL queries work on generated the RDF data, and return RDF data, not OML. To make OML more usable, we think a reasoner and query language should be created that works on the level of OML (possibly built around an OWL/RDF reasoner and query engine).

\paragraph{Expanding the framework.} 
We have implemented an infrastructure where knowledge can be properly stored and related to relevant concepts using knowledge graphs. So far, we are using this knowledge to support and improve the system development processes based on domain-specific necessities from the engineers. However, we can advance one step further by adding a cognitive layer, similar to what has been recently proposed for cognitive digital twins. This would provide even more refined services based on AI and reasoning techniques adding more capabilities for the end-user.

\paragraph{\added[id=r3c4]{Application in a complete system life cycle.}} 
\added[id=r3c4]{So far, we have applied our framework to parts of the system life cycle, more precisely, during system experimentation. Although we are aware that current PLM software provides support for model management, but usually, extensibility and reasoning are not easily achieved in these environments. We do not want to replace them but just show a more flexible and extensible approach to model management using knowledge graphs and ontologies. However, we have not yet applied our framework in a full system development context. Even though we believe it is possible because the life cycle can be supported by modelling the workflows explicitly. When these workflows are enacted, traceability information (e.g., versioning and refinement) can be added to the models automatically to record the evolution of models generated through a life cycle. Additional traceability information can be recorded in the knowledge graph according to the user's needs. Nevertheless, we expect to perform such an application and evaluation in the future.}

%% file: cas-refs.bib
@article{Codd70,
author = {Codd, E. F.},
title = {A Relational Model of Data for Large Shared Data Banks},
year = {1970},
issue_date = {June 1970},
publisher = {Association for Computing Machinery},
address = {New York, NY, USA},
volume = {13},
number = {6},
issn = {0001-0782},
url = {https://doi.org/10.1145/362384.362685},
doi = {10.1145/362384.362685},
journal = {Commun. ACM},
month = {jun},
pages = {377–387},
numpages = {11}
}

@article{Chen76,
author = {Chen, Peter Pin-Shan},
title = {The Entity-Relationship Model—toward a Unified View of Data},
year = {1976},
issue_date = {March 1976},
publisher = {Association for Computing Machinery},
address = {New York, NY, USA},
volume = {1},
number = {1},
issn = {0362-5915},
url = {https://doi.org/10.1145/320434.320440},
doi = {10.1145/320434.320440},
journal = {ACM Trans. Database Syst.},
month = {mar},
pages = {9–36},
numpages = {28}
}

@inproceedings{Elam1980,
  title={Model Management Systems: an Approach to Decision Support in Complex Organizations},
  author={Joyce J. Elam and John C. Henderson and Louis W. Miller},
  booktitle={International Conference on Interaction Sciences},
  year={1980}
}

@article{APPLEGATE1986,
title = {Model management systems: Design for decision support},
journal = {Decision Support Systems},
volume = {2},
number = {1},
pages = {81-91},
year = {1986},
issn = {0167-9236},
doi = {https://doi.org/10.1016/0167-9236(86)90124-7},
url = {https://www.sciencedirect.com/science/article/pii/0167923686901247},
author = {Lynda M Applegate and Benn R Konsynski and Jay F Nunamaker}
}

@article{Geoffrion87,
author = {Geoffrion, Arthur M.},
title = {An Introduction to Structured Modeling},
year = {1987},
issue_date = {May 1987},
publisher = {INFORMS},
address = {Linthicum, MD, USA},
volume = {33},
number = {5},
issn = {0025-1909},
journal = {Manage. Sci.},
month = {may},
pages = {547–588},
numpages = {42}
}

@article{KIMBROUGH198627,
title = {A graph representation for management of logic models},
journal = {Decision Support Systems},
volume = {2},
number = {1},
pages = {27-37},
year = {1986},
issn = {0167-9236},
doi = {https://doi.org/10.1016/0167-9236(86)90118-1},
url = {https://www.sciencedirect.com/science/article/pii/0167923686901181},
author = {Steven O Kimbrough}
}

@Article{Jones90,
  author={Christopher V. Jones},
  title={{An Introduction to Graph-Based Modeling Systems, Part I: Overview}},
  journal={ORSA Journal on Computing},
  year=1990,
  volume={2},
  number={2},
  pages={136-151},
  month={May},
  doi={10.1287/ijoc.2.2.136},
  url={https://ideas.repec.org/a/inm/orijoc/v2y1990i2p136-151.html}
}

@article{Jones91,
  title={An Introduction to Graph-Based Modeling Systems, Part II: Graph-Grammars and the Implementation},
  author={Christopher V. Jones},
  journal={ORSA Journal on Computing},
  year={1991},
  volume={3},
  pages={180-206}
}

@article{BALDWIN1991,
title = {The evolution and problems of model management research},
journal = {Omega},
volume = {19},
number = {6},
pages = {511-528},
year = {1991},
issn = {0305-0483},
doi = {https://doi.org/10.1016/0305-0483(91)90002-B},
url = {https://www.sciencedirect.com/science/article/pii/030504839190002B},
author = {AA Baldwin and D Baldwin and TK Sen}
}

@article{Bharadwaj92,
author = {Bharadwaj, Anandhi and Choobineh, Joobin and Lo, Amber and Shetty, Bala},
title = {Model Management Systems: A Survey},
year = {1992},
issue_date = {1992},
publisher = {J. C. Baltzer AG, Science Publishers},
address = {USA},
volume = {38},
number = {1–4},
issn = {0254-5330},
url = {https://doi.org/10.1007/BF02283650},
doi = {10.1007/BF02283650},
journal = {Ann. Oper. Res.},
month = {dec},
pages = {17–67},
numpages = {51}
}

@article{GRUBER1993,
title = {A translation approach to portable ontology specifications},
journal = {Knowledge Acquisition},
volume = {5},
number = {2},
pages = {199-220},
year = {1993},
issn = {1042-8143},
doi = {https://doi.org/10.1006/knac.1993.1008},
url = {https://www.sciencedirect.com/science/article/pii/S1042814383710083},
author = {Thomas R. Gruber}
}

@book{plm,
  title={Product lifecycle management},
  author={Saaksvuori, Antti and Immonen, Anselmi},
  year={2008},
  publisher={Springer Science \& Business Media}
}

@article{Mostefai2006,
author = {Mostefai, Sihem and Bouras, Abdelaziz and Batouche, M.},
year = {2006},
month = {01},
pages = {},
title = {Effective Collaboration in Product Development via a Common Sharable Ontology},
volume = {2},
journal = {International Journal of Computational Intelligence}
}

@article{BOCK2010,
title = {Ontological product modeling for collaborative design},
journal = {Advanced Engineering Informatics},
volume = {24},
number = {4},
pages = {510-524},
year = {2010},
note = {Construction Informatics},
issn = {1474-0346},
doi = {https://doi.org/10.1016/j.aei.2010.06.011},
url = {https://www.sciencedirect.com/science/article/pii/S1474034610000558},
author = {Conrad Bock and XuanFang Zha and Hyo-won Suh and Jae-Hyun Lee}
}

@book{Hitzler2010,
  author = {Hitzler, Pascal and Krötzsch, Markus and Rudolph, Sebastian},
  ee = {https://www.crcpress.com/Foundations-of-Semantic-Web-Technologies/Hitzler-Krotzsch-Rudolph/9781420090505},
  isbn = {9781420090505},
  publisher = {Chapman and Hall/CRC Press},
  title = {Foundations of Semantic Web Technologies},
  year = 2010
}

@article {Hitzler2012,
	title = {OWL 2 Web Ontology Language: Primer (Second Edition)},
	year = {2012},
	month = {12/11/2012},
	pages = {W3C Recommendation},
	url = {http://www.w3.org/TR/owl2-primer},
	author = {Pascal Hitzler and Markus Kr{\"o}tzsch and Bijan Parsia and Peter F. Patel-Schneider and Sebastian Rudolph}
}

@techreport{sparql2013,
  author={W3C},
  editor = {Prud'hommeaux, Eric and Harris, Steve and Seaborne, Andy},
  institution = {W3C},
  title = {{SPARQL 1.1 Query Language}},
  url = {http://www.w3.org/TR/sparql11-query},
  year = 2013
}

@inproceedings{webgme,
  author       = {Mikl{\'{o}}s Mar{\'{o}}ti and
                  Tam{\'{a}}s Kecsk{\'{e}}s and
                  R{\'{o}}bert Keresk{\'{e}}nyi and
                  Brian Broll and
                  P{\'{e}}ter V{\"{o}}lgyesi and
                  L{\'{a}}szl{\'{o}} Jur{\'{a}}cz and
                  Tihamer Levendovszky and
                  {\'{A}}kos L{\'{e}}deczi},
  editor       = {Daniel Balasubramanian and
                  Christophe Jacquet and
                  Pieter Van Gorp and
                  Sahar Kokaly and
                  Tam{\'{a}}s M{\'{e}}sz{\'{a}}ros},
  title        = {Next Generation (Meta)Modeling: Web- and Cloud-based Collaborative
                  Tool Infrastructure},
  booktitle    = {Proceedings of the 8th Workshop on Multi-Paradigm Modeling co-located
                  with the 17th International Conference on Model Driven Engineering
                  Languages and Systems, MPM@MODELS 2014, Valencia, Spain, September
                  30, 2014},
  series       = {{CEUR} Workshop Proceedings},
  volume       = {1237},
  pages        = {41--60},
  publisher    = {CEUR-WS.org},
  year         = {2014},
}

@misc{RDF2014,
  author = {Schreiber, Guus and Raimond, Yves},
  editor = {Schreiber, Guus and Raimond, Yves},
  howpublished = {Online},
  title = {RDF 1.1 Primer W3C Working Group Note},
  url = {https://www.w3.org/TR/rdf11-primer/},
  year = 2014
}

@techreport{bibSHACL,
  editor = {Knublauch, Holger and Kontokostas, Dimitris},
  institution = {W3C},
  month = jul,
  title = {{Shapes constraint language (SHACL)}},
  url = {https://www.w3.org/TR/shacl/},
  year = 2017
}

@article{protege,
  author       = {Mark A. Musen},
  title        = {The prot{\'{e}}g{\'{e}} project: a look back and a look
                  forward},
  journal      = {{AI} Matters},
  volume       = {1},
  number       = {4},
  pages        = {4--12},
  year         = {2015},
  url          = {https://doi.org/10.1145/2757001.2757003},
  doi          = {10.1145/2757001.2757003},
  bibsource    = {dblp computer science bibliography, https://dblp.org}
}

@article{Xiao2019,
    author = {Xiao, Guohui and Ding, Linfang and Cogrel, Benjamin and Calvanese, Diego},
    title = "{Virtual Knowledge Graphs: An Overview of Systems and Use Cases}",
    journal = {Data Intelligence},
    volume = {1},
    number = {3},
    pages = {201-223},
    year = {2019},
    month = {06},
    issn = {2641-435X},
    doi = {10.1162/dint_a_00011},
    url = {https://doi.org/10.1162/dint\_a\_00011},
    eprint = {https://direct.mit.edu/dint/article-pdf/1/3/201/683759/dint\_a\_00011.pdf}
}

@article{Noy2019,
author = {Noy, Natasha and Gao, Yuqing and Jain, Anshu and Narayanan, Anant and Patterson, Alan and Taylor, Jamie},
title = {Industry-Scale Knowledge Graphs: Lessons and Challenges},
year = {2019},
issue_date = {August 2019},
publisher = {Association for Computing Machinery},
address = {New York, NY, USA},
volume = {62},
number = {8},
issn = {0001-0782},
url = {https://doi.org/10.1145/3331166},
doi = {10.1145/3331166},
journal = {Commun. ACM},
month = {jul},
pages = {36–43},
numpages = {8}
}

@book{Allemang2020,
author = {Allemang, Dean and Hendler, Jim and Gandon, Fabien},
title = {Semantic Web for the Working Ontologist: Effective Modeling for Linked Data, RDFS, and OWL},
year = {2020},
isbn = {9781450376174},
publisher = {Association for Computing Machinery},
address = {New York, NY, USA},
edition = {3},
volume = {33}
}

@article{Hitzler21,
author = {Hitzler, Pascal},
title = {A Review of the Semantic Web Field},
year = {2021},
issue_date = {February 2021},
publisher = {Association for Computing Machinery},
address = {New York, NY, USA},
volume = {64},
number = {2},
issn = {0001-0782},
url = {https://doi.org/10.1145/3397512},
doi = {10.1145/3397512},
journal = {Commun. ACM},
month = {jan},
pages = {76–83},
numpages = {8}
}

@inproceedings{OpenFlexoFederation,
  author       = {Fahad Rafique Golra and
                  Antoine Beugnard and
                  Fabien Dagnat and
                  Sylvain Gu{\'{e}}rin and
                  Christophe Guychard},
  editor       = {Lidia Fuentes and
                  Don S. Batory and
                  Krzysztof Czarnecki},
  title        = {Addressing modularity for heterogeneous multi-model systems using
                  model federation},
  booktitle    = {Companion Proceedings of the 15th International Conference on Modularity,
                  M{\'{a}}laga, Spain, March 14 - 18, 2016},
  pages        = {206--211},
  publisher    = {{ACM}},
  year         = {2016},
  url          = {https://doi.org/10.1145/2892664.2892701},
  doi          = {10.1145/2892664.2892701},
}

@misc{W3C,
  author={World Wide Web Consortium},
  title = {Semantic Web},
  howpublished = {\url{https://www.w3.org/standards/semanticweb/}},
  note = {Accessed: 2023-06-12}
}

@misc{ArasPLM,
  author={Aras},
  title = {Aras Innovator},
  howpublished = {\url{https://www.aras.com/en/technology}},
  note = {Accessed: 2023-01-25}
}

@misc{ArasArchitecture,
  author={Mark Reisig},
  title = {Architecture Matters - Blogpost},
  howpublished = {\url{https://www.aras.com/community/b/english/posts/architecture-matters}},
  note = {Accessed: 2023-01-29}
}

@misc{3DX,
  author={Dassault Syst\`{e}mes},
  title = {3DEXPERIENCE},
  howpublished = {\url{https://www.3ds.com/3dexperience}},
  note = {Accessed: 2023-01-25}
}

@misc{neo4j,
    author = {Neo4j},
    title = {From Graph to Knowledge Graph: How a Graph Becomes a Knowledge Graph},
    howpublished = {\url{https://neo4j.com/blog/from-graph-to-knowledge-graph-how-a-graph-becomes-a-knowledge-graph/}},
    year={2021},
  note = {Accessed: 2023-06-12}    
}

@misc{oml,
    author = {Maged Elaasar and Nicolas Rouquette},
    title = {Ontological Modeling Language},
    howpublished = {\url{http://www.opencaesar.io/oml/}},
    year={2023},
  note = {Accessed: 2023-06-12}   
}

@misc{rosetta,
    author = {Maged Elaasar},
    title = {An Eclipse IDE that supports OML natively - Rosetta},
    howpublished = {\url{https://github.com/opencaesar/oml-rosetta}},
    year={2023},
  note = {Accessed: 2023-06-12}   
}

@misc{jena,
    author = {Apache Software Foundation},
    title = {Apache Jena},
    howpublished = {\url{https://jena.apache.org/}},
    year = 2021,
  note = {Accessed: 2023-06-12}   
}

@misc{ontop,
    author = {Ontop},
    title = {Ontop - A Virtual Knowledge Graph System},
    howpublished = {\url{https://ontop-vkg.org/}},
  note = {Accessed: 2023-06-12}   
}

@misc{siemens,
    author = {Siemens},
    title = {Teamcenter PLM Siemens Software},
    howpublished = {\url{https://plm.sw.siemens.com/en-US/teamcenter/}},
  note = {Accessed: 2023-06-12}   
}

@misc{oracle,
    author = {Oracle},
    title = {Product Lifecycle Management Software Oracle},
    howpublished = {\url{https://www.oracle.com/scm/product-lifecycle-management/}},
  note = {Accessed: 2023-06-12}   
}

@misc{ptc,
    author = {PTC},
    title = {Arena PLM and QML Solutions PTC},
    howpublished = {\url{https://www.ptc.com/en/products/arena}},
  note = {Accessed: 2023-06-12}   
}

@misc{openmbee,
    author = {OpenMBEE},
    title = {OpenMBEE - Open Model-based Engineering Environment},
    howpublished = {\url{https://www.openmbee.org/}},
  note = {Accessed: 2023-06-12}   
}

@misc{syndeia,
  author={Intercax},
  title = {Syndeia},
  howpublished = {\url{https://intercax.com/products/syndeia/}},
  note = {Accessed: 2023-08-28}
}

@misc{gremlin,
  author={Apache},
  title = {Gremlin Query Language},
  howpublished = {\url{https://tinkerpop.apache.org/gremlin.html}},
  note = {Accessed: 2023-08-28}
}

@misc{oslcwebsite,
  author={OSLC Open Project},
  title = {Open Services for Lifecycle Collaboration},
  howpublished = {\url{https://open-services.net/}},
  note = {Accessed: 2023-08-28}
}

@misc{openflexo,
  author={OpenFlexo},
  title = {OpenFlexo},
  howpublished = {\url{https://www.openflexo.org}},
  note = {Accessed: 2023-09-05}
}

@misc{multi,
  author={Jo\~{a}o Paulo A. Almeida and
          Thomas K\"{u}hne and
           Adrian Rutle and
             Manuel Wimmer},
  title = {The MULTI Process Challenge – EMISAJ Special Issue Version},
  howpublished = {\url{https://homepages.ecs.vuw.ac.nz/foswiki/pub/Groups/MultiLevelModeling/MultiPublications/EMISAJ_Process_Modeling_Challenge.pdf}},
  note = {Accessed: 2023-09-05}

}

@article{openflexo_multi,
  author       = {Sylvain Gu{\'{e}}rin and
                  Jo{\"{e}}l Champeau and
                  Jean{-}Christophe Bach and
                  Antoine Beugnard and
                  Fabien Dagnat and
                  Salvador Mart{\'{\i}}nez Perez},
  title        = {Multi-Level Modeling with Openflexo/FML {A} Contribution to the Multi-Level
                  Process Challenge},
  journal      = {Enterp. Model. Inf. Syst. Archit. Int. J. Concept. Model.},
  volume       = {17},
  year         = {2022},
  url          = {https://doi.org/10.18417/emisa.17.9},
  doi          = {10.18417/emisa.17.9},
  bibsource    = {dblp computer science bibliography, https://dblp.org}
}

@inproceedings{oslcpaper14,
  author       = {Mehrdad Saadatmand and
                  Alessio Bucaioni},
  title        = {{OSLC} Tool Integration and Systems Engineering - The Relationship
                  between the Two Worlds},
  booktitle    = {40th {EUROMICRO} Conference on Software Engineering and Advanced Applications,
                  {EUROMICRO-SEAA} 2014, Verona, Italy, August 27-29, 2014},
  pages        = {93--101},
  publisher    = {{IEEE} Computer Society},
  year         = {2014},
  url          = {https://doi.org/10.1109/SEAA.2014.64},
  doi          = {10.1109/SEAA.2014.64},
}

@ARTICLE{8010352,
  author={Forrier, Bart and Naets, Frank and Desmet, Wim},
  journal={IEEE Transactions on Industrial Electronics}, 
  title={Broadband Load Torque Estimation in Mechatronic Powertrains Using Nonlinear Kalman Filtering}, 
  year={2018},
  volume={65},
  number={3},
  pages={2378-2387},
  doi={10.1109/TIE.2017.2739709}}

@misc{incose,
    author = {INCOSE},
    title = {Systems Engineering Vision 2035},
    howpublished = {\url{https://www.incose.org/about-systems-engineering/se-vision-2035}},
    year=2022,
  note = {Accessed: 2023-06-12}   
}

@INPROCEEDINGS{randy22,
  author={Paredis, Randy and Exelmans, Joeri and Vangheluwe, Hans},
  booktitle={2022 Annual Modeling and Simulation Conference (ANNSIM)}, 
  title={Multi-Paradigm Modelling For Model Based Systems Engineering: Extending The FTG + PM}, 
  year={2022},
  volume={},
  number={},
  pages={461-474},
  doi={10.23919/ANNSIM55834.2022.9859391}
}

@article{dsct,
  author = {Amerongen, J.},
  year = {2010},
  month = {01},
  pages = {},
  title = {Dynamical Systems for Creative Technology}
}

@article{ryan11,
  author = {Monroe, Ryan and Shaw, Steven},
  year = {2011},
  month = {03},
  pages = {},
  title = {On the transient response of forced nonlinear oscillators},
  volume = {67},
  journal = {Nonlinear Dynamics},
  doi = {10.1007/s11071-011-0174-4}
}

@article{juan07,
  author = {Seck Tuoh Mora, Juan and Gonzalez-Hernandez, M. and Hernández Romero, Norberto and Trejo, Aaron and Chapa, Sergio},
  year = {2007},
  month = {05},
  pages = {833-848},
  title = {Modeling Linear Dynamical Systems by Continuous-Valued Cellular Automata},
  volume = {18},
  journal = {International Journal of Modern Physics C - IJMPC},
  doi = {10.1142/S0129183107010589}
}

@article{moody,
  author={Moody, Daniel},
  journal={IEEE Transactions on Software Engineering}, 
  title={The ``Physics'' of Notations: Toward a Scientific Basis for Constructing Visual Notations in Software Engineering}, 
  year={2009},
  volume={35},
  number={6},
  pages={756-779},
  doi={10.1109/TSE.2009.67}
}

@article{designspace,
    author = {Tr\"{o}ls, Michael Alexander and Marchezan, Luciano and Mashkoor, Atif and Egyed, Alexander},
    title = {Instant and Global Consistency Checking during Collaborative Engineering},
    year = {2022},
    issue_date = {Dec 2022},
    publisher = {Springer-Verlag},
    address = {Berlin, Heidelberg},
    volume = {21},
    number = {6},
    issn = {1619-1366},
    url = {https://doi.org/10.1007/s10270-022-00984-4},
    doi = {10.1007/s10270-022-00984-4},
    journal = {Softw. Syst. Model.},
    month = {dec},
    pages = {2489–2515},
    numpages = {27}
}

@inproceedings{designspace-efficient-consistency,
  title={Efficient detection of inconsistencies in a multi-developer engineering environment},
  author={Demuth, Andreas and Riedl-Ehrenleitner, Markus and Egyed, Alexander},
  booktitle={Proceedings of the 31st IEEE/ACM International Conference on Automated Software Engineering},
  pages={590--601},
  year={2016}
}

@INPROCEEDINGS{Shiva2019,
  author={Tavallaey, Shiva Sander and Ganz, Christopher},
  booktitle={2019 24th IEEE International Conference on Emerging Technologies and Factory Automation (ETFA)}, 
  title={Automation to Autonomy}, 
  year={2019},
  volume={},
  number={},
  pages={31-34},
  doi={10.1109/ETFA.2019.8869329}}

@ARTICLE{Tao2019,
  author={Tao, Fei and Zhang, He and Liu, Ang and Nee, A. Y. C.},
  journal={IEEE Transactions on Industrial Informatics}, 
  title={Digital Twin in Industry: State-of-the-Art}, 
  year={2019},
  volume={15},
  number={4},
  pages={2405-2415},
  doi={10.1109/TII.2018.2873186}}

@INPROCEEDINGS{Abburu2020a,
  author={Abburu, Sailesh and Berre, Arne J. and Jacoby, Michael and Roman, Dumitru and Stojanovic, Ljiljana and Stojanovic, Nenad},
  booktitle={2020 IEEE International Conference on Engineering, Technology and Innovation (ICE/ITMC)}, 
  title={COGNITWIN – Hybrid and Cognitive Digital Twins for the Process Industry}, 
  year={2020},
  volume={},
  number={},
  pages={1-8},
  doi={10.1109/ICE/ITMC49519.2020.9198403}
}

@article{Zheng2022,
author = {Xiaochen Zheng and Jinzhi Lu and Dimitris Kiritsis},
title = {The emergence of cognitive digital twin: vision, challenges and opportunities},
journal = {International Journal of Production Research},
volume = {60},
number = {24},
pages = {7610-7632},
year  = {2022},
publisher = {Taylor \& Francis},
doi = {10.1080/00207543.2021.2014591},

URL = {https://doi.org/10.1080/00207543.2021.2014591},
eprint = {https://doi.org/10.1080/00207543.2021.2014591}
}

@INPROCEEDINGS{Abburu2022b,
  author={Abburu, Sailesh and Berre, Arne J. and Jacoby, Michael and Roman, Dumitru and Stojanovic, Ljiljana and Stojanovic, Nenad},
  booktitle={2020 IEEE International Conference on Engineering, Technology and Innovation (ICE/ITMC)}, 
  title={COGNITWIN – Hybrid and Cognitive Digital Twins for the Process Industry}, 
  year={2020},
  volume={},
  number={},
  pages={1-8},
  doi={10.1109/ICE/ITMC49519.2020.9198403}
}

@INPROCEEDINGS{Faruque2021,
  author={Al Faruque, Mohammad Abdullah and Muthirayan, Deepan and Yu, Shih-Yuan and Khargonekar, Pramod P.},
  booktitle={2021 Design, Automation \& Test in Europe Conference \& Exhibition (DATE)}, 
  title={Cognitive Digital Twin for Manufacturing Systems}, 
  year={2021},
  volume={},
  number={},
  pages={440-445},
  doi={10.23919/DATE51398.2021.9474166}
}

@ARTICLE {Ali2021,
author = {M. Intizar Ali and P. Patel and J. G. Breslin and R. Harik and A. Sheth},
journal = {IEEE Intelligent Systems},
title = {Cognitive Digital Twins for Smart Manufacturing},
year = {2021},
volume = {36},
number = {02},
issn = {1941-1294},
pages = {96-100},
doi = {10.1109/MIS.2021.3062437},
publisher = {IEEE Computer Society},
address = {Los Alamitos, CA, USA},
month = {mar}
}

@article{Eirinakis2022, 
title={Cognitive Digital Twins for Resilience in Production: A Conceptual Framework}, volume={13}, ISSN={2078-2489}, url={http://dx.doi.org/10.3390/info13010033}, DOI={10.3390/info13010033}, number={1}, journal={Information}, publisher={MDPI AG}, author={Eirinakis, Pavlos and Lounis, Stavros and Plitsos, Stathis and Arampatzis, George and Kalaboukas, Kostas and Kenda, Klemen and Lu, Jinzhi and Rožanec, Jože M. and Stojanovic, Nenad}, year={2022}, month={Jan}, pages={33} 
}

@INPROCEEDINGS{Eirinakis2020,
  author={Eirinakis, Pavlos and Kalaboukas, Kostas and Lounis, Stavros and Mourtos, Ioannis and Rožanec, Jože M. and Stojanovic, Nenad and Zois, Georgios},
  booktitle={2020 IEEE International Conference on Engineering, Technology and Innovation (ICE/ITMC)}, 
  title={Enhancing Cognition for Digital Twins}, 
  year={2020},
  volume={},
  number={},
  pages={1-7},
  doi={10.1109/ICE/ITMC49519.2020.9198492}
}

@article{DAMICO2022,
title = {Cognitive digital twin: An approach to improve the maintenance management},
journal = {CIRP Journal of Manufacturing Science and Technology},
volume = {38},
pages = {613-630},
year = {2022},
issn = {1755-5817},
doi = {https://doi.org/10.1016/j.cirpj.2022.06.004},
url = {https://www.sciencedirect.com/science/article/pii/S1755581722001158},
author = {Rosario Davide D’Amico and John Ahmet Erkoyuncu and Sri Addepalli and Steve Penver}
}

@article{Jinzhi2022a,
      title = {Exploring the concept of Cognitive Digital Twin from  model-based systems engineering perspective},
      author = {Lu Jinzhi and Yang Zhaorui and Zheng Xiaochen and Wang  Jian and Dimitris, Kiritsis},
      publisher = {SPRINGER LONDON LTD},
      journal = {International Journal Of Advanced Manufacturing  Technology},
      address = {London},
      year = {2022},
      url = {http://infoscience.epfl.ch/record/295469},
      doi = {https://doi.org/10.1007/s00170-022-09610-5},
}

@ARTICLE{Jinzhi2022b,
  author={Lu, Jinzhi and Chen, Dejiu and Wang, Guoxin and Kiritsis, Dimitris and Törngren, Martin},
  journal={IEEE Transactions on Systems, Man, and Cybernetics: Systems}, 
  title={Model-Based Systems Engineering Tool-Chain for Automated Parameter Value Selection}, 
  year={2022},
  volume={52},
  number={4},
  pages={2333-2347},
  doi={10.1109/TSMC.2020.3048821}}

@ARTICLE{Jinzhi2018,
  author={Lu, Jinzhi and Wang, Jiqiang and Chen, Dejiu and Wang, Jian and Törngren, Martin},
  journal={IEEE Access}, 
  title={A Service-Oriented Tool-Chain for Model-Based Systems Engineering of Aero-Engines}, 
  year={2018},
  volume={6},
  number={},
  pages={50443-50458},
  doi={10.1109/ACCESS.2018.2868055}}

@inproceedings{owlreasoners,
title = "A Survey of Current, Stand-alone OWL Reasoners",
author = "Nicolas Matentzoglu and Jared Leo and Valentino Hudhra and Uli Sattler and Bijan Parsia",
year = "2015",
language = "English",
pages = "68--79",
booktitle = "Informal Proceedings of the 4th International Workshop on OWL Reasoner Evaluation (ORE-2015) co-located with the 28th International Workshop on Description Logics (DL 2015), Athens, Greece, June 6, 2015.",
}

@article{Noguchi2020,
author = {Noguchi, Ryan and Wheaton, Marilee and Martin, James},
year = {2020},
month = {07},
pages = {1716-1730},
title = {Digital Engineering Strategy to Enable Enterprise Systems Engineering},
volume = {30},
journal = {INCOSE International Symposium},
doi = {10.1002/j.2334-5837.2020.00815.x}
}

@techreport{DOD2018,
  title={Department of Defense Digital Engineering Strategy},
  author={Office of the Deputy Assistant Secretary of Defense for Systems Engineering},
  year={2018}
}

@article{ExelmansPRVT23,
  author       = {Joeri Exelmans and
                  Jakob Pietron and
                  Alexander Raschke and
                  Hans Vangheluwe and
                  Matthias Tichy},
  title        = {A new versioning approach for collaboration in blended modeling},
  journal      = {J. Comput. Lang.},
  volume       = {76},
  pages        = {101221},
  year         = {2023},
  url          = {https://doi.org/10.1016/j.cola.2023.101221},
  doi          = {10.1016/J.COLA.2023.101221},
  bibsource    = {dblp computer science bibliography, https://dblp.org}
}

@inproceedings{roques2016,
  TITLE = {{MBSE with the ARCADIA Method and the Capella Tool}},
  AUTHOR = {Roques, Pascal},
  URL = {https://hal.science/hal-01258014},
  BOOKTITLE = {{8th European Congress on Embedded Real Time Software and Systems (ERTS 2016)}},
  ADDRESS = {Toulouse, France},
  YEAR = {2016},
  MONTH = Jan,
  HAL_ID = {hal-01258014},
  HAL_VERSION = {v1},
}

@techreport{ibmharmony,
    author = {Hoffmann, Hans-Peter },
    title = {Model-Based Systems Engineering with Rational Rhapsody and Rational Harmony for Systems Engineering},
    institution = {IBM},
    year = {2022}
}
